\documentclass{amsbook}
\usepackage{amsmath,amssymb,graphicx}
\usepackage{amsfonts}
\usepackage{hyperref}
\usepackage{ifthen}
\newtheorem{Theorem}{Theorem}[chapter]

\newtheorem{Lemma}[Theorem]{Lemma}
\newtheorem{Proposition}[Theorem]{Proposition}

\newtheorem{Definition}[Theorem]{Definition}
\newtheorem{rem}[Theorem]{Remark}
\numberwithin{equation}{chapter}
\newtheorem{example}[Theorem]{Example}

\newcommand{\R}{\mathbb R}
\newcommand{\K}{\mathbb{K}}
\newcommand{\N}{\mathbb N}
\newcommand{\Z}{\mathbb Z}
\newcommand{\eps}{\varepsilon}

\newcommand{\C}{\mathcal{C}}
\newcommand{\F}{\mathcal{F}}
\newcommand{\p}{\mathcal{P}}
\newcommand{\A}{\mathcal{A}}
\newcommand{\D}{\mathcal{D}}

\newcommand{\Hol}{\mathcal{H}}

\newcommand{\pol}{\mathfrak{Pol}}
\newcommand{\normal}{\mathcal{N}}
\newcommand{\res}{\hbox{\rm res}}
\newcommand{\tr}{\hbox{\rm tr}}
\newcommand{\mbbc}{\mathbb{C}}
\newcommand{\dd}{\partial}

\newcommand{\mcc}{{\mathcal C}}

\newcommand{\mbc}{{\mathbb C}}

\newcommand{\mcf}{{\mathcal F}}

\begin{document}

\author{Jean-Pierre Magnot}
\title{Contributions to infinite dimensional differential geometry and analysis}

\frontmatter
\maketitle
\tableofcontents

\mainmatter
\chapter*{Acknowledgements}
L'exercice qui consiste \`a remercier les personnes ayant contribu\'e, de pr\`es ou de loin, aux travaux de recherche pr\'esent\'es dans ce m\'emoire, tient de la gageure. En effet, comment ne pas oublier dans un seul document l'une ou l'autre des multiples personnes que j'ai pu croiser, en vingt ans, et plus encore celles qui, non investies en recherche math\'ematique, m'ont aid\'e \`a cro\^itre dans cette activit\'e? Aussi, je prie les personnes que j'oublie ici, malgr\'e l'approche m\'ethodique que je vais avoir, de bien vouloir \^etre indulgentes.

Mes premiers remerciements iront bien entendu aux membres du jury, et en particulier aux rapporteurs, qui ont accept\'e cette t\^ache et l'ont men\'ee avec rigueur, portant des remarques enrichissantes et encourageantes. Chaque membre du jury, \`a sa mani\`ere, a montr\'e son int\'er\^et pour mes travaux et m'a soutenu. Ces quelques lignes, pour certains membres du jury, me semblent bien maigres tant ma gratitude envers eux est grande. 

Je peux \'elargir ce premier groupe restreint en incluant dans mes remerciements tous les chercheurs avec qui j'ai pu interagir, pass\'es et pr\'esents. Je tiens en particulier \`a mentionner les ``diff\'eologues'', non seulement Paul Donato mais aussi Alireza Ahmadi, Patrick Iglesias-Zemmour et Jordan Watts, qui portent des probl\'ematiques \`a mon sens porteuses d'avenir que j'ai d\^u restreindre en terme de place dans ce m\'emoire pour des raisons expliqu\'ees un peu plus loin. Il y a \'egalement des chercheurs tels que Sergio Albeverio, Waldemar Koczkodaj, et quelques uns de leurs collaborateurs: Simona Bernabei, Viera Cernanova, Yuri Kondratiev, Alexandre Kosyak, Gerald Goldin, Eugene Lytvynov, Jiri Mazurek, Bakhrom Omirov, James F. Peters, Ambar Sengupta, Horst Thaler, et tous ceux que j'ai pu \'ecouter ou avec qui j'ai pu \'echanger dans leurs sillages. 

Bien entendu, je dois aussi mentionner Enrique G.Reyes avec qui ma collaboration scientifique très fructueuse s'est doubl\'ee d'une amiti\'e tr\`es sinc\`ere, mais aussi d'autres personnes, comme Akira Asada, Saad Baaj, Augustin Batubenge, Alexander Cardona, Anahita Eslami-Rad, Alessandra Frabetti, Helge Gl\"ockner, Piotr Graczyk, Hong Van Le, Dominique Manchon, Motohico Mulase, Karl-Hermann Neeb, Claude Roger, Steve Rosenberg, qui eurent chacun \`a un moment ou \`a un autre une influence certaine.  Je tiens \'egalement \`a mentionner ceux qui, non encore cit\'es, sont disparus ou ont cess\'e la recherche universitaire, en particulier Paul de Haas,  Atle Hahn, Pierre-Yves Jeanne, Paul Malliavin, Thierry Robart et Krzysztof Wojciechowski.

Je dois \'egalement mentionner mes professeurs, non seulement Sylvie Paycha qui m'a encadr\'e en thèse, mais aussi tous ceux qui ont contribu\'e \`a faire grandir en moi les math\'ematiques. Dans une liste tr\`es incompl\`ete qui contient elle aussi quelques noms du pass\'e, je peux citer Charles Auque, Simone Chevet, Andr\'e Desnoyer, G\'erard Fleury, Labib Haddad, Alex Petitcolin, Le Thahn Phong, Bernard Saramito, mais sans d\'enigrer tous ceux qui, non cit\'es ici, ont \'et\'e mes professeurs \`a un moment ou un autre.     

Je souhaite terminer par les plus importants: les non chercheurs, tous ceux qui m'ont entour\'e positivement, les uns de leur amour, de leur bienveillance, ou de leur amiti\'e, les autres de leur professionalisme. Devant cette immense multitude de personnes de bonne volont\'e mettant le prochain au centre de leurs pr\'eoccupations, je ne peux que citer les plus proches d'entre eux, mon \'epouse et mes enfants. 
 
\chapter*{Presentation}
My research works are centered on infinite dimensional differential geometry and analysis, including its possible ramifications. There are many ways to get an infinite dimensional object: algebraic (e.g. by various -infinite- series), via analysis (e.g. by topological completion) or via probability (e.g. via cylindrical measures). These approaches all have applications. Mostly, function spaces, their duals and operator algebras furnish the basic tools to define geometric objects. In this field of research, motivating examples are themselves carried from applied problems, in particular the theory of (nonlinear) partial differential equations (PDEs) or the first and second quantizations in quantum physics.
For example, algebraic infinite dimensional groups apply in the theory of integrable systems and symmetries, topological completion is a key tool in numerical methods for solving partial differential equations, and the probabilistic approach of partial differential equations relies in stochastic PDEs while the Feynman-Kac formula expresses a solution of the Schr\"odinger equation in terms of an integral over an infinite dimensional space.  Other numerous examples in analysis and geometry, such as the meaning of the renormalized tangent bundle in the stochatstic geometry on the path space and the correspondence between classical, statistical, first and second quantized mechanics raise difficult and still unsolved conceptual or technical problems which cannot be all considered in only one lifetime.

Even more, in any of the suitable approaches for infinite dimensional analysis or geometry, the intuitions modeled from finite dimensional examples have served in early times to generalize by the same approaches the finite dimensional objects to infinite dimensional settings, and have failed quite quickly on purely infinite dimensional effects.   
Indeed, infinite dimensional objects carry so many technical difficulties that one often needs to bypass direct generalization of finite dimensional techniques in order to get efficient results. One famous example is Kuiper's contractibility theorem \cite{Kui}, which implies that any Hilbert vector bundle is parallelizable, while Fredholm bundles can be non trivial. Again actually, Kuiper's result  still represents  in the opinion of some non-specialists (who are fewer an  fewer, very fortunately) the ``proof'' that infinite dimensional geometry is purely formal, since no obstruction can occur, whereas it only indicates that the setting of Hilbert manifolds cannot catch and express all the possible ``infinite dimensional features''. 

The problem of the setting is a big problem, and is not solved yet. This leads to necessary heuristic parts in applications. This is in particular true in physics. These heuristic parts need not to be understood as failures but as open questions, just like Feynman integrals which produced numerous results before being rigorously defined in an unified way.        
We expose these problems and considerations in the first chapter of this text, which has to be understood as a necessary introduction to the spirit and the settings in which I have worked during all these years. 

\vskip 6pt
In a so wide range of open questions and unknowns, I have then to more precise where my efforts have been concentrated. Instead of trying to define and develop general settings, in the spirit of Omori \cite{Om}, Neeb \cite{Neeb2007},  Kriegl and Michor \cite{KM} or Chen, Souriau and Iglesias \cite{Sou,Igdiff}, since I did not (and still do not) have their deep insight, I tried to consider only restricted examples and problems on which I feel that I understand clearly the importance and the key features that make them important. This lead me naturally to the following open questions:
\begin{enumerate}
	\item The Ambrose-Singer theorem \cite{AS,Li} describes the Lie algebra of the holonomy group of a connection. In terms of integrable systems, it shows 
	\begin{itemize}
		\item if a system is formally integrable (which is equivalent to a zero curvature condition), then it is globally integrable in a convex domain (ie there exists a global solution) and well-posed (i.e. with continuous or smooth dependence on the initial conditions). 
		\item if a system is not formally integrable, it measures the obstruction for obtaining a global  solution.
	\end{itemize}
Moreover, the finite dimensional Ambrose-Singer theorem is (almost) equivalent to the Frobenius theorem. For all these reasons, it was important to extend it to infinite dimensional settings as far as possible.
\item From the problem of the Ambrose Singer theorem, we get the problem of the existence of an exponential map on an infinite dimensional Lie group, as a necessary tool for the construction of the holonomy group. The question fo the existence of an exponential map is also crucial in many infinite dimensional linear first order equations. In the existing literature, authors often wanted to show that infinite dimensional Lie groups have an exponential map, which we did too in some examples, but no-one showed, to our knowledge, that the exponential maps do not exist in some examples of interest, which we performed  for two examples.
\item Neighbor to the first two points we can find:
\begin{itemize}
	\item the theory of geometric invariants for infinite dimensional principal bundles, which was the central topic of my PhD thesis, that tries to show that some bundles are not trivial. The first known example of such invariant comes from the index of Fredholm operators, which I adapted (mostly in my PhD thesis) on loop groups by the use of spectral properties and so-called renormalized traces of pseudo-differential operators.  These approaches are motivated by the study of a class of line bundles called determinant bundles, in heuristic link with the theory of Feynman integrals. It was then natural to extend the search for geometric invariants to other infinite dimensional Lie groups and principal bundles.
	\item The study of the Kadomtsev-Petviashvili (KP) hierarchy, an infinite dimensional integrable system which is not only derived from the so-called KP equation (also called 2d-KdV equation), and not only a hierarchy of equations from which we can derive integrals of the motion and special solutions for KP, KdV, Gelfand-Dickey, Boussinesq equations, but which is also linked with the determinant bundles already mentioned, and from more recent results has connexions with Hodge theory and combinatorics of triangulations.  
	
	The integration of the KP hierarchy was purely algebraic, based on formal Lie groups on which a mild application of the theory of r-matrices was the more serious way to prove existence and uniqueness of the solutions while a heuristic link with an Ambrose-Singer like approach was already mentioned in the literature. It was then natural to try to clarify the geometric aspects of this algebraic integration, both in the classical settings and in some generalized versions of this hierarchy, concentrating my efforts on differential geometry related to the system with in mind the problem of well-posedness.
	\item The theory of symmetries of PDEs, well developed for so-called projectable symmetries, i.e. symmetries inherited from the finite dimensional space on which the functions, solutions of the PDE, are defined. Symmetries and their dual counterpart, the integrals of the motion, basically, play an important role in solving difficult equations by reducing the set of possible solutions, or deducing a class of solutions from a trivial one. But actually, while the theory of symmetries for (full, strong) solutions of PDEs is well-developped, the same approach cannot be performed for weak solutions, which appear non-trivially in the context of equations of hydrodynamic type. It was then natural to propose a geometric framework for the study of weak solutions. 
\end{itemize}  
\item In the three last points, one important class of operators is the class of pseudodifferential operators (PDOs), formal or non-formal. Another class of operators of interest remain on groups of diffeomorphisms. In these two classes of operators, and their ``generalizations'', we find a central topic, which then has to be studied in its own right. I developed groups of Fourier-integral operators, called $Diff-$pseudodifferential operators, that gather these two classes of examples, and studied their geometric properties. 
\item Finally, I also considered one side problem to Feynman's integral formula: the existence (or not) of an infinite dimensional Lebesgue measure and its normalizations. It was already knwon that such a ``measure'' did not have well-suited properties for an efficient use in quantum physics, then I proposed two approaches:
\begin{itemize} 
\item define a class of infinite dimensional integrals which mimick the Monte-carlo method, obtaining means instead of measures. Generalizing this approach by replacing the Dirac measures of the Monte-Carlo method by suitable sequences of probabilities, I got a notmalized infinite dimensional Lebesque-like mean.
\item considering more precisely the space of connections of a principal bundle and the Whitney discretization through triangulations, which classically defines infinite dimensional integrals, I proposed to discretize connections through their holonomy instead, in the spirit of quantum gravity approaches. This lead to unexpected developments in decision theory.    
\end{itemize}
\end{enumerate}

{\it In view of the various topics present in this text, I feel necessary to give more precise background, with a selection of relevant references, in (in principle short) introductory parts of each chapter, for the ease of the reader and for better reading of separate chapters.} 

\section*{Some notes on non-presented works}

Cited but non-presented, my published works developed for my PhD thesis are under the references \cite{CDMP,Ma2003,Ma2006}.  Due to the late publication of \cite{Ma2006} (4 years after the completion of the PhD thesis), there is one application of the presented article \cite{Ma2004}, developed just after the thesis, which is mentioned in \cite{Ma2006}. A short description of this application is given in the introductory notes of Chapter 6.

There are also three other classes of works which I do not wish to present.
\begin{itemize}
	\item Prepublished but unpublished works \cite{Ma2015-3,MR2020,Ma2021-2}: Some of them can be discussed in the ``open problems'' disseminated in the text, but they are not strictly speaking well-established works since not yet published after a peer-review process.
	\item Works in topics related to applications of mathematics, that contain (to my opinion) not enough mathematical background to be suitable for a habilitation thesis in mathematics \cite{Ma2017-1,Ma2018-0,SSCM}
	\item Works on diffeologies and Fr\"olicher spaces \cite{Ma2006-3,Ma2015,Ma2015-1, Ma2017-0,MW2016,Ma2015-2}. These two settings for generalized differential geometry and related topics consist in novel and easy approaches of differential geometry without atlas. They have been developed since the 80's but the actual envolved mathematical community is mostly specialized on their theoretical aspects, still under development. My own contributions on diffeologies and Fr\"olicher spaces are poor compared with those of full specialists. Therefore, I only keep for this exposition results where the envolved diffeologies and Fr\"olicher spaces are easy to understand for non-specialists (i.e. encode ``natural differentiation'').
\end{itemize}
\section*{Organization of the habilitation thesis}

The presented works are referenced as the items from  \cite{Ma2004} to \cite{MRu2021-1} in the bibliography. Results are re-organized according to selected topics of interest, therefore, one article can be cited at several places in the exposition. As announced, Chapter one is dedicated to methodological and preliminary considerations.

\vskip 12pt
$\bullet$ \underline{\bf \it Contents of Chapter 2}

The intuitive idea of the formal implicit function theorem present in my PhD thesis was that, in the ILB setting \cite{Om} for the Nash-Moser inverse functions theorem (see e.g. \cite{KM,Om}), the additional uniform norm estimates assumed in the hypothesis could be replaced by mild considerations on the functions considered. At this time, 
differentiation on a non open domain (which is not a sub submanifold), along the lines of e.g. \cite{KM}, was too exotic for me. Many years later a groundbreaking idea came: the control on the domain of the implicit functions must be deduced from the framework. This lead to an infinite dimensional implicit functions theorem with additional norm estimates \cite{Ma2020-3} and some of its consequences: the related inverse functions theorem \cite{Ma2020-3} and the related Frobenius theorem \cite{Ma2016-4}. 

This re-examination of the classical proofs of inverse theorems came in a more general context: the global framework of numerical schemes that are used in the field of numerical analysis in order to determine approximate solutions of functional equations, and in particular PDEs. The notion of well-posedness is generalized to the notion of smooth differential schemes \cite{Ma2016-4} where ``smoothness'' is considered in a generalized context adapted to the problem and where initial conditions are only part of the data on which smootness is established. Natural definitions for symmetries of weak solutions are also given. In the same reference, the (classical) finite elements method for the Dirichlet problem is shown to be smooth on the border function and on the sequence of triangulations chosen, in a way where smooth dependence of both exact and approximated solutions is obtained.

\noindent
\vskip 12pt
$\bullet$ \underline{\bf \it Contents of Chapter 3}
It presents results in the direction of integration theory, in particular integration on infinite dimensional spaces. I gave the definition of normalized infinite dimensional integrals as means which are limits of Dirac measures \cite{MaICM,Ma2016-3} as well as the existence of a non-trivial (normalized) Lebesgue mean which differs from the existing ones \cite{Ma2016-3}.
With this approach the main results presented is the following:
On any Hilbert space $H$, there exists a translation invariant, scale invariant mean, defined on a domain which contains cylindrical functions, called Lebesgue mean. 
	
$\bullet$ \underline{\bf \it Contents of Chapter 4}
An infinite dimensional group is called regular if it admits an exponential map. The theory of regularity of infinite dimensional Lie groups gathered the efforts of top researchers since the 70's till now, from Omori, Milnor and Ratiu to Neeb and Michor among others. My contributions in this field are more humble. 
I showed non-regularity of the (diffeological Lie) group of diffeomorphisms of an open manifold equipped with the $C^\infty-$compact open topology \cite{Ma2013-2}. With E.Reyes \cite{MR2016}, we showed non-regularity of the (locally convex) group of invertible elements of $\mathbb{R}((X))$, and as a consequence, non regularity of the group of invertible (maybe unbounded) formal pseudo-differential operators.

Then I describe the groups of operators generated by pseudo-differential operators (formal or non formal, bounded or unbounded) and by diffeomorphisms. Following \cite{Ma2016-1,Ma2021-1} I investigate their relationship with the restricted linear group group $GL_{res}$ \cite{PS} as well as their intrinsic structure. More precisely, I showed regularity (i.e. existence and smoothness of the exponential maps) of groups of bounded  $Diff(M)-$pseudodifferential operators, when $M$ is a boundaryless compact manifold \cite{Ma2016-1}. They are subgroups of $GL_{res}$ (defined along the lines of \cite{PS}) when $M=S^1$ and considering only orientation preserving diffeomorphisms and bounded PDOs, but they are not Lie subgroups of $GL_{res}$ \cite{Ma2021-1}.

Finally, the need of an infinite dimensional Ambrose-Singer theorem raised many times in the literature, see e.g. \cite{F1,F2,Pen}. A first result came in \cite{Vas1978} assuming strong restrictions that enable the application of a global Banach Frobenius theorem on a Banach principal bundle, while \cite{KM} showed an analogous result in the $c^\infty$ setting for flat connections and regular structure groups. The last result suggested that only  integrability of the holonomy Lie algebra is necessary to prove the Ambrose-Singer theorem. This is the result that I got first in the context of Fr\'echet principal bundles in \cite{Ma2004}, and it was completed in \cite{Ma2013} in order to consider more general contexts. In these works, the holonomy algebra is spanned by curvature elements (Ambrose-Singer theorem) but the word "spanned" depends on the category of Lie groups considered, while the holonomy group is the smallest group to which a reduction theorem of the prescribed connection applies. These groups can be considered intuitively as ``completions'' of the holonomy group classically generated by horizontal paths.

\noindent
\vskip 12pt
$\bullet$ \underline{\bf \it Contents of Chapter 5}
The quantum gravity approach of Yang-Mills action functional considers holonomies along the edges of a triangulation instead of connections. Curvatures read as holonomies of a loop. I precised in \cite{Ma2018-3} the formal description present in e.g. \cite{RV2014} by giving a rigorous procedure to discretize a connection along its holonomies, and made some remarks on possible integrals on the space of connections and on the expression of the Yang-Mills action functional. This work still has to be developed. 

During the first investigations that led to \cite{Ma2018-3} I have been invited to participate to \cite{Ma2017-1,Ma2017-2} to precise technical (i.e. mathematical) aspects, work where I discovered operations research and decision theory through the lights of pairwise comparisons. The first evidence to me was that pairwise comparisons coincide exactly with the holonomy of paths in quantum gravity, inconsistency coincides with non-trivial curvature, and some so-called inconsistency indicators are Yang-Mills type functionals. These remarks, announced in \cite{Ma2018-1} have developed to the globalized work \cite{Ma2016-5} where (quantum gravity) Yang-Mills fields appear as an interdisciplinary topic. In this work, I generalize pairwise comparisons to coefficients in a group, and I describe basic mathematical aspects of this framework.
\noindent
\vskip 12pt
$\bullet$ \underline{\bf \it Contents of Chapter 6}
Algebras of non-formal classical (maybe unbounded) pseudo-differential operators (PDOs) carry a family of linear functionals, called ($\zeta-$)renormalized traces, that extend the trace of trace-class operators. These traces are present in the Radul cocycle, which stands formally as a non formal realization of the Kravchenko-Khesin cocycle. After clarifying the relations of these cocycles with the Schwinger cocycle \cite{CDMP}, the sign of the Dirac operator over $S^1$ appears as a way to describe a rigorous correnspondence between these cocyles, first for multiplication operators \cite{Ma2003}, for unbounded classical PDOs \cite{Ma2006-2}, and for non-classical PDOs \cite{Ma2008}. 

Groups of bounded classical PDOs can serve as a structure group for frame bundles over manifolds of smooth maps \cite{CDMP,Ma2006} while groups of $Diff(M)-$PDOs,  appear naturally in the geometry of  spaces of non-parametrized embeddings \cite{Ma2016-1}. Both of them enable a non-trivial Chern-Weil theory \cite{Ma2016-1,Ma2021-1} and the second class of examples enjoy surprising properties enhancing renormalized traces \cite{Ma2021-1}.

When considering unbounded operators, the same constructions are valid and groups of bounded  $Diff(M)-$pseudodifferential operators are generalized Lie groups. On these groups for $M = S^1,$ $\zeta$-renormalized traces enable to generalize the formula of the Hermitian metric of groups of matrices $(a,b) \mapsto tr(ab^*)$ to get pseudo-Hermitian metrics on groups of $Diff(S^1)-$pseudodifferential operators. These metrics have interesting properties, such as the existence of pseudo-Hermitian connections with curvature with values in smoothing operators. For these metrics, multiplication operators and vector fields are isotropic, and for one of these ``smoothing'' connections, the first Chern from is in the cohomology class of the Schwinger cocycle \cite{Ma2021-1}. 
\noindent
\vskip 12pt
$\bullet$ \underline{\bf \it Contents of Chapter 7}
My works concerned two types of KP hierarchies: the standard one, and a new \textit{deformed} KP hierarchy, obtained from the usual one by an adequate time scaling. In the second one, I first showed how the Ambrose-Singer theorem  fits with formal integrability \cite{Ma2013}. Then, the same approach is applied  in a colloboration with E.G. Reyes and A.Eslami Rad \cite{ERMR2017,MR2016}, where the main result states well-posedness of KP hierarchy by re-analyzing Mulase's bicross product of groups. In \cite{ERMR2017}, more abstract algebras of operators are considered for ``generalized'' KP hierarchies which are shown to be well-posed. Part of these results is summarized in \cite{MR2019}. Finally, in \cite{MRu2021-1} that presents the results of a collaboration with V. Rubtsov, we extend the KP hierarchy by the study of a new integrable almost complex structure derived from the sign of the Dirac operator over $S^1.$ The same results of well-posedness are stated in most cases considered. 

\chapter{Some preliminary considerations}
\section{Mathematical physics: heuristics as a dialectic approach}
Mathematical physics often appear as a hybrid topic between two distinct disciplines: physics and mathematics, which have their own motivations and ways of thinking. 
\begin{itemize}
	\item \underline{Physics} intend to describe the physical world, and to predict physical effects. Most quantities have ther significance only on one part of the (still non unified) theories, and mathematical concepts appear as a language to encode physics in modelizations. This modelling is sometimes based on rigorous mathematical constructions. In that case, the consequences of the mathematical properties of the model enable to test its accuracy through experiments. When modelling is not rigorously based, it intends to describe heuristically a physical realm. The heuristical computation rules are then confirmed by the experiments. 
	\item \underline{Mathematics} are based on a chosen logical framework, described by a coherent finite system of axioms. These axioms intend to encode facts which are obviously true. Heuristic considerations lead to conjectures, which have to be proved to be true or false by logical arguments. By G\"odel's incompleteness theorem, a finite set of axioms cannot make decidable all possible statements. Moreover, some axioms can be questioned. One of the most famous examples of such a questioned statement remains the axiom of choice. Other controversies appear on the accuracy of non standard analysis.
\end{itemize}
Let us illustrate now how these different ways of thinking arise in three examples. 
\begin{enumerate}
	\item The set of rational numbers $\mathbb{Q},$ considered as describing the real world by e.g. Plato, has been completed to $\mathbb{R}$ and to $p-$adic fields. Concentrating on $\R,$ it is basically represented by the ``real line'' which looks like a ``physical'' line. Real numbers encode classical mechanics. However, 
	\begin{itemize}
		\item in mathematics, one can give a name (i.e. a denomination with a word made of a finite sequence of letters in a finite alphabet) to only a countable set of real numbers,
		\item in physics, the Heisenberg uncertainty principle and quantum gravity theory (see e.g \cite{RV2014})seem to contradict the space time continuum hypothesis at a critical scale, in terms of modelization of observables in the phase space.
		\item besides these restrictions, non-standard analysis has itself enlarged real numbers with infinitesimals with success in simplifying some proofs in mathematical analysis and in giving rigorous descriptions of standard objects, see e.g. \cite{AFKHKL1986}.  
	\end{itemize} 
\item The Adler symplectic structure on formal pseudo-differential operators $\mcf Cl(\R,\mbc)$ is derived from the Adler residue \cite{Adl} $$ res : \sum_{n < N} a_n \partial^n \in \mcf Cl(\R,\mbc) \mapsto \int a_{-1}$$
is not well-defined for any smooth partial symbol $a_{-1} \in C^\infty(\R,\mbc)$ but only for integrable ones. This integral is then a formal integral, with classical properties of the (well-defined, Riemann) integral. This formula becomes rigorous changing $\R$ to $S^1,$ and the Wodzicki residue \cite{W} is often proposed as an extension of the Adler residue for pseudo-differential operators acting on smooth functions on a smooth compact boundaryless manifold $M,$ while this generalization is not straightforward even for $M= S^1$ as analyzed in \cite{MRu2021-1}.  

\item The Feynman-Kac integral
$$ f \mapsto \frac{1}{Z(S)}\int_H e^{-iS} f \mathcal{D}\lambda$$
is certainly one of the most famous heuristic formulas in physics from the twentieth century which leads to various non-equivalent rigourous definitions, see e.g. \cite{AHM,F2017}. In this expression:
\begin{itemize}
	\item $\mathcal{D}\lambda$ is a ``heuristic'' (translation invariant) infinite dimensional measure on the Hilbert space $H.$ Such a ``product measure'' exists, but has very few measurable sets \cite{Ba1,Ba2} with finite non-zero measure.  
	\item The ``normalization term'' $Z(S) = \int_H s^{-iS} \mathcal{D}\lambda$ stands as the (heuristic) total mass of $H.$ 
	\item The total formula stands as a mean value formula, and is understood this way for heuristic calculations in mathematical physics, see e.g. \cite{KW} for an example leading to knot invariants.
\end{itemize}  
\end{enumerate}

In these chosen examples, there does not appear an opposition, but a dialectic, between heuristics and mathematical rigor. The minimal requirement in mathematics is to identify where heuristics are, and one difficult task of the mathematician is to get a way to describe rigourously what heuristics describe brightfully. 

\section{From Hilbert manifolds to diffeological spaces and Gelfand's formal geometry}
The description of spaces of mappings between two $C^\infty,$ finite dimensional manifolds $M$ and $N$ is known since Riemann's inaugural lecture, and the notion of infinite dimensional manifold has been developped actually around the most proeminent examples arising from physics:

\begin{itemize}
	\item mapping spaces, jets spaces etc...
	
	\item manifolds of metrics
	
\end{itemize} 
and concerning infinite dimensional Lie groups:
\begin{itemize}
	\item loop and current groups
	
	\item groups of diffeomorphisms
	
	\item groups of symmetries (of ODEs, PDEs etc...)
	
	\item groups of operators (bounded or unbounded), groups of the units of algebras.
\end{itemize}

 The tentatives to generalize the finite dimensional settings have been developped starting from the most easy settings, that is, the settings nearest from the finite dimensions. There are two:
 
 - direct limits of finite dimensional manifolds. This approach is useful mostly in probability theory \cite{Bog}, in frameworks linked with cylindrical measures (e.g. Wiener measure) and also linked with the theory of Fredholm operators (see e.g. \cite{Kub2020} for a basic review on spectral theory), since $\varinjlim_{n \in \N^*} M_n(\R)$ is a dense subgroup of the ideal of compact operators acting on a real Hilbert space.
 
 - Strong Hilbert manifolds, which carry structures that are very similar to finite dimensional manifolds: Riemannian metrics, existence of smooth partitions of the unit, existence of Gaussian measures among others. 
 
 However, these "nice structures" fail to describe the detailed structure of many objects. For example, a strong Hilbert manifold is parallelizable \cite{Kui}. Another example relies on the group of diffeomorphisms of a compact boundaryless manifold which does not carry any strong Riemannian metric as a ILH Lie group \cite{Om}. The same way, groups of bounded operators on a Hilbert space carry a natural Banach norm, but no such natural Hilbert norm. Thus, the first "nice" settings leads to consider also 
 \begin{itemize}
 	\item Banach manifolds, which e.g. no longer carry strong Riemannian metrics, nor $C^\infty$ partition of the unit \cite{Bou}
 	\item Fr\'echet manifolds, which carry no "canonical" structure Lie group, on which many authors consider additional structures of projective limits of Banach or Hilbert spaces \cite{Om}.
 \end{itemize} 

As one can see we have a classical logical effect: less properties enable to deal with more examples, but with less technical abilities. Therefore, one can investigate two opposite but complementary viewpoints:
\begin{itemize}
	\item one considers only Fr\'echet spaces which have a prescribed structure projective limit of Banach or Hilbert spaces. This idea was first developped by Omori \cite{Om} motivated by the example of groups of diffeomorphisms, and has been recently re-investigated in \cite{DGV}. With this approach, one can deal, more or less superficially for applications, with the most popular examples of infinite dimensional manifolds arising in mathematical physics: mapping spaces, groups of diffeomorphisms, spaces of exterior forms and connections etc. over a compact boundaryless manifold. One can refer to \cite{KW} for an overview of classical applications in this field. 
	\item One can also generalize the framework once again in order to be able to consider all necessary objects such as e.g. the bundle of frames.  Indeed, analyzing deeply the works \cite{GV,DGV} for generalized frame bundles as well as in \cite{Om} for so-called generalized Lie groups, there appears topological Lie groups, where one can differentiate, but which carry no atlas. Let us mention that, apparently, this is not, in these two cases, a lack of knowledge to build an adequate atlas, but that this is impossible to find one. The same question rose in \cite{Ma2006-3} (work not presented in this habilitation) which led us to consider diffeological spaces \cite{Sou,IgPhD,Igdiff}. The question of a "ground" minimal setting for infinite dimensional geometry is then raised, and our "candidate" is diffeological spaces, as in \cite{Neeb2007} while many generalizations of the notion of differentiability have been developped (for a sample non-exhaustive list with references, see \cite[page 243]{BS2017}). We quote the opinion given by Neeb in \cite{Neeb2007}: diffeological spaces are, by some aspects, too general to be useful in any case, especially for applications, and very often one needs stronger settings. On one hand, this is completely true from the viewpoint of K.H. Neeb's work, who intends to develop results about pure settings for geometry and topology of infinite dimensional groups. On the other hand, following an approach represented in our references by the book of Khesin and Wendt \cite{KW}, one can focuse on examples and applications ignoring sometimes some technicalities in their description, refering if necessary to Gelfand's formal geometry \cite{GK1971}.   
\end{itemize} 
The choice of the adequate minimal setting for differentiability, and hence for differential geometry is the reason for the study \cite{BH}, where various generalized settings for differential geometry are considered. 
We have to recognize honestly that the use of a setting such as diffeologies appears as necessary in order to work safely in some circumstences. From this viewpoint, diffeological spaces offer a very easy-to-use setting, coherent with classical differential objects. The category of diffeological spaces contains all the other categories described before which are sub-categories. In the rest of the text, the setting considered will change, depending on the "strongest" setting which enables to deal with a situation or another. Even if our opinion is much more positive about diffeologies, maybe because diffeologies already helped us to deal with geometric objects, we agree with \cite{Neeb2007} that there is no adequate universal setting for such objects, at least at the actual state of knowledge. We recall the "tower" of categories in which we shall work: 
$$\begin{array}{c}
\hbox{finite dimensional manifolds \cite{KN,Olv}} \\
\Downarrow\\
\hbox{Hilbert manifolds \cite{Bou,Lang}} \\
\Downarrow\\
\hbox{Banach manifolds \cite{Bou}} \\
\Downarrow\\
\hbox{Fr\'echet manifolds, ILB manifolds \cite{DGV,Om} }\\
\Downarrow\\
\hbox{locally convex manifolds (not necessarily locally complete) \cite[section I.1]{Neeb2007}} \\
\Downarrow\\
c^\infty\hbox{manifolds \cite{KM}} \\
\Downarrow\\
\hbox{Fr\"olicher spaces \cite{FK,KM}} \\
\Downarrow\\
\hbox{diffeological spaces \cite{Igdiff}} 
\end{array}
$$
The two last settings are actually the less known (and the most general). They do not assume the presence of an atlas. This is exactly the same idea as Gelfand's formal geometry \cite{GK1971}, or so-called ``smoothness in non-open domains'' in \cite{KM}: consider "reasonable" smooth objects in difficult settings. Our choice for diffeologies and Fr\"olicher spaces is also the choice of works such as \cite{Can2015,Can2020} in quantum physics and \cite{We2017} in shape analysis.

\vskip 12pt

\textbf{In the presented works hereafter, diffeologies and Fr\"olicher spaces appear in parts of the statements and results. They are here only technical tools, used to talk rigorously about smooth objects such as Fr\"olicher algebras (with smooth operations and inversion), diffeological and Fr\"olicher Lie groups (with smooth multiplication, inversion, and Lie algebra) etc. but nowhere  the presented work carry novelty about these theoretical settings. They are only considered as a necessary technical features to get rigorous statements and proofs.}

\vskip 12pt
In view of the relative novelty of these two last notions, we feel the need to recall basic definitions for the reader who is a beginner in diffeologies.  We begin with the notion of a diffeological space:

\begin{Definition} Let $X$ be a set.
	
	\noindent $\bullet$ A \textbf{p-parametrization} of dimension $p$ 
	on $X$ is a map from an open subset $O$ of $\R^{p}$ to $X$.
	
	\noindent $\bullet$ A \textbf{diffeology} on $X$ is a set $\p$
	of parametrizations on $X$ such that:
	
	- For each $p\in\N$, any constant map $\R^{p}\rightarrow X$ is in $\p$;
	
	- For each arbitrary set of indexes $I$ and family $\{f_{i}:O_{i}\rightarrow X\}_{i\in I}$
	of compatible maps that extend to a map $f:\bigcup_{i\in I}O_{i}\rightarrow X$,
	if $\{f_{i}:O_{i}\rightarrow X\}_{i\in I}\subset\p$, then $f\in\p$.
	
	- For each $f\in\p$, $f : O\subset\R^{p} \rightarrow X$, and $g : O' \subset \R^{q} \rightarrow O$, 
	in which $g$ is  
	a smooth map (in the usual sense) from an open set $O' \subset \R^{q}$ to $O$, we have $f\circ g\in\p$.
	
	\vskip 6pt If $\p$ is a diffeology on $X$, then $(X,\p)$ is
	called a \textbf{diffeological space} and, if $(X,\p)$ and $(X',\p')$ are two diffeological spaces, 
	a map $f:X\rightarrow X'$ is \textbf{smooth} if and only if $f\circ\p\subset\p'$. 
\end{Definition} 

The notion of a diffeological space is due to J.M. Souriau, see \cite{Sou}, and \cite{Igdiff} for a contemporary point of view. Of particular interest 
to us is the following subcategory of the category of diffeological spaces. 

\begin{Definition} 
	A \textbf{Fr\"olicher} space is a triple $(X,\F,\C)$ such that
	
	- $\C$ is a set of paths $\R\rightarrow X$,
	
	- $\F$ is the set of functions from $X$ to $\R$, such that a function
	$f:X\rightarrow\R$ is in $\F$ if and only if for any
	$c\in\C$, $f\circ c\in C^{\infty}(\R,\R)$;
	
	- A path $c:\R\rightarrow X$ is in $\C$ (i.e. is a \textbf{contour})
	if and only if for any $f\in\F$, $f\circ c\in C^{\infty}(\R,\R)$.
	
	\vskip 5pt If $(X,\F,\C)$ and $(X',\F',\C ')$ are two
	Fr\"olicher spaces, a map $f:X\rightarrow X'$ is \textbf{smooth}
	if and only if $\F'\circ f\circ\C\subset C^{\infty}(\R,\R)$.
\end{Definition}

This definition first appeared in \cite{FK}; we use terminology 
borrowed from Kriegl and Michor's book \cite{KM}.
A short comparison of the notions of diffeological and Fr\"olicher spaces is in \cite{Ma2006-3}; the reader can 
also see \cite{Ma2013,MR2016,Wa} for extended, but not so long, expositions.
In particular, it is explained in \cite{MR2016} that 
{\em Fr\"olicher and Gateaux smoothness are the same notion if we 
	restrict ourselves to a Fr\'echet context.} We refer to \cite{MR2019} for an overview of the necessary notions in this document. 

\chapter{On infinite dimensional inverse theorems and numerical schemes}
In order to prove the existence of charts and atlases on submanifolds, one of the main tools rely on the well-known implicit functions theorem, for submanifolds defined by an implicit condition of the type $F(x)=0.$ Such equations arise in almost all fields of mathematics and specially in the class of partial differential equations (PDE). Both for implicit functions and for PDEs, many proofs rely on a contraction principle, or on the use of compactness, and in almost any situation, on the construction of a sequence that converges to a solution of the equation considered. Let us precise a little more: 
  
  \begin{itemize}
  	\item Results that are called inverse theorems, namely the inverse function theorem, the implicit functions theorem, the Frobenius theorem, are equivalent statements in Banach spaces assuming presence of topological complements to closed vector subspaces, see \cite{Pen} (for example, on reflexive Banach spaces). In this context, the essential proofs are based on a contractive map which build up the solution as a fixed point. The control of  solution(s) is then given by additional estimates on a differential. Estimates are also crucial in the Nash-Moser theorem \cite{Ham} which serves in the proof of the existence and uniqueness of several PDEs, and stands as motivating point for the construction of ILB manifolds \cite{Om} which includes diffeomorphisms groups.  
  	\item Among the numerous methods for solving PDEs, principally because of our own lack of knowledge in a wider range of methods, we concentrated on the classical finite elements method which relies on a \emph{fixed} triangulation, a covering of the domain considered by simplexes isomorphic to  \begin{equation} \label{embtriangulation}\Delta_n= \left\{(x_0,...,x_n)\in \R_+^{n+1} | x_0 +... +x_n = 1\right\}
  	\end{equation} and refining the triangulation by triangulating each copy of $\Delta_n,$ the sequence of approximate solutions converge to the exact solution of the PDE considered, see e.g. \cite{BS1994}. 
  \end{itemize}
In our contributions to this very wide topic, we will analyse firstly  a first order finite elements method (Galerkin method), and an ILB implicit functions theorem with weakened hypothesis.

\section{On the space of triangulations of a manifold, based on \cite{Ma2016-2,Ma2016-4}}

As we just mentioned, the base triangulation is always fixed in numerical methods. However, one can precise better the differential structures that remain implicit (and assumed natural) in many geometric applications of triangulation, see e.g. \cite{Wh,Dup1978} and primarily exhibited in \cite{CW2014-2}. Nevertheless, we will expose here how the space of triangulations a smooth manifold $M$ carries itself a Fr\"olicher structure. Part of the presented results have been recently performed independently by other authors in \cite{HL2020}, from a perspective based on numerical analysis and hence less geometric.  
We wish to highlight the smooth structures inherited from the space of triangulations of a smooth manifold, and consider the refinement map of a mesh of triangulations as a smooth map on the space os triangulations. This is exposed in \cite{Ma2016-4} as a final version of this work, with preliminary results given in \cite{Ma2016-2}.
For this, we need to refine very basic notions, for a fixed $n-$dimensional manifold $M$ possibly with boundary.

\begin{Definition} \label{smtriang} A \textbf{smooth triangulation} of $M$ is a family $\tau = (\tau_i)_{i \in I}$
	where $I \subset \N$ is a set of indexes, finite or infinite, each $\tau_i$ is a smooth map $\Delta_n \rightarrow M,$ and such that:
	\begin{enumerate}
		\item $\forall i \in I, \tau_i$ is a (smooth) embedding, i.e. a smooth injective map such that {{}$(\tau_i)_*\left(\p({\Delta_n})\right)$} is also the subset diffeology of $\tau_i(\Delta_n)$ as a subset of $M.$
		\item $\bigcup_{i \in I }\tau_i(\Delta_n) = M.$ (covering)
		\item $\forall (i,j) \in I^2,$ $\tau_i(\Delta_n) \cap \tau_j(\Delta_n) \subset  \tau_i(\partial \Delta_n) \cap \tau_j(\partial \Delta_n).$ (intersection along the borders)
		\item  $\forall (i,j) \in I^2$ such that $ D_{i,j}=\tau_i(\Delta_n) \cap \tau_j(\Delta_n) \neq \emptyset,$ for each $(n-1)$-face $F$ of $D_{i,j},$ the ``transition maps" $``\tau_j^{-1} \circ \tau_i'' : \tau_i^{-1}(F) \rightarrow \tau_j^{-1}(F)$ are affine maps.  
	\end{enumerate} 
\end{Definition} 
Under these  conditions, we equip the triangulated manifold $(M,\tau)$ with a Fr\"olicher structure $(\F_I,\mcc_I),$ generated by the smooth maps $\tau_i$ that encode the natural differentiation inherited from the maybe infinite cartesian product $C^\infty(\Delta_n,M)^I.$
 Maps in $\F_I$ can be intuitively identified as some piecewise smooth maps $M \rightarrow \R,$ which are of class $C^0$ along the $(n-1)-$skeleton of the triangulation.
	We have proved also that $\mcc_I \subset \p_\infty(M).$ Some characteristic elements of $\mcc_I$ can be understood as paths which are smooth (in the classical sense) on the interiors of the domains of the simplexes of the triangulation, and that fulfill some more restrictive conditions while crossing transerversaly the $(n-1)-$skeleton of the triangulation, such as the vanishing of their infinite jet. 

\begin{rem}
	While trying to define a Fr\"olicher structure from a triangulation, one could also consider $$\C_{I,0} = \left\{ \gamma \in C^{0}(\R,M)\, | \, \forall i \in I, \forall f \in C^\infty_c(\phi_i(\Delta_n),\R), f \circ \gamma \in C^\infty(\R,\R) \right\}$$ where $C^\infty_c(\phi_i(\Delta_n),\R)$ stands for compactly supported smooth functions $M \rightarrow \R$ with support in $\phi_i(\Delta_n).$ Then define $$\F_I' = \left\{f : M \rightarrow \R \, | \, f \circ \C_{I,0} \in C^\infty(\R,\R)\right\}$$
	and $$\C_I' = \left\{C : \R \rightarrow M \, | \, \F_I' \circ c \in C^\infty(\R,\R)\right\}.$$
	We get here another construction, but which does not understand as smooth maps $M \rightarrow \R$ the maps $\delta_k$ mentionned in next sections.
\end{rem}
Now, let us fix the set of indexes $I$ and fix a so-called \textbf{model triangulation} $\tau.$ {{} This terminology is justified by two ideas: 
	\begin{itemize} 
		\item Anticipating next constructions, this model triangulation $\tau$ will serve at defining a sequence of refined triangulations. This is our ``starting triangulation'' for the refinement procedure in the finite elements method. 
		\item Changing $\tau$ into $g \circ \tau,$ where $g$ is a diffeomorphism, we get another model triangulation, which has merely the same properties as $\tau.$ But each ``starting'' trinagulation cannot be obtained by transforming a fixed triangulation by using a diffeomorphism. For example, on the 2-sphere, a tetrahedral triangulation $\tau_1$ and an octahedral triangulation $\tau_2$ separately generate two sequences of refined triangulations, and there is a topological obstruction for changing $\tau_1$ into $\tau_2$ by the action of a diffeomorphism of the sphere. This leads to an \underline{\bf open problem}: classify triangulations with respect to the action of groups of diffeomorphisms, in other words, describe the orbits (and their topology) of the group of diffeomorphisms acing on the space of triangulations.    
\end{itemize} } We {{} denote} by $\mathcal{T}_\tau$ the set of triangulations $\tau'$ of $M$ such that the corresponding $(n-1)-$skeletons are diffeomorphic to the $(n-1)-$skeleton of $\tau$ (in the Fr\"olicher category). {{} The set $\mathcal{T}_\tau$ contains, but is not reduced to, the orbit of $\tau$ by the action of the group of diffeomorphisms.
\begin{Definition} \label{d3}
	Since $\mathcal{T}_\tau \subset C^\infty(\Delta_n, M)^I,$ we can equip  $\mathcal{T}_\tau$ with the subset Fr\"olicher structure, in other words, the Fr\"olicher structure on $\mathcal{T}_\tau$ whose generating family of contours $\mcc$ are the contours in $C^\infty(\Delta_n, M)^I$ which lie in $\mathcal{T}_\tau.$
\end{Definition} 
Therefore it is possible to define classical methods of refinement of triangulations as a  map $$\mu: \mathcal{T} \rightarrow \mathcal{T}.$$ 

\begin{Theorem} \cite{Ma2016-4}
	The map $\mu: \mathcal{T} \rightarrow \mathcal{T}$ is smooth.
\end{Theorem}

Therefore generating a sequence of triangulations adapted to $H^1_0-$approximation can be reformulated the following way: 

\begin{Definition}
	Let $\tau \in \mathcal{T}.$ We define the $\mu-$refined sequence of triangulations $\mu^\N(\tau) = (\tau_n)_{n \in \N}$ by $$ \left\{ \begin{array}{ccl} \tau_0 & = & \tau \\ \tau_{n+1} & = & \mu(\tau_n) \end{array} \right.$$ 
\end{Definition}

\begin{Proposition} \label{seqref} \cite{Ma2016-4}
	The map $$\mu^\N : \mathcal{T} \rightarrow \mathcal{T}^\N$$ is smooth (with $\mathcal{T}^\N$ equipped with the infinite product Fr\"olicher structure).
\end{Proposition}
In the case of the Dirichlet problem, we consider a subspace of $\mathcal{T}_\tau.$

Let $\Omega$ be a bounded connected open subset of $\R^n,$ and assume that the border $\partial \Omega = \bar{\Omega} - \Omega$ is a polyhedra. Since $\R^n$ is a vector space, we can consider the space of affine triangulations: 
$$Aff\mathcal{T}_\tau = \left\{ \tau' \in \mathcal{T}_\tau | \forall i , \tau_i' \hbox{ is (the restriction to } \Delta_n \hbox{ of) an affine map } \right\}.$$
We define $Aff\mathcal{T}$ from $Aff\mathcal{T}_\tau$ the same way we defined $\mathcal{T}$ from $\mathcal{T}_\tau,$ via disjoint union.
We equip $Aff(\mathcal{T}_\tau)$ {{}and} $Aff(\mathcal{T})$ with their subset diffeology. We use here the obvious notations for the index $i$ of a simplex in a triangulation anf for the coordinate $x_j$ in a simplex..
\begin{Theorem}
	Let $$c : \R \rightarrow Aff(\mathcal{T}_\tau)$$ be a path on $Aff(\mathcal{T}_\tau).$ Then $$ c \hbox{ is smooth } \Leftrightarrow \forall (i,j) \in I \times \N_{n+1}, t \mapsto x_j(c(t)_i) \hbox{ is smooth. }$$
\end{Theorem}
This construction is recovered in \cite{HL2020}.
\begin{Proposition}
	Let {{}$\mu$} be a fixed affine triangulation of $\Delta_n.$
	The map $\mu^\N$ restricts to a smooth map from the set of affine triangulations of $\Omega$ to the set of sequences of affine triangulations of $\Omega.$
\end{Proposition}
\section{On the way to the geometry of numerical schemes, based on \cite{Ma2016-4}}

Let {{} $X$ and $Z$ be a (LCTVS) and let $Y$ be Fr\'echet spaces.\begin{itemize}
		\item {{} Assume that the inclusion map $X \rightarrow Y$ is smooth.}
		\item Let us consider the space of Cauchy sequences $\mcc(X,Y)$ that are Cauchy sequences in $X$ with respect to the uniform structure on $Y.$
	\end{itemize} Then following \cite{Ma2016-4} it is possible to encode smoothness of the limit and of each evaluation map $(u_n)_\N \mapsto u_n$  on  $\mcc(X,Y)$} by a diffeology called Cauchy diffeology.  Let $Q$ be a diffeological space of parameters. 
 
\begin{Definition}

	A \textbf{smooth functional equation} is defined by a smooth map $F : X\times Q \rightarrow Z$ and by the condition
	\begin{equation} \label{eq}
	F(u,q)=0
	\end{equation}
	{{} The set $Num_F(Y)$ of $Y-$\textbf{smooth numerical schemes} is the set of smooth maps $$x: Q \rightarrow \mcc(X,Y)$$
		such that, if $x(q) = (x_n)_{n \in \N} \in Num_F(Y)(q)\subset \mcc(X,Y)$ for $q \in Q,$ $$\lim_{n \rightarrow +\infty }F(x_n,q) = 0.$$}
	We call the image space  {{}$$\mathcal{S}_Y(F)= \left\{ \left(\lim_{n \rightarrow +\infty} x\right) \in C^\infty(Q,Y) \, |\, x \in Num_F(Y)\right\}$$} the space of {{} $Q-$parametrized} solutions of (\ref{eq}) with respect to $Num_F(Y).$ \end{Definition}

{{}
	\begin{rem}
		In the definition of the space $\mathcal{S}_Y(F)$, we consider the image of $Num_F(Y)$ with respect to the limit map. This means that, for a fixed parameter $q \in Q$, the space of $Y-$solutions to $F(.,q)=0$  is $\mathcal{S}_Y(F)(q).$ 
\end{rem}}

This setting fits with many numerical methods, especially those that envolve fixed point (and in particular gradient) methods in producing weak solutions. Since smooth dependence on the parameters is ensured, a notion of (smooth) symmetries can be derived by extension of the classical notions of symmetries. In this setting the approach described in e.g. \cite{Olv} provides some restricted class symmetries (called projectable symmetries), see e.g. \cite{Vin2013}. We proposed an extension of them in \cite{Ma2016-4} by considering the full group of diffeomorphisms of the set of solutions. This proposal even if abstract, has the particular advantage to set a maximal framework for symmetries of solutions of any kind, weak or strong, while projectable symmetries are naturally included in our setting.  

\vskip 6pt
\noindent
\underline{\bf Open problem:} Compare this setting with so-called approximate symmetries in e.g. \cite{SGO2021} and in references therein, and with the very recent work \cite{CvS2021}. Indeed, weak solutions are limits of approximate solutions and their inter-related construction suggests that their geometric properties can be compared.  Actually, the intrinsic geometry of solutions of PDEs has to be clarified in most cases, exspecially for weak and approximate ones. Some interesting situations, from the viewpoint of weak solutions, are given in \cite{Ma2016-4} including paradoxal solutions of fluid equations from \cite{dLS1,dLS2,dLS3,Vil2006,Sch1993,Sh1,Sh2}, and the recent review \cite{dLS2021} as well as the example \cite{PS2021} show both the wide variety of interesting examples and the non-triviality of geometric problems here raised.

\section{Smoothness of the finite elements method based on \cite{Ma2016-4}}
 One classical way to solve the Dirichlet problem is to approximate $u$ by a sequence $(u_n)_{n \in \N}$ in the Sobolev space $H^1_0(\Omega,\R)$ which converges to $u$ for the $H^1_0-$convergence by the (degree 1) finite elements method, see e.g. \cite{BS1994}. For this, based on a triangulation $\tau_0$ with $0-$vertices $(s^{0}_k)_{k \in K_0},$ where $K_0$ is an adequate set of indexes, and we consider the $H^1_0-$orthogonal family $\left(\delta_{s_k^{(0)}}\right)_{k \in K_0}$ of continuous, piecewise affine maps on each interior domain of triangulation, defined by $$\delta_{s_k^{(0)}} (s_j^{(0)}) = \delta_{j,k} \hbox{ (Kronecker symbol).}$$   
With this setting, $u_0$ is a linear combination of $\left(\delta_{s_k^{(0)}}\right)_{k \in K_0}$ such that
$$\forall k \in K_0, \left(\Delta u_0, \delta_{s_k^{(0)}}\right)_{H^{-1}\times H^1_0} = \left( f, \delta_{s_k^{(0)}}\right)_{L_2} .$$

 With a sequence of affine triangulations $(\tau_n)_{n \in \N}$ defined as before on a suitable domain $\Omega$ of $\R^n,$ we wish to establish smoothness of the family of maps $\delta$ defined before with respect to the underlying triangulation (with the notations of the first section). 

\begin{Theorem} \cite{Ma2016-4}
	Let $\tau \in \mathcal{T}.$
	The map $$\delta: \mathcal{T}_\tau \rightarrow \left(H^1_0 \cap C^0(\Omega)\right)^I$$ is smooth. 
\end{Theorem}

Now, let us fix $\mu$ a triangulation of $\Delta_n,$ then under midl conditions $\mu$ defines refinement sheme in $\mathcal T,$ by dividing simplexes of the initial triangulation, which introduces, for each $\tau \in \mathcal T,$ a sequence $\tau_n,$ and a family of functions $\delta^{\tau_n}.$

\begin{Theorem} \cite{Ma2016-4}
	The map $(\tau_0',f ) \in \mathcal{T} \times C^\infty(\Omega,\R)\mapsto (u_n)_{n \in \N}$ is a smooth $H^1_0-$ numerical scheme for the Dirichlet problem.
\end{Theorem}  
In other words, $\forall n \in \N,$ the approximate solution $u_n,$ piecewise affine  on $\tau_n,$ depends smoothly on $\tau_0$ and $f,$ while we alrealy know  that the solution $u = \lim u_n$ does not depend on $\tau_0$ and depends smoothly on $f.$ 
\vskip 6pt
\noindent
\underline{\bf Open problem:} The Dirichlet problem is a toy example compared with the full range of applications of finite elements methods. Applications of the smooth structure of the space of triangulations are in progress for shape analysis problems. Well-posedness of other related problems of discretization via triangulation, such as Whitney discretization of connections on a trivial principal bundle \cite{Wh}, can be considered.  

\section{New lights on implicit functions and related results, based on \cite{Ma2020-3,Ma2016-4}}
 We set the following notations, following \cite{Om}:
Let  $\hbox{\bf E} = (E_i)_{i \in \N}$ and  
$\hbox{\bf F} = (F_i)_{i \in \N}$ be two sequences of Banach spaces, decreasing for inclusion and with dense embedding, called ILB vector spaces (ILB for Inverse Limit of Banach), i.e. 
$\forall i > j, E_i \subset E_j$ and $F_i \subset F_j,$ with smooth inclusion and density, and set $E_\infty = \bigcap_{i \in \N}E_i,$ $F_\infty = \bigcap_{i \in \N}F_i$ with projective limit topology.
Let $O_0$ be an open neighborhood of $(0;0)$ in $E_0 \times F_0,$ let
$\hbox{\bf O} = (O_i)_{i \in I}$ with $O_i = O_0 \cap ({E_i \times F_i})$, for $i \in \N \cup \{\infty\}.$
\begin{Theorem} \cite[Theorem 2.2]{Ma2020-3}\label{1.6} There exists a non-empty domain $D_\infty \subset U_\infty,$ possibly non-open
	in $U_\infty,$ and a function $$u_\infty : D_\infty \rightarrow V_\infty$$ such that
	$$\forall x \in D_\infty, \quad f_\infty(x; u_\infty(x)) = 0.$$
	{{} Moreover, there exists a sequence $(c_i)_{i \in \N} \in (\R_+^*)^\N$ and a Banach space $B_{f_\infty}$ such that
		\begin{itemize}
			\item $B_{f_\infty} \subset E_\infty$ (as a subset)
			\item the canonical inclusion map $B_{f_\infty} \hookrightarrow E_\infty$ is continuous
		\end{itemize} 
		which is the domain of the following norm (and endowed with it): $$||x||_{f_\infty} = \sup
		\left\{ \frac{||x_i||}{c_i}| i \in\N \right\}. $$ Then $D_\infty$ contains $\mathcal{B},$ the unit ball (of radius $1$ centered at 0) of  $B_{f_\infty}.$} \end{Theorem}   
In \cite{Ma2020-3}, the question of the regularity of the implicit function is left open, because the domain $D_\infty$ is not a priori open in $O_\infty.$ Moreover, the presence of the Banach space $B_{f_\infty}$ suggests that the properties of the implicit function $u_\infty$ may depend on the properties of   the function $f_\infty$ under consideration. This lack of regularity induces a critical breakdown in generalizing the classical proof of the Frobenius theorem to this setting. We fill this gap in the sequel, by completing the proof of Theorem \ref{1.6} from \cite{Ma2016-4}  using the Cauchy diffeology,   
under the light of numerical schemes. 
\begin{Theorem} \label{IFTh} \cite{Ma2016-4}
{{}Let $$f_i: O_i \rightarrow F_i, \quad i \in \N \cup \{\infty \}$$ be a family of maps, 
	let $u_\infty$ the implicit function defined on the domain $D_\infty$,  as in Theorem \ref{1.6}.
	Then, there exists a domain $D$ such that $\mathcal{B} \subset D \subset D_\infty$ such that the function $u_\infty$  is smooth for the subset diffeology of $D.$}
\end{Theorem}
 The same way, we can state the corresponding Frobenius theorem, denoting by $L(E,F)$ the space of bounded linear maps between two Banach spaces $E$ and $F:$

\begin{Theorem}\label{lFrob} \cite{Ma2016-4}
	
	Let 
	$$ f_i : O_i \rightarrow L(E_i,F_i), \quad i \in {{}\N}$$ 
	be a collection of smooth maps satisfying the following condition: 
	$$ i > j \Rightarrow f_j|_{O_i} = f_i$$ and such that, $$\forall (x,y) \in O_i, \forall a,b \in E_i$$
	$$(D_1f_i(x,y)(a)(b) + (D_2f_i(x,y))(f_i(x,y)(a))(b) =$$
	$$(D_1f_i(x,y)(b)(a) + (D_2f_i(x,y))(f_i(x,y)(b))(a) .$$

	Then,
	$\forall (x_0, y_0) \in O_{\infty}$, there exists a diffeological subspace  $ D $ of $O_\infty$ that contains $(x_0, y_0)$ and a smooth map
	$J : D \rightarrow  F_\infty$
	such that
	$$ \forall (x,y) \in D, \quad D_1J(x,y) = f_i(x, J(x,y)) $$
	and, if {{}$D_{x_0}$ is the connected component of $(x_0,y_0)$ in $\{(x,y) \in D \, | \, x = x_0  \},$ }
	$$J_i(x_0,.) = Id_{D_{x_0}}.
	$$ 
	{{} Moreover, there exists a sequence $(c_i)_{i \in \N} \in (\R_+^*)^\N$ and a Banach space $B_{f_\infty}$ such that
		\begin{itemize}
			\item $B_{f_\infty} \subset E_\infty\times F_\infty$ (as a subset)
			\item the canonical inclusion map $B_{f_\infty} \hookrightarrow E_\infty\times F_\infty$ is continuous
		\end{itemize} 
		which is the domain of the following norm (and endowed with it): $$||x||_{f_\infty} = \sup
		\left\{ \frac{||x||_{E_i \times F_i}}{c_i}| i \in\N \right\}. $$ Then $D_\infty$ contains $\mathcal{B},$ the unit ball (of radius $1$ centered at 0) of  $B_{f_\infty}.$}
	
\end{Theorem}

\vskip 6pt
\noindent
\underline{\bf Open problem:} There exists many extended versions of implicit functions theorems, and among them the version in \cite{HN1971} based on bornologies. Natural bornologies defined by diffeologies may give new clues  and new ways to get smooth implicit functions. 
\chapter{Contributions to infinite dimensional integrals and means}

Let $(X,\mu)$ be a measured space. Following \cite{Pa}, \cite{Pes}, let us fix a vector subspace $\mathcal{F}\subset L^\infty(X,\mu)$ such that $1_X \in \mathcal{F}.$ A \textbf{mean} on $\mathcal{F}$ is a linear map $\phi:\mathcal{F}\rightarrow \mathbb{C}$
such that $\phi(1_X)=1.$ Alternately, if $(X,d)$ is a metric space, 
given $\mathcal{F}\subset C^0_b(X)$ (space of continuous bounded maps), 
a \textbf{mean} on $\mathcal{F}$ is a linear map 
$\phi:\mathcal{F}\rightarrow \mathbb{C}$ such that $\phi(1_X)=1.$
These two terminologies come from the basic example where $\mu$ is a Borel
probability measure on a compact metric space $(X,d),$ for which the mean of a continuous integrable map $f$ is its expectation value $$\int_X f d\mu,$$
and can be approximated by sequences of barycenters of Dirac measures via Monte Carlo methods.
For functions on infinite dimensional spaces derived from Feynman-Kac formulas, normalized integrals of the type $$\frac{1}{Z(S)} \int f e^{-iS} d \lambda,$$ where $$Z(S) = \int  e^{-iS} d \lambda,$$ and where $d\lambda$ is a formal infinite dimensional Lebesgue measure and $S$ is an action functional, need to be defined rigourously, which is actually performed by vaious non-equivalent ways, among which Fresnel oscillatory integrals \cite{AB,AHM,F2017} are the most known of us. Underlying these formulas, we identify here the problem of the extension of the definition of an integral to infinite dimensional spaces. This problem has been independently addressed in \cite{AM2016,MoSm2016,AM2018} for non-necessarily normalized integrals 
while the results \cite{MaICM,Ma2016-3} presented hereafter intend to describe normalized linear functionals extending the notion of integral of functions defined on infinite dimensional spaces.
    
\section{Means spanned by probability measures, based on \cite{MaICM,Ma2016-3}}

Let $X$ be a complete metric space and let $C^0_b(X)$ be the space of bounded $\K-$valued continuous maps on $X.$ We note by $\mathbb{P}(X)$ the space of Borel probability measures on $X.$ Let us first set $V = \K.$
When $X$ is a compact metric space, it is well-known that $\mathbb{P}(X)$ is a convex set with extremal points the Dirac measures. one can generalize this construction for non compact, maybe infinite dimensional spaces, by producing this way {means}. More precisely: 
\begin{itemize}
	\item A $\K-$\textbf{probability mean} is a linear map $\tau:\mathcal{D}_\tau \subset C^0_b(X)\rightarrow \K $ which is defined as the limit of barycenters with $\K-$weights of a sequence of Borel probability measures on $X,$ i.e.
	
	$$\exists (\mu_n, \alpha_n)_{n \in \N}\in (\mathbb{P}(X) \times \mathbb{K})^\N, \forall m\in \N^*, $$ $$\left\{ \sum_{n = 0}^{m} \alpha_n \neq 0 \right\} \quad \wedge \quad \left\{ \forall f \in C^0_b(X), \tau(f) = \lim_{m \rightarrow +\infty} \frac{1}{\sum_{n = 0}^{m} \alpha_n} \left( \sum_{n = 0}^{m} \alpha_n {\mu_n}(f) \right)\right\}.$$
	\item 	Following \cite{MaICM}, a $\K-$\textbf{Dirac mean} is a linear map $\tau:\mathcal{D}_\tau \subset C^0_b(X)\rightarrow \K $ which is defined as the limit of barycenters with $\K-$weights of a sequence of Dirac measures on $X,$ 
	$$\exists (x_n, \alpha_n)_{n \in \N}\in (X \times \mathbb{K})^\N, \forall m\in \N^*, $$ $$\left\{ \sum_{n = 0}^{m} \alpha_n \neq 0 \right\} \quad \wedge \quad \left\{ \forall f \in C^0_b(X), \tau(f) = \lim_{m \rightarrow +\infty} \frac{1}{\sum_{n = 0}^{m} \alpha_n} \left( \sum_{n = 0}^{m} \alpha_n \delta_{x_n}(f) \right)\right\}.$$
\end{itemize} .

We note by $\widetilde{\mathcal{PM}}_\mathbb{K}(X)$ the space of $\mathbb{K}-$probability means, by ${\mathcal{PM}}_\mathbb{K}(X)$ the set of probability means $\tau$ such that $\mathcal{D}_\tau = C^0_b(X),$ by $\widetilde{\mathcal{PM}}_\mathbb{R}^+(X)$ the means $\tau$ obtained by a sequence $(\alpha_n)_{n \in \N}\in \R_+^*$ and we set $${\mathcal{PM}}_\mathbb{R}^+(X)={\mathcal{PM}}_\mbbc(X)\cap\widetilde{\mathcal{PM}}_\mathbb{R}^+(X). $$ 
We note by $\widetilde{\mathcal{DM}}_\mathbb{K}(X),$  ${\mathcal{DM}}_\mathbb{K}(X), $   $\widetilde{\mathcal{DM}}_\mathbb{R}^+(X),$  ${\mathcal{DM}}_\mathbb{R}^+(X) $ the sets of Dirac means corresponding respectively to  $\widetilde{\mathcal{PM}}_\mathbb{K}(X),$  ${\mathcal{PM}}_\mathbb{K}(X), $   $\widetilde{\mathcal{PM}}_\mathbb{R}^+(X),$  ${\mathcal{PM}}_\mathbb{R}^+(X) $

\begin{Definition} \cite{Ma2016-3}
	Let $\mathcal{X}= (X_n)_\N $ be an exhaustive sequence of compact subsets of nonzero finite measure in a mm-space $(X,d,\mu).$ Let $$\mu_n = \frac{1}{\mu(X_n)} \hbox{\bf 1}_{X_n} \mu.$$
	Let $f: X \rightarrow \mathbb{C}$ be a map
	such that for each $ n \in \N,$ the restriction of $f$ to 
	$X_n$ is $\mu_n-$integrable. Then, the \textbf{mean value} 
	of $f$ with respect to the family $\mathcal{X}$ is
	$$WMV^{\mathcal{X}}(f)=\lim_{n \rightarrow +\infty} \int_{X_n} f d\mu_n$$
	if the limit exists. 
\end{Definition}
\begin{Definition}\label{lmean} \cite{Ma2016-3}
	Let $X=(X_n,\tau_n)_{n \in \N}$ be a sequence of probability spaces such that 
	
	- $\forall n \in \N, $ $X_n$ is a metric space.
	
	- $\forall n \in \N, X_n \subset X_{n+1},$ and the topology of $X_{n+1}$ restricted to $X_n$ co\"incides with the topology of $X_n.$
	
	- $\forall n \in \N, \tau_n\in \widetilde{\mathcal{PM}}_\mbbc (X_n).$
	Then, we define, for the maps $f$ defined on $\bigcup_{n \in \N} X_n,$ if $\forall n \in \N, f_{|X_n} \in \mathcal{D}_{\tau_n}$ and if the limit converges, 
	$$ LM^X(f) =\lim_{n\rightarrow +\infty} \tau_n (f)$$
	called
	\textbf{limit mean} of $f$ with respect to $X.$  
	
\end{Definition}

Following \cite{Gro}
and \cite{Pes}, a
	\textbf{space with metric and measure}, or a \textbf{metric measured space}
	(mm-space for short) is a triple $(X, d, \mu)$ where $(X,d)$
	is a metric space and $\mu$ is a probability measure on the
	Borel algebra on $X.$

Let $A \subset X,$ let $\varepsilon >0.$ We note by
$$A_\eps = \{ x \in X | d(A,x)<\eps \}.$$

In the sequel, we shall assume that 
$$ \forall n \in \N, X_n \subset X_{n+1}$$
with continuous injection. Notice that we do not assume that
$d_n$ is the restriction of $d_{n+1}$ hich allows us some 
freedom on metric requirements. The technical necessary condition is the following: let $n \in \N$ and let $B_{n+1}$ be a Borel subset of $X_{n+1}.$ Then $B_{n+1} \cap X_n$ is a Borel subset of $X_n.$ 
We have here a priori a class of limit means following the terminology of Definition \ref{lmean}.
Let us quote first the classical (and historical) example of a Levy family see e.g. \cite{Gro}, section 3$\frac{1}{2}$.19,
which gives an example of mean value:

\textbf{The Levy family of spheres and the concentration phenomenon}
	
	Let us consider the seuquence of inclusions
	$$ S^1 \subset S^2 \subset ... \subset S^n \subset S^{n+1} ... \subset S^\infty=\bigcup_{n = 1}^\infty S^n$$
	equipped with the classical Euclid (or Hilbert) distance and (except for $S^\infty$) the normalized spherical measure $\mu$ (we drop the index for the measure in sake of clear notations). Then, for any $\R -$valued 1-Lipschitz function on $S^\infty,$ there exists $a\in \R$ such that:
	$$\forall \epsilon >0, \quad \mu\left\{ x \in S^n | ||f(x) - a||>\epsilon \right\} < 2e^{-\frac{(n-1)\epsilon^2}{2}}.$$
\vskip 12pt

In a more intuitive formulation, one can say that any 1-Lipschitz function concentrates around a real vaule $a$ with respect to $\mu.$ We leave the reader with the reference \cite{Gro} for more on the metric geometry of this example.
We can reformulate: 
\begin{Proposition} \cite{Ma2016-3}
	Let $\mathcal{X}=(S^n; ||.||; \mu)_{n \in \N^*}$. Then 
	for any 1-Lipschitz function $f$ defined on $S^\infty,$
	and with the notations used before,
	$$LM^\mathcal{X}(f)=a.$$
\end{Proposition} 
\vskip 12pt
\noindent
\begin{rem}[L\'evy families induced by Lebesgue measures] \label{Levylebesgue} \cite{Ma2016-3}
	Let $m, n\in (\N^*)^2.$ 
	Take $K_m \subset`\mathbb{R}^n.$ 
	For each $m \in \N^*,$ we equip $K_m$ with the usual 
	distance $d$ induced by $\R^n$ and with the probability measure
	$$\mu_n = \frac{\mathbf{1}_{K_n}}{\lambda(K_n)} \lambda.$$
	Setting $\mathcal{K}= (K_m,d, \mu_m)_{m \in \N^*},$
	we get that $\mathcal{K}$ is a L\'evy family, but there is no concentration property.
\end{rem}

\section{Limit means and infinite dimensional integrals, based on \cite{Ma2016-3}}

This definition intends to fit with the procedure of integration of cylindrical functions in Hilbert spaces. This enables to analyze two well-known classes of infinite dimensional integrals. 
\begin{itemize}
	\item \textbf{Daniell integral} \cite{Ma2016-3} consists in intergation with respect to the infinite dimensional product probability measure over $[0;1]^\N.$ Then, adequate sequences for the Monte Carlo method are those whose
	push-forward on $[0; 1]^{k}$
	are also adequate for this method. The condition on the
	sequence $(x_n)$
	is that for each
	$k \in \mathbb{N}
	,$
	the push-forwards of the sequences $(P_k(x_n))$
	on
	$[0; 1]^{k}$
	fit with the desired conditions: the sequence $(P_k(x_n))$ is a Monte Carlo sequence for the cube $[0; 1]^{k}$ equipped with the (trace of) the Lebesgue measure. It is well-known that such a sequence $ (x_n)$
	exists,
	through e.g. the powers of
	$\pi$: $$\forall n \in \mathbb{N}^*, \quad x_n = (n\pi^{l+1} - int(n \pi^{l+1}))_{l \in \N} \in [0;1]^\N,$$ where $int(x)$ is the integer part of the real number $x.$
	Thus, Daniell integral appears by its definition as a limit mean for the sequence $(X_n)_{\N^*}$ defined by $X_n = [0;1]^n,$ equipped with the classical Lebesque measure. But Daniell integral appears also as a Dirac mean which domain contains cylindrical functions. 
	\item \textbf{Fresnel integrals:} Let $\Phi \in C^\infty(\R^n, \R)$ be a fixed function. Following \cite{ET} (see e.g. \cite{AHM,AMaz1,Dui}), we define:
	\begin{Definition} \label{Fresnel}
		Let $f$ be a measurable function on $\R^n.$
		Let $\varphi \in \mathcal{S}(\R^n)$ be a weight function such that $\varphi(0)=1.$
		if the limit 
		$$\lim_{\epsilon \rightarrow 0}\int_{\R^n}e^{i\Phi(x)} f(x) \varphi(\epsilon x) dx$$
		exists and is independent of the fixed function $\varphi,$
		then this limit is called oscillatory integral of $f$ with respect to $\Phi,$
		noted $$\int_{\R^n}^o e^{i\Phi(x)} f(x) dx .$$
	\end{Definition} 
	
	The choice $\Phi(x) = \frac{i}{2h} |x|^2$ is of particular interest, and is known under the name of \textbf{Fresnel integral}. This choice gives us a mean, up to normalization by a factor $(2i\pi h)^{-\frac{d}{2}},$
	and can be generalized to a Hilbert space $\mathcal{H}$ the following way:
	
	\begin{Definition}
		A Borel measurable function $f: \mathcal{H}\rightarrow \mbbc$ is called $h-$ integrable in the sense of Fresnel is for each increasing 
		sequence of projectors $(P_{n})_{n \in \N}$ such that $\lim_{n \rightarrow +\infty} P_n = Id_\mathcal{H},$ the finite dimensional approximations of the oscillatory integrals of $f$ 
		$$\left\{\int_{Im P_n}^o e^{\frac{i}{2h}|P_n(x)|^2} f(P_n(x)) d(P_n(x))\right\}
		\left\{\int_{Im P_n}^o e^{\frac{i}{2h}|P_n(x)|^2}  d(P_n(x))\right\}^{-1}$$
		are well-defined and the limit as $n\rightarrow +\infty$ does not depend on the sequence $(P_{n})_{n \in \N}.$ In this case, it is called infinite dimensional Fresnel integral of $f$ and noted 
		$$\int_{\mathcal{H}}^o e^{\frac{i}{2h}|x|^2} f(x) d(x).$$
	\end{Definition}
	
	The invariance under the choices of the map $\varphi$ and the projections $P_n$ is assumed mostly to enable stronger analysis on these objects, which intend to be useful to describe physical quantities and hence can be manipulated in applications where one sometimes works ``with no fear on the mathematical rigor'' in calculations. But we can also remark, following \cite{Ma2016-3} that:
	\begin{itemize}
		\item for functions $f$ defined on $\mathbb{R}^n,$ the map $$f \mapsto \int_{\R^n}^o e^{i\Phi(x)} f(x) dx \in \widetilde{\mathcal{PM}}_\mathbb{\mathbb{C}}(\mathbb{R}^n),$$
		\item the map $$f \mapsto \lim_{n \rightarrow +\infty} \int_{\R^n}^o e^{i\Phi(x)} f(x) dx $$ is a limit mean through the sequence $\R \subset...\subset \R^n \subset \R^{n+1} \subset ... \subset \mathcal{H}.$
	\end{itemize}
	The limit mean obtained is got through the classical trick of cylindrical functions, which we shall also use in the sequel. But we have no way to define some adequate sequence of Dirac means which could approximate the oscillatory integral, even in the finite dimensional case actually.  
\end{itemize}
\vskip 6pt
\noindent
\underline{\bf Open problem:} In the space $\D$ of sequences $(d_n)$ of Dirac means, for which $\lim d_n f $ exists for a family $\F$ of test functions $f$, there is a natural relation of equivalence $$(d_n) \sim (d'_n) \Leftrightarrow \forall f \in \F, \,\lim d_n f = \lim d_n' f.$$
When $X$ is a compact metric space and when $\F = C^0(X,\R),$ the quotient space is exactly the space of Radon measures.  

In any of these situations, the quotient $$\D \rightarrow \D/\sim$$ has geometric or topological properties that are unknown, only a sketch of adequate diffeology for its study is given in \cite{Ma2016-4}. The same question of geometric properties can be raised for the framework for infinite dimensional (cylindrical) integrals described in \cite{AM2016}.

\section{Infinite products and normalized infinite dimensional Lebesgue measure following \cite{Ma2016-3}}
It is almost straightforward to extend means to an infinite product  by the procedure of cylindrical functions. The same way, one can define \textbf{admissible domains} on the infinite product by generalizing open dense subsets \cite{Ma2016-3}. This leads to the following applications.

\subsection{Application: the mean value on marked infinite configurations}

Let $X$ be a locally compact and paracompact manifold, 
orientable, and let $\mu$ be a measure on $X$ 
induced by a volume form.
In the following, we have either

- if $X$ is compact, setting $x_0 \in X,$ and following \cite{Str},
$$ \Gamma = \{ (u_n)_{n\in \N} \in X^\N | \lim u_n = x_0 \hbox{ and } \forall (n,m)\in \N^2, n\neq m \Rightarrow u_n \neq u_m\} $$ 

- if $X$ is not compact, setting $(K_n)_{n\in\N}$ an exhaustive sequence of compact subspaces of $X,$ 

$$O\Gamma = \{ (u_n)_{n\in \N} \in X^\N |  \forall p \in \N,  |\{u_n;n\in\N\} \cap K_p|<+\infty \hbox{ and } \quad \quad \quad \quad \quad \quad$$ 
$$ \quad \quad \quad \quad \quad \quad \quad \quad \quad \forall (n,m)\in \N^2, n\neq m \Rightarrow u_n \neq u_m\} $$
The first setting was first defined by Ismaginov, Vershik, Gel'fand and Graev,
see e.g. \cite{Ism} for a recent reference, and the 
second one has been extensively studied 
by Albeverio, Daletskii, Kondratiev, Lytvynov, see e.g. \cite{A}.
Alternatively, $\Gamma$ can be seen as a set of countable sums of Dirac measures
equipped with the topology of vague convergence.

For the following, we also need the set of ordered finite $k-$configurations: 
$$O \Gamma^k = \{ (u_1,...,u_k) \in X^k | \forall (n,m)\in \N^2, (1\leq n < m \leq k) \Rightarrow (u_n \neq u_m) \} $$
$O\Gamma$ is an admissible domain in $X^\N$ in the sense there exists an increasing sequence of ``admissible subsets''on which $O\Gamma$ can be ``assimilated almost everywhere''to a cartesian product, see \cite{Ma2016-3} for the details. We then define, for a bounded cylindrical function $f,$  

$$ WMV_\mu^U(f).$$
More precisely, the normalization sequence $U$ on $O\Gamma$ is induced from 
the normalization sequence on $X^\N .$ This implies heuristically that 
cylindrical functions with a weak mean value with respect to $U$ are in a sense
small perturbations of functions on $X^\N.$
This is why we can modify the sequence $U$ on $O\Gamma$ in the following way:
let $\varphi: \R_+ \rightarrow \R_+^*$ be a function such that 
$\lim_{x \rightarrow +\infty}\varphi =0.$ 
Then, if $f$ is a cylindrical function on $O\Gamma,$ we set 
$$U^n_\varphi = U^n - \{(x_i)_{1 \leq i \leq n} | \exists (i,j) \hbox{ such that } i < j \wedge d(x_i,x_j)<\varphi(n)\}.$$

\subsection{Normalized infinite dimensional Lebesgue integral}

\begin{Definition}
	A \textbf{normalized Fr\'echet space}  is a pair $(F,H),$ where 
	
	\begin{enumerate}
		\item $F$ is a Fr\'echet space,
		
		\item $H$ is a Hilbert space, 
		
		\item $F \subset H$ and
		
		\item $F$ is dense in $H.$
	\end{enumerate}
	
\end{Definition}

Another way to understand this definition is the following: 
we choose a pre-Hilbert norm on the Fr\'echet space $F.$ 
Then, $H$ is the completion of $F.$

\begin{Definition}\label{dh}
	Let $V$ be a complete locally convex topological vector space. A function $f : F \rightarrow V$ is \textbf{cylindrical} 
	if there exists $F_f , $  a finite dimensional affine subspace of $F,$ for which, if $\pi$ is the orthogonal projection, 
	$\pi : F \rightarrow F_f$ such that $$\forall x \in F, f(x) = f\circ \pi(x).$$  
	
\end{Definition}

\begin{Proposition}
	Let $(f_n)_{n \in \N}$ be a sequence of cylindrical functions. There exists an
	unique sequence $(F_{f_n})_{n\in \N}$ increasing for $\subset,$ for which 
	$\forall m \in \N,$, $F_{f_m}$ is the minimal affine space for which $$\forall n \leq m, f_n \circ \pi_m = f_n.$$ 
\end{Proposition}

Let $f$ be a bounded function which is the uniform limit of a sequence of 
cylindrical functions $(f_n)_{n \in \N}.$  
Here, an orthonormal basis 
$(e_k)_{k \in \N}$ is obtained by induction, completing at each 
step an orthonormal basis of $F_{f_n}$ by an orthonormal basis 
of $F_{f_{n+1}}.$ Thus we can identify $F$ with a subset $\mathcal{D}$ of $\R^\N$
which is invariant under change of a finite number of coordinates. 
This qualifies it as admissible
for any set of renormalization procedures in $\R^\N$ . 
We note by $$ WMV_\lambda(f)$$ the weak mean value here constructed, which stand for a normalized infinite dimensional Lebesgue integral, generalizing to an infinite product the Levy family in Remark \ref{Levylebesgue}.
This mean 
value does not depend on the sequence $(f_n)_{n \in \N}$ 
\textbf{only once the sequence} $(F_{f_n})_{n \in \N}$ \textbf{is fixed}. In other words, 
two sequences $(f_n)_{n \in \N}$ and $(f'_n)_{n \in \N}$ which converge 
uniformly to $f$ a priori lead to the same mean value if $F_{f_n}=F_{f'_n}$ 
(maybe up to re-indexation). From heuristic calculations, it seems to come from 
the choice of the renormalization procedure, 
which is dependent on the basis chosen, 
more than from the sequence $(F_{f_n})_{n \in \N}.$ 

We notice in \cite{Ma2016-3} three types of invariance for $WMV_\lambda:$
\begin{enumerate}
	\item \underline{\it Scale invariance:} Let $\alpha \in \N^*.$ Let $f$ be a function on an infinite dimensional vector space $F$ with mean value. 
	Let $f_\alpha: x\in F \mapsto f(\alpha x).$ Then $f_\alpha$ has a 
	mean value and $$WMV_\lambda(f_\alpha) = WMV_\lambda(f).$$
	\item \underline{\it Translation invariance:} Let $v \in F.$ Let $f$ be a function on $F$ with mean value. 
	Let $f_v: x\in F \mapsto f(x+v).$ 
	Then $f_v$ has a mean value and $$WMV_\lambda(f_v)=WMV_\lambda(f).$$
	\item \underline{\it Invariance under the orthogonal (or unitary) group:} Let $U_F$ be the group of unitary operators of $H$ 
	which restricts to a bounded map $F \rightarrow F$ 
	together with its inverse. Let $u \in U_F.$ Let $f$
	be a map with mean value. Then $f \circ u$ has a mean value and 
	$$ WMV_\lambda(f \circ u)=WMV_\lambda(f).$$ 
\end{enumerate}

These three fundamental properties qualifies $WMV_\lambda$ to be called normalized generalization of a Lebesgue measure. This construction is very heuristically a normalization of the classical  infinite dimensional Lebesgue measure \cite{Ba1,Ba2} for which:

- the Hilbert cube is of measure 1;

- its dilatations are of measure $+\infty;$

- its homothetic contractions are of measure $0.$
By its lack of increasing sequence of bounded subsets covering $\R^\infty,$ its normalized version had to be made before passing to the cylindrical limit. Let us now give an application of our "normalized Lebesgue integral" on a Hilbert space.  We also have to mention the work \cite{MoSm2016} which is to our knowledge the first paper to follow the approach that we initiated on infinite dimensional Lebesgue integration. For the authors,  
\begin{Definition} (intuitive translation of the definition in \cite{MoSm2016})
	Let $S(E)$ be some class of infinitely differentiable complex functions on the locally convex topological vector space $E.$ Assume also that $S(E)$, which is assumed stable under composition by translations, is equipped with a topology which makes differentiation and composition with a translation differentiable. Then any translation-invariant $\nu \in S'(E)$ is called \textbf{Lebesgue-Feynman measure}.
	 
\end{Definition}
\vskip 6pt
\noindent
\underline{\bf Open problem:} The normalized infinite dimensional Lebesgue integral has clearly to be studied under the lights of \cite{MoSm2016} on the one hand, and of \cite{AM2016} on the other hand . 
\subsection{Application: Lebesgue integral on spaces of n-differential forms and Hodge theory}

\vskip 12pt
Let $M$ be a finite dimensional manifold quipped with a 
Riemannian metric $g$ and the corresponding Laplace-Beltrami operator $\Delta,$
and with finite dimensional de Rham cohomology space $H^*(M,\R).$
One of the standard results of Hodge theory is the onto and one-to-one map 
between 
$H^*(M,\R)$ and the space of $L^2-$harmonic forms $\mathcal{H} $ made by integration over simplexes:
\begin{eqnarray*} I :& \mathcal{H} \rightarrow & H^*(M,\R)  \\
	& \alpha \mapsto & I(\alpha)
\end{eqnarray*}
where $$I(\alpha): s \hbox{ simplex } \mapsto I(\alpha)(s)=\int_{s}\alpha.$$ 
We have  assumed here that the simplex has the order of the harmonic form.
This is mathematically coherent stating  $\int_s\alpha = 0$ if $s$ and $\alpha$ do not have the same order. 
Let $\lambda$ be the Lebesgue measure on $\mathcal{H}$ with respect to the scalar product induced by 
the $L^2-$scalar product. Let $U=(U_n)_{n\in \N}$ be the sequence 
of Euclidian balls centered at $0$ such that, for each $n\in \N,$ the ball $U_n$ is of radius $n.$
\begin{Proposition} \cite{Ma2016-3}
	Assume that $H^*(M,R)$ is finite dimensional.
	Let $s$ be a simplex. Let $$\varphi_s = \frac{|I(.)(s)|}{1+|I(.)(s)|}.$$
	The cohomology class of $s$ is null if and only if  $$WMV^U_\lambda(\varphi_s) = 0.$$
\end{Proposition}

The map $s \mapsto WMV_\lambda^U(\varphi_s)$ is a $\{0;1\}-$ valued map.

	Moreover, the map $\varphi_s$  extends to a cylindrical function on $L^2-$forms, and hence the construction described above applies to this extension map.

\chapter{Contributions to infinite dimensional Lie groups and principal bundles}

 Let us review few definitions in order to fix the necessary vocabulary. For this, we have the very difficult task to summarize investigations led during decades by Omori, Milnor, Ratiu, and then Michor, Neeb, Gl\"ockner among others, without going too deeply into not necessary detailed descriptions and refined properties but sketching the key ideas existing in other works by other authors in order to motivate applications of our results.  Passing from finite dimensional settings  to infinite dimensional ones, there are, among others, fundamental properties which are difficult to state for the geometry of groups: 
 \begin{itemize}
 	\item \underline{\it Enlargibility:} Given a Lie algebra $\mathfrak{g},$ does there exist a Lig group $G$ with Lie algebra $\mathfrak{g} ?$ If so, $\mathfrak{g}$ is called \textbf{enlargible}. Non-enlargible Lie algebras are known since \cite{vEK1964,Om3}. For example, the Lie algebra of smooth vector fields $Vect(M)$ over a non-compact, paracompact manifold $M.$ This Lie algebra stands heuristically as the Lie algebra of groups of diffeomorphisms (see the ``euristique'' \cite[section 9]{Arn1966}), but considering more rigourously these groups, the groups of diffeomorphisms of a non-compact manifold can be endowed with very various topologies which lead naturally to different deduced Lie algebras that are all subalgebras of $Vect(M).$ In order to finish with technical difficulties of the setting of the group of diffeomorphisms, we have to mention that this group (and its various topologies) was one of Souriau's motivations to define so-called ``groupes diff\'erentiels'' which became the actual diffeological Lie groups.   
 	\item \underline{\it Integrability:} Given $G$ a Lie group with Lie algebra $\mathfrak{g},$ given $\mathfrak{g}_1$ a Lie subalgebra of $\mathfrak{g}$ (in a terminology to be precised depending on the category of Lie groups considered), does there exist $G_1$ a Lie subgroup of $G$ with Lie algebra $\mathfrak{g}_1 ?$ If so, $\mathfrak{g}_1$ is called \textbf{integrable} in $G.$ Non-integrable Lie algebras are known also from Omori's pioneering work, and investigated by many authors, see e.g. \cite{Neeb2007} since then.
 	\item \underline{\it Regularity:} Given $G$ a Lie group with Lie algebra $\mathfrak{g},$ the group is called \textbf{regular} if there exists an exponential map \[
 	Exp:C^{\infty}([0;1],\mathfrak{g})\rightarrow C^{\infty}([0,1],G)\]
 	which, roughly speaking, integrates the differential equation on logarithmic derivatives  \begin{equation} \label{log-der} \frac{dg(t)}{dt}g(t)^{-1}=v(t). \end{equation}
 	Pragmatic conditions to integrate this equation are gathered in the conditions for Omori regularity \cite{Om} while an extended notion which avoid technical conditions for the integration of equation (\ref{log-der}) are given in the $c^\infty-$setting \cite{KM}. Other related works in various contexts are e.g. \cite{Rob,Les} and a non-exhaustive review on this notion is given in \cite{Neeb2007}.
 \end{itemize}
 In order to deal with infinite dimensional generalizations of principal bundles, better is to gather these three properties on the structure group and on the Lie algebras considered. This happens in the category of Banach principal bundles. Moreover, in the $c^\infty-$setting, when the structure group  is regular, the basic theory of connections is quite similar to the one known in the finite dimensional setting, once one has fixed the regular Lie group which serves as a structure group following the discussion in \cite{Ma2006}, see details in \cite{KM}.
 \vskip 6pt
 \centerline{\it We have now finished the panoramic view  which can serve as a backdrop} \centerline{\it for the exposition of our work.}
\section{On regular Fr\"olicher Lie groups, based on \cite{Ma2013,Ma2013-2,MR2019,MR2016}}

The following definitions are first given in \cite{Ma2013} and in \cite{MR2019}, based on the observation and results of \cite{Les} in the context of diffeologies.
\begin{Definition} \label{reg1} \cite{Ma2013} A Fr\"olicher Lie group $G$ with Lie algebra $\mathfrak{g}$
	is called \textbf{regular} if and only if there is a smooth map \[
	Exp:C^{\infty}([0;1],\mathfrak{g})\rightarrow C^{\infty}([0,1],G)\]
	such that $g(t)=Exp(v(t))$ is the unique solution
	of the differential equation \begin{equation}
	\label{loga}
	\left\{ \begin{array}{l}
	g(0)=e\\
	\frac{dg(t)}{dt}g(t)^{-1}=v(t)\end{array}\right.\end{equation}
	We define the exponential function as follows:
	\begin{eqnarray*}
		exp:\mathfrak{g} & \rightarrow & G\\
		v & \mapsto & exp(v)=g(1) \; ,
	\end{eqnarray*}
	where $g$ is the image by $Exp$ of the constant path $v.$ \end{Definition}

\begin{Definition} \label{reg2} \cite{Ma2013}
	Let $(V,\F, \C)$ be a Fr\"olicher vector space, i.e. a vector space $V$ equipped with a Fr\"olicher structure compatible
	with  vector space addition and  scalar multiplication. The space $(V,\F, \C)$ is \textbf{regular} if there is a smooth map
	$$ \int_0^{(.)} : C^\infty([0;1];V) \rightarrow C^\infty([0;1],V)$$ such that $\int_0^{(.)}v = u$ if and only if $u$ is the unique solution of
	the differential equation
	\[
	\left\{ \begin{array}{l}
	u(0)=0\\
	u'(t)=v(t)\end{array}\right. .\]
	
\end{Definition}

\begin{Definition} \cite{MR2019}
	Let $G$ be a Fr\"olicher Lie group with Lie algebra $\mathfrak{g}.$ Then, $G$ is \textbf{regular with regular Lie algebra} or \textbf{fully regular}
	if both $G$ and $\mathfrak{g}$ are regular in the sense of definitions \ref{reg1} and \ref{reg2} respectively.
\end{Definition}
This definition fits with the terminology of \textbf{fully regular Lie group} due to E. Reyes while writing \cite{MR2019}.

\begin{Theorem} \cite{Ma2013}, inspired from the remarks of \cite{Rob}.
	Let $G$ be a regular Fr\"olicher Lie group with Lie algebra $\mathfrak{g}.$
	Let $\mathfrak{g}_1$ be a Lie subalgebra of $\mathfrak{g}$, and
	set $G_1 = Exp(C^\infty([0;1];\mathfrak{g}_1))(1).$
	If $Ad_{G_1 \cup G_1^{-1}}(\mathfrak{g_1})=\mathfrak{ g}_1,$
	$$i.e. \quad \forall g \in Exp(C^\infty([0;1];\mathfrak{g}_1))(1), \forall v \in \mathfrak{g}_1, \quad Ad_gv \in \mathfrak{g}_1 \hbox{ and } Ad_{g^{-1}}v \in \mathfrak{g}_1,$$ then $G_1$ is a Fr\"olicher subgroup of $G.$
\end{Theorem}

The following example fits with the so-called ``structure group'' of an ILB manifold $GL_\infty = \bigcap_{n \in \N} GL(E_n)$ where $(E_n)_\N$ is an ILB chain. Omori calls it ``topological group with natural differentiation'' \cite{Om} certainly by lack of adequate setting. This example also arises in \cite{GV,DGV} and is treated as a topological group in these references.

\begin{example} \label{omo} \cite{Ma2013}
   Let $(G_{n})_{n\in\N}$ be a sequence of Banach
   Lie groups increasing for $\supset$ (that is, $G_{n+1} \subseteq G_n$ for
   $n \in \N$), and such that the inclusions
   are Lie group morphisms. Let $G=\bigcap_{n\in\N}G_{n}.$ Then, $G$
   is a Frölicher regular Lie group with regular Lie algebra $\mathfrak{g}=\bigcap_{n\in\N}\mathfrak{g}_{n}.$
\end{example}

\begin{Theorem} \label{regulardeformation}\cite{Ma2013}
	Let $(A_n)_{n \in \N^*} $ be a sequence of complete locally convex (Fr\"olicher)
	vector spaces which are regular,
	equipped with a graded smooth multiplication operation
	on $ \bigoplus_{n \in \N^*} A_n ,$ i.e. a multiplication such that for each $n,m
	\in \N^*$,
	$A_n .A_m \subset A_{n+m}$ is smooth with respect to the corresponding Fr\"olicher structures.
	Let us define the (non unital) algebra of formal series:
	$$\A= \left\{ \sum_{n \in \N^*} a_n | \forall n \in \N^* , a_n \in A_n \right\},$$
	equipped with the Fr\"olicher structure of the infinite product.
	
	Then, the set
	$$1 + \A = \left\{ 1 + \sum_{n \in \N^*} a_n | \forall n \in \N^* , a_n \in A_n \right\} $$
	is a Fr\"olicher Lie group with regular  Fr\"olicher Lie algebra $\A.$
	
	Moreover, the exponential map defines a smooth bijection $\A \rightarrow 1+\A.$
\end{Theorem}

\noindent
\textbf{Notation:} for each $u \in \A,$ we note by $[u]_n$ the $A_n$-component of $u.$
\begin{Theorem}\label{exactsequence} \cite{Ma2013}
	Let
	$$ 1 \longrightarrow K \stackrel{i}{\longrightarrow} G \stackrel{p}{\longrightarrow}  H \longrightarrow 1 $$
	be an exact sequence of Fr\"olicher Lie groups, such that there is a smooth section $s : H \rightarrow G,$ and such that
	the trace diffeology  from $G$ on $i(K)$ coincides with the push-forward diffeology from $K$ to $i(K).$
	We consider also the corresponding sequence of Lie algebras
	$$ 0 \longrightarrow \mathfrak{k} \stackrel{i'}{\longrightarrow} \mathfrak{g} \stackrel{p}{\longrightarrow}  \mathfrak{h} \longrightarrow 0 . $$
	Then,
	\begin{itemize}
		\item The Lie algebras $\mathfrak{k}$ and $\mathfrak{h}$ are regular if and only if the
		Lie algebra $\mathfrak{g}$ is regular;
		\item The Fr\"olicher Lie groups $K$ and $H$ are regular if and only if the Fr\"olicher Lie group $G$ is regular.
	\end{itemize}
	
\end{Theorem}

The question:
\vskip 12pt
\centerline{\it Does there exist any non regular Lie group in the sense of Omori ?}
\vskip 12pt

is natural and is raised e.g. in \cite{KM}.
Let us comment more this question.
There are actually many groups for which we cannot prove the existence, or the non existence, of the exponential map. 

There are groups of  units of a (``nice'') algebra, modelled on a complete, Mackey complete, locally convex topological vector space, which are proved to be regular \cite{Glo2002} see also \cite{GN2012} for refined results and open questions.  Let us now exhibit a Lie group, modelled on a locally convex topological vector space, which is not Omori-regular, from \cite{MR2016}.

$\mathbb{R}((X))^*$ is an open subset of $\R((X)), $ and multiplication and inversion are smooth. As a consequence, $\mathbb{R}((X))^*$ is a Lie group modelled on a locally convex topological vector space.
This gives the following theorem:
\begin{Theorem} \label{series3}\cite{MR2016}
	$\mathbb{R}((X))^*$ is not regular in the sense of Omori.
\end{Theorem}
From this example, the substutition map $ X \mapsto \partial^{-1}$ embeds the commutative $\R((X)) = \R((\partial^{-1}))$ in algebras of formal pseudo-differential operators. This enables to state
\begin{Theorem} \cite{MR2016}
	The group of the invertible formal pseudodifferential operators is not regular.
	\end{Theorem}

The second example of non-regular group is historically the first one that we exhibited in a preprint of 2011, published in \cite{Ma2013-2}. Our first contribution is centered on what was one motivating examples for the definition of diffeologies, that we treat here as a Fr\"olicher Lie group.
Let us consider $$Diff_+(]0;1[) = \left\{ f \in C^\infty(]0;1[,]0;1[) \, | \, \lim_{0^+} f = 0, \, \lim_{1^-} f = 1 \hbox{ and } f'>0\right\}$$ equipped with its functionnal diffeology, is not actually a Lie group because there is no known atlas on it. This functional diffeology is the nebulae diffeology associated to the smooth compact-open topology, which assumes uniform convergence on any compact subspace of $]0;1[$ of derivatives at any order.   

\begin{Theorem}\cite{Ma2013-2}
		The Fr\"olicher Lie group $Diff_+(]0;1[)$ is non regular in the sense of definition \ref{reg1}.
	\end{Theorem} 
and as a consequence, we get the same result for $Diff(M),$ when $M$ is a connected, non compact manifold, and when $Diff(M)$ is equipped with the compact-open topology.

\begin{Theorem}\cite{Ma2013-2}
		The Fr\"olicher Lie group $Diff(M)$ is non regular in the sense of definition \ref{reg1}.
	\end{Theorem}

\begin{rem}
		These results are in apparent contradiction with the results in \cite{KM,KMR,GN2017} which state that a group of diffeomorphisms of some example of non compact manifold is regular. This is where we have to mention that the topology is important when considering the group of diffeomorphisms. In \cite{KM}, the $C^\infty$ Whitney topology is considered. In \cite{KMR}, exotic choices of model spaces. In \cite{GN2017}, the open manifold $M$ is a convexe, relatively compact subset of an Euclidian space. In each of these works, the ``asymptotic'' or ``border'' control of the diffeomorphisms is intrisically present in the chosen topology. This is not the case in the topology that we chose in \cite{Ma2013-2}, which may explain the difference between the results obtained. This topology is very classical (see e.g. \cite{Hir} where it is given the name of ``weak topology'') which justifies the geometric study of the infinite dimensional group here produced. 
	
	\end{rem}

\vskip 6pt
\noindent
\underline{\bf Open question:}
After the investigations on regular semi-direct products, the following open question on bicross products of two groups $G$ and $H,$ i.e. when neither $G$ nor $H$ are normal subgroups of the full group that they generate along the lines of the description of \cite{Maj}, denoted by $G \bowtie H,$ is quite natural: 
\vskip 12pt
\centerline{ If $G$ and $H$ are regular, is $G \bowtie H$ regular? }
\vskip 12pt
and, as a related investigation:
\vskip 12pt
\centerline{If $\mathfrak{g}$ and $\mathfrak{h}$ are enlargible, is  $\mathfrak{g} \bowtie \mathfrak{h}$ enlargible?}
\vskip 6pt
\noindent
\underline{\bf Open question:}{On the topologies and the geometries of $Diff(M)$}: Finite configurations \cite{Str} and infinite configurations \cite{A,Alb} appears as a useful phase space for representations of $Diff(M)$ \cite{Ism} and the various topologies of $Diff(M)$ may be classified by their ``push forward'' topology on finite and infinite configurations. Regularity, depending on the topology of $Diff(M),$ may also be classified this way. 

\section{$Diff(M)-$pseudodifferential operators: a restricted class of Fourier integral operators}

We now consider a smooth, boundaryless, compact Riemannian manifold $M.$ The algebra $DO(M),$ graded by the order, is a subalgebra of the algebra of classical pseudo-differential operators
$Cl(M)$ 
which contains some trace-class operators on $L^2(M,\R).$ 
An exposition  of basic facts on pseudo-differential operators defined 
on a vector bundle $E \rightarrow M$ can
be found in \cite{Gil} for definition of pseudo-differential operators 
and of their order, 
(local) definition of symbols and spectral properties. 
We assume known the definition of the algebra of pseudo-differential operators
$PDO(M,E)$,  
classical pseudo-differential operators $Cl(M,E)$. When the vector bundle $E$ is assumed trivial, i.e. $E = M\times V$ or $E = M \times \K^p$ with $\K = \R$ or $\mbbc,$ we use the notation $Cl(M,V)$ or $Cl(M,\K^p)$ instead of $Cl(M,E).$ 
A global symbolic calculus has been described 
by two authors in \cite{BK}, \cite{Wid}, where we can see how the 
geometry of the base manifold $M$ furnishes an obstruction to generalize 
local formulas of composition and inversion of symbols.
\vskip 12pt
\noindent
\textbf{Notations.} 
We note by  $ PDO (M, \mbbc) $
(resp.  $ PDO^o (M, \mbbc)
$, resp. $Cl(M,\mbbc)$) the space of
pseudo-differential operators (resp.
pseudo-differential operators of order o, resp. classical pseudo-differential operators) acting on smooth
sections of $E$, and by $Cl^o(M,\mbbc)= PDO^o(S^1,\mbbc) \cap Cl(S^1,\mbbc)$ the space of classical 
pseudo-differential operators of order $o$.
If we set
$ PDO^{-\infty}(M,\mbbc) = \bigcap_{o \in \Z} PDO^o(M,\mbbc),$
we notice that it is a two-sided ideal of $PDO(M,\mbbc)$, and we define the quotient algebra of formal PDOs:
$$\mathcal{F}PDO(M, \mbbc) = PDO(M,\mbbc) / PDO^{-\infty}(M,\mbbc),$$
with analogous notations with script $\mathcal{F}$ for other algebras of PDOs. 
With the notations that we have set before, a scalar
Fourier-integral operator of order $o$ 
is an operator $$ A : C^\infty(M, \mbbc) \rightarrow C^\infty(M,\mbbc)$$
such that, for any smooth partitions of the unit $(s_i)_I$ indexed by a finite set $I,$ $\forall (i,j) \in I^2,$
\begin{eqnarray} \label{localization} A_{k,j}(f) &  = & \int_{supp(s_j)} e^{-i\phi(x,\xi)}\sigma_{k,j}(x,\xi) \hat{(s_j.  
		f)} (\xi) d\xi \end{eqnarray}
where $\sigma_{k,j} \in C^\infty( supp(s_j) \times \R, \mbbc)$ satisfies classical estimates on symbols of PDOs
and where, on any domain $U$ of a chart on $M$, 
$$\phi(x,\xi) : T^*U-U\approx U\times \R^{dim M}-\{0\}\rightarrow \R$$
is a smooth map, positively homogeneous of degree 1 fiberwise and such that $$\det\left(\frac{\partial^2\phi }{\partial x \partial_\xi} \right) \neq 0.$$
Such a map is called \textbf{phase} function.
(In these formulas, the maps are read on local 
charts but we preferred to only mention this aspect and not 
to give heavier formulas and notations) 
Notice that, in order to define an operator $A,$ the choice of 
$\varphi$ and $\sigma_{k,l}$ is not a priori unique for 
general Fourier integral operators.
Let $E = S^1 \times \mbbc^k$ be a trivial smooth vector bundle over $S^1$.
An operator acting on $C^\infty(M,\mbbc^n)$ is Fourier integral operator 
(resp. a pseudo-differential operator) 
if it can be viewed as a $(n \times n)$-matrix of Fourier integral operators with same phase function 
(resp. scalar pseudo-differential operators).

The topological structures can be derived both from 
symbols and from kernels, as we have quoted before
but principally because there is the exact sequence described below with slice.
At the level of units of these sets, i.e. of groups of invertible operators, 
the existence of the slice is also crucial.  
In the papers \cite{ARS,ARS2,OMY1,OMY2,OMYK3,OMYK4,OMYK5,OMYK6,OMYK7,OMYK8,RS}, 
the group of invertible Fourier integral operators receives first 
a structure of topological group, 
with in addition a differentiable structure, e.g. a Fr\"olicher structure,
which recognized as a structure of generalized Lie group, see e.g. \cite{Om}. 

We have
to say that, with the actual state of knowledge, using \cite{KM}, we can give  
a manifold structure (in the convenient setting described by Kriegl and Michor or in the category of Fr\"olicher spaces following \cite{Ma2013}) 
to the corresponding Lie groups.

\begin{rem}
	In \cite{ARS,ARS2,OMY1,OMY2,OMYK3,OMYK4,OMYK5,OMYK6,OMYK7,OMYK8,RS}, the group $K$ considered is the group of 1-positively homogeneous symplectomorphisms $Diff_\omega(T^*M - M)$ where $\omega$ is the canonical symplectic form on the cotangent bundle. The local section considered enables to build up the phase function of a Fourier integral operator from such a symplectic diffeomorphism inside a neighborhood of $Id_M.$ There is a priori no reason to restrict the constructions to classical pseudo-differential operators of order 0, and have groups to Fourier integral operators with symbols  in wider classes. This remark appears important to us because the authors cited before restricted themselves to classical symbols, and principaly to bounded operators.
\end{rem}

\subsection{$PDO(M,E),$ $Aut(E),$ $Diff(M)$ and a restricted class of FIOs, based on \cite{Ma2016-1,Ma2021-1}}
We get now to another group:
\begin{Theorem} \label{FIOdiff} \cite{Ma2016-1}
	Let $H$ be a regular Lie group of pseudo-differential operators acting on smooth sections of  a trivial bundle $E \sim V \times M \rightarrow M.$
	The group $Diff(M)$ acts smoothly on  $C^\infty(M,V),$ and is assumed to act smoothly on $H$ by adjoint action.
	If $H$ is stable under the $Diff(M)-$adjoint action, then there exists a corresponding regular Lie group $G$ of Fourier integral operators through the exact sequence:
	$$ 0 \rightarrow H \rightarrow G \rightarrow Diff(M) \rightarrow 0.$$ If $H$ is a Fr\"olicher (resp. a Fr\'echet) Lie group, then $G$ is a Fr\"olicher (resp. a Fr\'echet) Lie group. 
\end{Theorem}

\begin{rem}\cite{Ma2016-1}
	The pseudo-differential operators can be classical, log-polyhomogeneous, or anything else. Applying the formulas of ``changes of coordinates'' (which can be understood as adjoint actions of diffeomorphisms) of e.g. \cite{Gil}, one easily gets the result.
\end{rem}

\begin{rem} \cite{Ma2016-1} The composition operator $$ f \in C^\infty(M,E) \mapsto f\circ g,$$ for $g \in C^\infty(M,M),$ is a linear operator with distributional kernel $$K_g(x,y)= \delta_{g(x),y}\in \mathcal{D}'(M \times M)$$ where $\delta$ is the Dirac distribution. This is never the kernel of a pseudo-differental operator, unless $g=Id_M,$ since the kernel of a pseudo-differential operator must be smooth off-diagonal \cite{Dieu2}.\end{rem}

\vskip 6pt
\noindent\underline{\bf Open problem :}
	One can compare the condition ``$H$ is stable by the $Diff(M)$-adjoint action'' with similar results of \cite{ARS,ARS2,BK,Om} e.g.  replacing $g \in Diff(M)$ by a symplectic diffeomorphism $g \in Diff_\omega(T^*M-M).$ In \cite{Ma2016-1}, the group under consideration seems different. However, a Fourier integral operator does not have a unique phase function \cite{Horm}.
	Some restricted classed of such operators are already considered in the literature under the name of $G-$pseudo-differential operators, see e.g.\cite{SS}, but the groups considered are discrete (amenable) groups of diffeomorphisms. All these classes have to be compared.

\begin{Definition} \cite{Ma2016-1}
	Let $M$ be a compact manifold and $E$ be a (finite rank) trivial vector bundle over $M.$
	We define $$ FIO_{Diff}(M,E) = \left\{ A \in FIO(M,E) | \phi_A(x,\xi) = g(x).\xi; g \in Diff(M) \right\}.$$
\end{Definition}
The subset of invertible operators $FIO_{Diff}^*(M,E)$ is obviously a group, that decomposes as 
$$ 0 \rightarrow PDO^*(M,E) \rightarrow FIO_{Diff}^*(M,E) \rightarrow Diff(M) \rightarrow 0$$
with global smooth section $$ g \in Diff(M) \mapsto (f \in C^\infty(S^1,E) \mapsto f \circ g ).$$
Hence, Theorem \ref{FIOdiff} applies trivially to the following context:

\begin{Proposition} \cite{Ma2016-1}
	Let $FCl^{0,*}_{Diff}(M,E) $ be the set of operators $ A \in FIO_{Diff}^*(M,E)$ such that $A$ has a 0-order classical symbol. Then we get the exact sequence:
	$$ 0 \rightarrow Cl^{0,*}(M,E) \rightarrow FCl_{Diff}^{0;*}(M,E) \rightarrow Diff(M) \rightarrow 0$$
	and $FCl^{0,*}_{Diff}(M,E) $ is a regular Fr\"olicher Lie group, with Lie algebra isomorphic, as a vector space, to $Cl^0(M,E) \oplus Vect(M).$
\end{Proposition}

This setting can be extended to a trivial complex vector bundle $E \rightarrow M.$ We remark that the group $Diff(M)$ 
cannot be recovered in this group of operators. 
On a non trivial bundle $E,$ let us consider the group of bundle automorphism $Aut(E).$ 
The gauge group, which can be identified with the gourp of invertible $0-$ordre differential operators $DO^{0^*}(M,E)$ is naturally embedded in $Aut(E)$ and the bundle projection $E \rightarrow M$ 
induces a group projection $\pi:Aut(E) \rightarrow Diff(M).$
Therefore we get a short exact sequence 
$$ 0 \rightarrow DO^{0,*}(M,E) \rightarrow Aut(E) \rightarrow Diff(M) \rightarrow 0.$$
Moreover, there exists a local slice $U \subset Diff(M) \rightarrow Aut(E),$ where $U$ is a $C^0-$open neighborhood on $Id_M,$ 
which shows that $Aut(E)$ is a regular Fr\'echet Lie group \cite{ACMM1989}.
Therefore, the smallest group spanned by $PDO^*(M,E)$ and $Aut(E)$ is such that:
\begin{itemize}
	\item the projection $E \rightarrow M$ induces a map $Aut(E) \rightarrow Diff(M)$ with kernel $DO^0(M,E) = Aut(E) \cap PDO(M,E)$
	\item $Ad_{Aut(E)}(PDO(M,E)) = PDO(M,E)$  
\end{itemize} 
therefore we can consider the space of operators on $C^\infty(M,E)$
$$ FIO_{Diff}^*(M,E) = Aut(E) \circ PDO^*(M,E).$$

According to \cite{Ma2016-1},
the map $$ (B,A) \in Aut(E) \times PDO^*(M,E) \mapsto \pi(B) \in Diff(M)$$ induces a ``phase map'' $$\tilde{\pi}:FIO_{Diff}^*(M,E) \rightarrow Diff(M).$$

\begin{Theorem}\label{splitfiodiff} \cite{Ma2016-1}
	There is a short exact sequence of groups : 
	$$ 0 \rightarrow PDO^*(M,E) \rightarrow FIO^*_{Diff}(M,E) \rightarrow Diff(M) \rightarrow 0$$
	and, if $H\subset PDO^*(M,E)$ is a regular Fr\'echet or Fr\"olicher Lie group of operators that contains the gauge group of $E,$
	if $K$ is a regular Fr\'echet or Fr\"olicher Lie subgroup of $Diff(M)$ such that there exists a local section  $K \rightarrow Aut(E),$
	the subgroup $G = K \circ H$ of $FIO^*_{Diff}(M,E)$ is a regular Fr\'echet Lie group from the short exact sequence:
	$$ 0 \rightarrow H \rightarrow G \rightarrow K \rightarrow 0.$$
\end{Theorem}

A similar defnition is given in \cite{Pay2013} in order to motivate the study of Chern-Weil forms for connections with curvature valued in unbounded (first order pseudodifferential) operators. However, the full development of these groups are results of our own to our knowledge. 
\subsection{Formal and non-formal $Diff(S^1)-$pseudodifferential operators, based on \cite{Ma2021-1}}
Now we present a restricted class of groups of 
Fourier integral operators which we will call 
$Diff_+(S^1)$-pseudodif\-fer\-en\-tial operators following \cite{Ma2016-1}.
These groups appear as central extensions of $Diff_+(S^1)$ by groups of (often bounded) pseudodifferential 
operators.
Using odd class PDOs, we define the following group:

\begin{Definition}
	The group $FCl_{Diff(S^1),odd}^{0,*}(S^1,V)$ is the regular 
	Fr\'echet Lie group $G$ obtained in Theorem 
	$\ref{splitfiodiff}$ with $H=Cl^{0,*}_{odd}(S^1,V),$ the group of invertible, bounded and odd class pseudodifferential operators (in the Kontsevich and Vishik terminology)  
\end{Definition}

Following \cite{Ma2016-1}, we remark that operators $A$ in this group 
can be understood as operators in $Cl^{0,*}_{odd}(S^1,V)$ twisted 
by diffeomorphisms, this is, 
\begin{equation} \label{aux}
	A = B \circ g 
\end{equation}
for unique $g \in Diff(S^1)$ and unique 
$B \in Cl^{0,*}_{odd}(S^1,V)$, 
and also that its Lie algebra is isomorphic as a vector space to
$Cl^0_{odd}(S^1,V)\oplus Vect(S^1)$, in which $Vect(S^1)$ is the 
space of smooth vector fields on $S^1$.

Now we note that 
the group $Diff(S^1)$ decomposes into two connected components 
$Diff(S^1) = Diff_+(S^1) \cup Diff_-(S^1),$
where the connected component of the identity, $Diff_+(S^1)$, is 
the group of orientation preserving diffeomorphisms 
of $S^1$. We make the following definition:

\begin{Definition}
	The group $FCl_{Diff_+(S^1),odd}^{0,*}(S^1,V)$ is the regular 
	Fr\'echet Lie group of all operators in 
	$FCl_{Diff(S^1),odd}^{0,*}(S^1,V)$ whose phase diffeomorphisms
	lie in the group $Diff_+(S^1).$
\end{Definition}

Let us describe additional structures of decompositions that arise only when $M=S^1.$
\begin{Theorem}\label{SY} \cite{Ma2021-1,MR2020}
	There is a short 
	exact sequence of Lie groups:
	$$
	{ 1} \rightarrow Cl^{-1,*}_{odd}(S^1,V) \rightarrow 
	FCl_{Diff_+(S^1),odd}^{0,*}(S^1,V) \rightarrow 
	DO^0(S^1,V)\rtimes Diff_+(S^1) \rightarrow { 1},$$
	where $Cl^{-1,*}_{odd}(S^1,V)$ is the group od invertible classical PDOs that are equal to $Id$ up to an operator of order $-1.$ 
\end{Theorem}

Let us summarize our constructions. The descriptionsemi-direct product of 
{Fr\'echet} Lie groups 
$$
FCl^{0,*}_{Diff_+(S^1),odd}(S^1,V) = 
Cl^{0,*}_{odd}(S^1,V)\rtimes Diff_+(S^1)
$$
%
%
can be completed by the following diagram in which vertical and 
horizontal lines are short exact sequences of Lie groups:
$$\begin{array}{ccccccccc}
	&&&&1&&1&&\\
	&&&&\downarrow&&\downarrow&&\\
	1 & \rightarrow & Cl^{-1,*}_{odd}(S^1,V) & \rightarrow & Cl^{0,*}_{odd}(S^1,V) & \rightarrow & DO^{0,\ast}(S^1,V)  
	& \rightarrow & 1\\
	&&\|&&\downarrow &&\downarrow &&\\
	1 & \rightarrow & Cl^{-1,*}_{odd}(S^1,V) & \rightarrow & FCl^{0,*}_{Diff_+(S^1),odd}(S^1,V) & \rightarrow & 
	DO^{0,\ast}(S^1,V)\rtimes Diff_+(S^1) & \rightarrow & 1\\
	&&&&\downarrow&&\downarrow&&\\
	&&&&Diff_+(S^1)&=&Diff_+(S^1)&&\\
	&&&&\downarrow&&\downarrow&&\\
	&&&&1&&1&&\\
\end{array}
$$

The corresponding diagram of Lie algebras, all of them embedded in 
$Cl_{odd}(S^1,V)$ is:

$$\begin{array}{ccccccccc}
	&&&&0&&0&&\\
	&&&&\downarrow&&\downarrow&&\\
	0 & \rightarrow & Cl^{-1}_{odd}(S^1,V) & \rightarrow & Cl^{0}_{odd}(S^1,V) & \rightarrow & DO^{0}(S^1,V)  & 
	\rightarrow & 0\\
	&&\|&&\downarrow &&\downarrow &&\\
	0 & \rightarrow & Cl^{-1}_{odd}(S^1,V) & \rightarrow & Cl^{0}_{odd}(S^1,V) \rtimes Vect(S^1) & \rightarrow & 
	DO^{0}(S^1,V)\rtimes Vect(S^1) & \rightarrow & 0\\
	&&&&\downarrow&&\downarrow&&\\
	&&&&Vect(S^1)&=&Vect(S^1)&&\\
	&&&&\downarrow&&\downarrow&&\\
	&&&&0&&0&&\\
\end{array}
$$

After this long description of new groups, let us gather notations for groups of (non-formal) $Diff(S^1)-$pseudodifferential operators from \cite{Ma2021-1}.
\begin{Definition}
	\begin{enumerate}
		\item The group $FCl_{Diff(S^1)}^{*}(S^1,V)$ is the infinite dimensional group defined by 
		$$FCl_{Diff(S^1)}^{*}(S^1,V) = \left\{ A = B \circ g \, | \, B \in Cl^*(S^1,V) \hbox{ and } g \in Diff(S^1) \right\}.$$
		\item The group $FCl_{Diff(S^1)}^{0,*}(S^1,V)$ is the infinite dimensional group defined by 
		$$FCl_{Diff(S^1)}^{0,*}(S^1,V) = \left\{ A = B \circ g \, | \, B \in Cl^{0,*}(S^1,V) \hbox{ and } g \in Diff(S^1) \right\}.$$
		\item The group $FCl_{Diff(S^1),odd}^{*}(S^1,V)$ is the infinite dimensional group defined by 
		$$FCl_{Diff(S^1),odd}^{*}(S^1,V) = \left\{ A = B \circ g \, | \, B \in Cl^*_{odd}(S^1,V) \hbox{ and } g \in Diff(S^1) \right\}.$$
		\item The group $FCl_{Diff(S^1),odd}^{0,*}(S^1,V)$ is the infinite dimensional group defined by 
		$$FCl_{Diff(S^1),odd}^{0,*}(S^1,V) = \left\{ A = B \circ g \, | \, B \in Cl^{0,*}_{odd}(S^1,V) \hbox{ and } g \in Diff(S^1) \right\}.$$
	\end{enumerate}
\end{Definition}

\begin{rem}
	The decomposition $A = B \circ g$ is unique \cite{Ma2016-1}, and the diffeomorphism appears as the phase of the Fourier integral operator. 	
\end{rem}

\begin{rem} the group $Diff(S^1)$ decomposes into two connected components 
	$Diff(S^1) = Diff_+(S^1) \cup Diff_-(S^1)\; ,$
	where the connected component of the identity, $Diff_+(S^1)$, is the group of orientation preserving diffeomorphisms 
	of $S^1$. By the way, we can replace $Diff(S^1)$ by $Diff_+(S^1)$ in the previous definition.
\end{rem} 


\begin{Definition} \cite{Ma2021-1}
	Let $(A,A') \in (FCl_{Diff(S^1)}^{*}(S^1,V))^2,$ with $A = B \circ g$ and $A' = B' \circ g'$ as before. Then $$A \equiv A' \quad \Leftrightarrow \quad \left\{\begin{array}{rcl} g = g' && \\
		B - B' & \in & Cl^{-\infty}(S^1,V) \end{array} \right.$$
	The set of equivalence classes with respect to $\equiv$ is noted as $\mathcal{F}FCl_{Diff(S^1)}^{*}(S^1,V)$ and is called the set of formal $Diff(S^1)-$pseudodifferential operators.
\end{Definition}

The same spaces of formal operators can be constructed using orientation-preserving diffeomorphisms of $S^1,$ odd class pseudodifferential operators and so on. We do not feel the need to give here redundant constructions, and obvious notations.

\begin{Theorem} 
	Let $$G = \left\{ A \in Cl^{0;*}(S^1,V) \, | \, A = Id + B, \, B \in Cl^{-\infty}(S^1,V)\right\}.$$
	Then \begin{itemize}
		\item  $G \vartriangleleft FCl^*_{Diff(S^1)}(S^1,V),$ 
		\item given $(A,A') \in FCl^*_{Diff(S^1)}(S^1,V)^2,$
		$$ A \equiv A' \Leftrightarrow AA'^{-1} \in G$$
	\end{itemize}
	which implies that $$\mathcal{F}FCl_{Diff(S^1)}^{*}(S^1,V)= FCl_{Diff(S^1)}^{*}(S^1,V)/G.$$ By the way, $\mathcal{F}FCl_{Diff(S^1)}^{*}(S^1,V)$ is a group. Moreover, \begin{equation} \label{str} \mathcal{F}FCl_{Diff(S^1)}^{*}(S^1,V) = \mathcal{F}Cl^{*}(S^1,V) \rtimes Diff(S^1).\end{equation}
\end{Theorem} 

We have the following commutative diagram:
	\begin{equation} \label{diagram1} \begin{array}{ccccccccc}
			&&  && 1 && 1 &&\\
			&&  && \downarrow && \downarrow &&\\
			1 & \rightarrow & G & \rightarrow & Cl^*(S^1,V) & \rightarrow & \F Cl^*(S^1,V) & \rightarrow & 1 \\
			&& \| && \downarrow && \downarrow &&\\ 
			1 & \rightarrow & G & \rightarrow & FCl^*_{Diff(S^1)}(S^1,V) & \rightarrow & \F FCl^*_{Diff(S^1)}(S^1,V) & \rightarrow & 1 \\
			&&  && \downarrow && \downarrow &&\\
			&&  && Diff(S^1) & = & Diff(S^1) &&\\
			&&  && \downarrow && \downarrow &&\\
			&&  && 1 && 1 &&\\
	\end{array} \end{equation}
	The three squares commute, the two horizontal lines are short exact sequences as well as the central culumn. 
%

\begin{Proposition} \cite{Ma2021-1}
	There is a natural structure of infinite dimensional Lie group on $\mathcal{F}FCl_{Diff(S^1)}^{*}(S^1,V)$, and
	its Lie algebra (defined by germs of smooth paths) reads as $$\mathcal{F}Cl(S^1,V) \rtimes Vect(S^1).$$ 
\end{Proposition}

\vskip 6pt
\noindent
\underline{\bf Open problem:} Study the cohomology of $Cl(S^1,V) \rtimes Vect(S^1)$ compared with the cohomology of $\mathcal{F}Cl(S^1,V) \rtimes Vect(S^1),$ that heuriscically can be deduced from the cohomology of each lie algebra. Make the same investigations with odd class PDOs. Associated investigations have been performed in e.g. \cite{PL2007} but still have to be completed. One technical difficulty remains on the non existence of any local slice $\mathcal{F}Cl(S^1,V) \rightarrow Cl(S^1,V)$ which is well-explained in \cite{Dieu2}.
\section{$GL_{res}$ and its subgroups of Fourier-integral operators}
This section is based on \cite{Ma2021-1} which extends the remarks made in \cite{Ma2016-1}. Let us now turn to the Lie group of bounded operators described in \cite{PS}:
$$ GL_{res}(S^1,\mathbb{C}^k) = \lbrace u \in GL(L^2(S^1,\mathbb{C}^k)) \hbox{ such that } [\epsilon(D),u] \hbox{ is Hilbert-Schmidt }\rbrace$$
with Lie algebra
$$ \mathcal{L}(S^1,\mathbb{C}^k) = \lbrace u \in \mathcal{L}(L^2(S^1,\mathbb{C}^k)) \hbox{ such that } [\epsilon(D),u] \hbox{ is Hilbert-Schmidt }\rbrace.$$

\begin{Proposition} \cite[Theorem 25-26]{Ma2016-1}
	$FCl_{Diff_+(S^1)}^{0,*}(S^1,{\mathbb{C}}^k) \subset GL_{res}(S^1,\mathbb{C}^k)$
\end{Proposition}

Let us now give a new light on an old result present in \cite{PS} from a topological viewpoint, expressed by remarks but not stated clearly in the mathematical litterature to our knowledge. We propose here an approach for the proof, more easy and much more fast, and adapted to our approach of (maybe generalized) differentiability prior to topological considerations. 
\begin{Lemma}\cite{Ma2021-1}
	The injection map $ Diff_+(S^1) \hookrightarrow GL_{res}(S^1,\mathbb{C}^k)$
	is not differentiable.
\end{Lemma}

From this Lemma, the next theorem is straightforward:
\begin{Theorem}\cite{Ma2021-1}
	The injection maps 		$FCl_{Diff_+(S^1)}^{0,*}(S^1,{\mathbb{C}}^k) \hookrightarrow GL_{res}(S^1,\mathbb{C}^k)$
	and $DO^{0,*}(S^1,\mathbb{C}^k) \rtimes Diff_+(S^1) \hookrightarrow GL_{res}(S^1,\mathbb{C}^k)$ are not differentiable.
\end{Theorem}

\vskip 6pt
\noindent
\underline{\bf Open problem:} $Cl(S^1,V) \rtimes Vect(S^1)$ is a potentially interesting candidate to replace $GL_{res},$ in particular for constructions related to the determinant bundle \cite{PS} and KdV equation \cite{SW1985}. An open problem consists in classifying all line bundles over this new group.

\section{Ambrose-Singer theorem on infinite dimensional principal bundles based on \cite{Ma2004,Ma2013}}
The construction of (locally trivializable) infinite dimensional principal bundles of frames (as they are described e.g. in \cite{Ma2006}) can lead, for example in the setting of ILB manifolds, to structure groups that are not infinite dimensional Lie groups modelled on locally convex topological vector spaces, but only Fr\"olicher Lie groups. These groups are typically of the kind of those described in Example \ref{omo}, see \cite{GV} for more details. From another viewpoint, even starting from a ``rather nice'' connection on a principal bundle with regular structure group, the classical construction of the holonomy group by path-lifting leads to a holonomy group with structure that a priori carries no atlas except in particular classes of connections in  \cite{Vas1978} in the context of Banach principal bundles. Indeed, the application of any Fr\"obenius type theorem that mimick the proofs of Ambrose and Singer \cite{AS} are far away from the actual state of knowledge. The question of the construction of the holonomy group is quoted in \cite{Pen,F2} and has a partial answer for flat connections in the $c^\infty-$setting in \cite{KM} (which inspired \cite{Ma2004}).

Let $p \in P$ and
$\gamma$ a smooth path in $P$ starting at $p,$ defined on $[0;1].$
Let $H\gamma (t) = \gamma(t)g(t)$ where $g(t) \in C^\infty([0;1];\mathfrak{g})$ is a path satisfying the differential equation:
$$\left\{ \begin{array}{c} \theta \left( \dd_t H\gamma(t) \right) = 0  \\ H\gamma(0)=\gamma(0) \end{array} \right.$$
The first line of the definition is equivalent to the differential equation
$g^{-1}(t)\dd_tg(t) = -\theta(\dd_t\gamma  (t))$ which is integrable, and the second to the initial condition $g(0)=e_G.$
The map $H(.)$ defines what \cite{Ig} calls a diffeological connection and what \cite{Ma2013} calls path lifting. This enables us to consider the
holonomy group of the connection.
Notice that a straightforward adaptation of the arguments of \cite{Li} shows that the
holonomy group is invariant (up to conjugation) under the choice of the basepoint $p.$
Now, we assume that $dim(M)\geq 2.$ We fix a connection $\theta$ on $P.$

\begin{Definition}
	Let $\alpha \in \Omega(P;\mathfrak{g})$ be a $G-$invariant form. Let $\nabla \alpha = d\alpha - {\frac{1}{ 2}}[\theta,\alpha]$ be the horizontal derivative of $\alpha.$
	We set $$ \Omega = \nabla \theta $$ the curvature of $\theta.$
\end{Definition}
We now turn to reduction of the structure group. A preliminary version, available with (``classical'' Fr\'echet) manifold settings was already given in \cite{Ma2004}. 

\begin{Theorem} \label{Courbure} \cite{Ma2013}
	We assume that $G_1$ and $G$ are regular Fr\"olicher groups
	with regular Lie algebras $\mathfrak{g}_1$ and $\mathfrak{g}.$
	Let $\rho: G_1 \mapsto G$ be an injective morphism of Lie groups.
	If there exists a connection $\theta$ on $P$, with curvature $\Omega$, such that, for any smooth 1-parameter family $Hc_t$ of horizontal paths starting at $p$, for any smooth vector fields $X,Y$ in $M$,
	\begin{eqnarray} s, t \in [0,1]^2 & \rightarrow & \Omega_{Hc_t(s)}(X,Y)  \label{g1}\end{eqnarray}
	is a smooth $\mathfrak g_1$-valued map (for the $\mathfrak g _1 -$ diffeology),
	\noindent
	and if $M$ is simply connected, then the structure group $G$ of $P$ reduces to $G_1,$ and the connection $\theta$ also reduces.
\end{Theorem}
We can now state the announced Ambrose-Singer theorem, using the terminology of \cite{Rob} for the classification of groups by properties of the exponential maps:

\begin{Theorem}
	\label{Ambrose-Singer} \cite{Ma2013}, partly from \cite{Ma2004}.
	Let $P$ be a principal bundle of basis $M$ with regular Fr\"olicher structure group $G$ with regular Lie algebra $\mathfrak{g}.$
	Let $\theta$ be a connection on $P$ .
	
	\begin{enumerate}
		\item For each $p \in P,$ the holonomy group $\Hol_p^L$ is a
		diffeological subgroup of $G$, which does not depend on the choice of
		$p$ up to conjugation.

		\item There exists a second holonomy group $H^{red},$ $\Hol \subset H^{red},$
		which is the smallest structure group for which there is a subbundle $P'$ to
		which $\theta$ reduces. Its Lie algebra is spanned by the curvature elements, i.e.
		it is the smallest integrable Lie algebra which contains the curvature elements.
		
		\item If $G$ is a Lie group (in the classical sense) of type I or II,
		there is a (minimal) closed Lie subgroup $\bar{H}^{red}$ (in the classical sense) such that $H^{red}\subset \bar{H}^{red},$
		whose Lie algebra is the closure in $\mathfrak{g}$ of the Lie algebra of $H^{red}.$ $\bar{H}^{red}$
		is the smallest closed Lie subgroup of $G$ among the structure groups
		of closed sub-bundles $\bar{P}'$ of $P$ to which $\theta$ reduces.
	\end{enumerate}
\end{Theorem}

From \cite{Ma2013} again, we have the following:
\begin{Proposition} \label{0-courbure}
	If the connection $\theta$ is flat and if $M$ is connected and simply connected, then , for any path $\gamma$
	starting at $p \in P,$ the map $$\gamma \mapsto H\gamma(1)$$ depends only on $\pi(\gamma(1))\in M$ and defines
	a global smooth section $M \rightarrow P.$ Therefore, $P = M \times G.$
\end{Proposition}

 \chapter{Contributions to simplicial gauge theories and decision theory}
 Quantum gravity theories \cite{RV2014} both quantize space and time and ignore the continuum limit. In their approach, connections are heuristiccally difference operators (from the viewpoint of affine connections) or holonomy elements (from the viewpoint of principal connections). The problem of discretization of connections is a problem that goes back a long way in time before investigations in quantum gravity. In the quantized approach, one way to define the Feynman integral over the space of connections is the use of a finite element method for discretizing connections, inspired by \cite{Wh}, see e.g. \cite{AHM,AZ1990,SSSA}. This standard discretization procedure leads to several technical problems and the convergence of integrals in the space of connections for e.g. Yang-Mills or Chern-Simons theories require heuristic gauge reductions generically called {\bf gauge fixing.}  The problem of gauge fixing in gauge theories is of fundamental importance for explicit calculations, before or after the quantization procedure, see e.g. \cite{AZ1990,Re1997,Hah2004,Lim2012,HCR2015} for a non-exhaustive list of references.   This approach works quite well for abelian gauge theories, but a well-chosen gauge-fixing actually produces with many difficulties some explicit results. Moreover, the invariance under gauge fixing is actually partially justified with heuristic arguments. 
 
 For all these reasons it is fully justified to search for discretization schemes that give alternate approaches to \cite{Wh}. The one proposed in \cite{RV2014} is here rigorously described, and has lead to unexpected developments in the field of decision theory, on astandard object defined as follows. 
  A \textbf{pairwise comparisons matrix} is a square matrix $A$ with $\R_+^*-$coefficients, such that $\forall i, a_{ii}=1$ 	and such that $a_{i,j} = a_{j,i}^{-1}.$ We note by $PC(\R_+^*)$ the space of pairwise comparisons matrices, and by $PC_n(\R_+^*)$ the space of $n \times n$-pairwise comparisons matrices.  
 Such matrices are used in information theory, with various applications: complex networks (they are used in wireless networks for example, see e.g. \cite{LLA2012}), decision (analysis of situations in e.g. nuclear or military projects, see e.g. \cite{CW1985}), medecine (symptomatic analysis, psychology), management (see e.g. \cite{Cd'AS2010,KKL2014}), economy, Brain modelization \cite{Ma2017-1} etc...  
 The use of these matrices do not envolve the classical algebraic structures on spaces of matrices, because they are understood as ``tables'' of scores when comparing states, objects or individuals. The coefficient $a_{i,j}$ compares the i-th state with the j-th state, which explains more clearly the rule $a_{i,j} = a_{j,i}^{-1}.$ A pairwise comparisons matrix is called \textbf{consistent} if and only if $\forall (i,j,k), a_{i,j} a_{j,k} = a_{i,k},$
 and one of the goals of applications is more to minimize, in some sense, inconsistency, than to obtain strict consistency. In order to evaluate inconsistency, one uses \textbf{inconsistency indicators}, which are mostly Saaty's \cite{S1977} and Koczkodaj's \cite{K1993}. 

The link that we now describe in more details is new, after the preliminary work \cite{Y1999} that links gauge theories with foreign exchange markets.
 \section{A new discretization scheme for gauge theories based on \cite{Ma2018-3}} \label{ss:discreteholonomy}

 The present approach of discretization is based on the following paradigm: the convergence of the discretization \textbf{ for a fixed connection } must be preserved, as well as gauge covariance of the curvature in the discretization scheme. 

 Let $\pi : P \rightarrow M$ be a principal bundle of connected Riemannian base, with structure group $G,$ equipped with a prescribed triangulation or cubification $\tau.$ The canonical maps induced by $\pi$ on relevant objects will be also noted by $\pi$ in the sequel when it carries no ambiguity. The nodes of this triangulation or cubification are assumed indexed by $\mathbb{N},$ noted by $(s_n)_\mathbb{N}$ (the manifold $M$ can be non compact). Recall that, for a fixed index $i_1,...i_n,$ $St(s_{i_1},...s_{i_n})$ is the domain described by the simplexes or the cubes with nodes $s_i.$  We note by $\mathcal{C}$ the space of connections on $P.$ Let $\theta \in \mathcal{C}.$ Fixing $s_0$ as a basepoint  and $p_0 \in \pi^{-1}(s_0),$ 
 \begin{enumerate}
 	\item Let $$j = \min \left\{ i \in \N^* | s_i \in St(s_0) \right\}.$$ We define $g_{0,j}=1$  and $p_j$ the endpoint of the horizontal path over $[s_0,s_j]$ with starting point $p_0.$ Let $I_2 = \{0;j\}.$
 	\item Assume that $I_n$ exists, and that, 
 	\begin{itemize}
 		\item $\forall i \in I_n, $ we have constructed $p_i \in \pi^{-1}(s_i)$ and 
 		\item $\forall (i,j) \in I_n^2,$ with $i<j$, $g_{i,j}$ is the holonomy of $[s_i,s_j]$, starting at $p_i,$ i.e.  $p_j. g_{i,j}$ is the endpoint of the horizontal path over $[s_i,s_j]$ with starting point $p_i.$ 
 	\end{itemize}
 	Let $$j = \min \left\{ i \in \N- I_n | s_i \in St(s_k; k \in I_n) \right\}.$$ We define 
 	\begin{itemize}
 		
 		\item  $k = \min\{ i \in I_n| s_i \in St(s_j)\}$ and let $p_j$ the endpoint of the horizontal path over $[s_i,s_j]$ starting at $p_i.$   
 		\item for $i \in I_n,$ $g_{i,j}$ is defined such as $p_j.g_{i,j}$ is the endpoint of $[s_i,s_j]$ starting at $p_i.$
 		\item $I_{n+1} = I_n \cup \{j\}.$
 	\end{itemize}
 	
 \end{enumerate}
 
 The discretization thus describes the holonomy of the connection along the 1-vertices. We have a first sequence $(p_n)_\N$ which stands as a slice of the pull-back $\left((s_n)_\N\right)^* P$ and if $\mathcal{K}_1$ is the 1-skeleton of $\tau,$ the family $(g_{i,j})_{i<j},$ expresses holonomy elements of the connection $\theta$ on the vertices of $\mathcal{K}_1.$ The holonomy of a smooth path $\gamma,$ for a fixed connection, can be approximated by the discretized holonomies computed along a piecewise smooth path along the vertices of the triangulation, close enough to $\gamma.$  In the sequel, we note by $|\sigma|$ the length (resp. $n-$dimensinal the Haussdorf volume) of the 1-vertex (resp. the $n-$simplex or $n-$cube) $\sigma.$
 Comparing to Whitney's discretization \cite{Wh}, this scheme does not depend on any exterior trivialization of the principal bundle $P.$ In the sequel, we work with triangulations but the same can be done for cubifications.
 These discretizations are used to define the new discrete analogs of connections and curvature. 
  Let $(\tau_n)$ be a sequence of triangulations of $M$ such that $\tau_{n} \subset \tau_{n+1}$ (subtriangulations) and such that the length of 1-vertices converge uniformly to $0.$ Then, for fixed $s \in \tau_n \subset... \subset \tau_{n+p} \subset...,$ we have the following 
  \begin{itemize}
  	\item Let $X$ be a germ of a path $\gamma: t \rightarrow P$ on $P$ at the parameter $t=0$ such that $\pi(\gamma) = \sigma_n$ is a 1-vertex of $\tau_n$ for $n$ large enough, and we assume with no loss of generality in the sequel that $\gamma$ is parametrized by arc-length of $\tau_n.$ Then  $\pi (\gamma_{[0;|\sigma_{n+p}|]})= \sigma_{n+p},$ for $p \in \N,$ and $$\theta(X) = \lim_{p \rightarrow +\infty} \frac{g(p) - I}{|\sigma_p|}$$ where $g_p \in G$ is defined by $\gamma(|\sigma_{n+p}|) = H_{\gamma(0)}\sigma_{n+p}(|\sigma_{n+p}|). g_p \in P.$    
  	\item Let $X,Y$ be germs of paths $\gamma_0, \gamma_1 : t \rightarrow P$ on $P$ at the parameter $t=0,$ with $\gamma(0)=\gamma'(0),$ such that there exists a 2-simplex $ \sigma_n \in \tau_n$ for $n$ large enough, where $\gamma$ and $\gamma'$ project on 1-simplexes of $\partial\sigma_n, $ and we also assume arc-length parametrisation as in the previous item. Then $\partial\sigma_{n+p}$ is a piecewise smooth loop parametrized by arc-length, staring from $\pi(\gamma(0))$ along $\pi(\gamma),$ and ending along $\pi(\gamma').$ Then we have $$\Omega(X,Y) =  C\lim_{p \rightarrow +\infty} \frac{Hol_{\gamma(0)}\partial \sigma_{n+p}(|\partial \sigma_{n+p}|) - I}{|\sigma_p|},$$ where $C$ is a constant.
  	\item These two items  correspond heuristically to a directional derivative of the holonomy at the continuum limit.          
  \end{itemize}
  
 \vskip 6pt
 \noindent
 \underline{\bf Open problems:} in the proposed approach, the ``natural'' continuum limit of the quantized model does not involve cylindrical functions on a vector space of connection forms equipped with a Lebesgue measure, but cylindrical functions defined on products of unimodular groups aquipped with their Haar measure.  Even if there exists an obvious conceptual link between Haar measure and Lebesgue measure, the two
 cylindrical approximations that are produced are not a priori equivalent.  We conclude this presentation conjecturing that the gauge anomalies 
 which can appear in classical discretized models may find an expression in this measure defect.  
 \section{Pairwise comparisons in decision making as a discrete Yang-Mills theory, based on \cite{Ma2017-2,Ma2018-1,Ma2016-5}}
 \subsection{Changing the comparisons structure to arbitrary groups}
 Classicaly, the comparisons coefficients are $a_{i,j}$ are scaling coefficients. This means that,if the PC matrix $A$ is consistent, given a state $s_k,$ we can recover all the other states $s_j$ by something assimilated to scalar mutiplication: $$s_j = a_{j,k}s_k.$$
 In other words, even if the states $s_j$ are driven by more complex rules, we reduce them to a ``score'' or an ``evaluation'' in $\R_+^*.$ 
 The states $s_j$ have to belong to a more complex state space $S,$ and in order to have pairwise comparisons, a straightforard study shows that we define matrices with coefficients in a group \cite{Ma2016-5}. Let $I$ be a set of indexes and let $(k,+,.,|.|)$be a field with absolute value and $V_k$ a normed $k-$vector space.
 
 \begin{Definition} \label{pcig} \cite{Ma2018-1}
 	Let $(G,.)$ be a group. A \textbf{pairwise comparisons matrix} is a matrix $$A= (a_{i,j})_{(i,j)\in I^2} $$
 	such that 
 	
 	\begin{enumerate}
 		\item $\forall(i,j) \in I^2, a_{i,j}\in G.$

 		\item  \label{inverse} $ \forall (i,j)\in I^2, \quad a_{j,i}=a_{i,j}^{-1}. $
 		
 		\item \label{diagonal} $a_{i,i} = 1_G.$ 
 		
 	\end{enumerate}
 	
 \end{Definition}
 We note by $PC_I(G)$ the set of pairwise comparisons matrices indexed by $I$ and with coefficients in $G.$ When $G$ is not abelian, there are two notions of inconsistency covariantly or contravariantly consistent, which corresponds merely to left or right actions.
 The two notions are dual \cite{Ma2016-5}. Contravariant consistency appears in the geometric realization of $PC_I(G)$ via the holonomy of a connection on a simplex $\Delta:$

 \begin{Proposition}\cite{Ma2018-1}
 	If $G$ is exponential, the map \begin{eqnarray*}
 		\Omega^1(\Delta, \mathfrak{g}) & \rightarrow & \{ \hbox{ PC matrices }\} \\
 		\theta & \mapsto & \hbox{ the holonomy matrix }
 	\end{eqnarray*}
 	is onto.
 \end{Proposition}  
 We note by $CPC_I(G)$ the set of consistent PC-matrices.
 
 \begin{Definition}
 	
 	A (non normalized, non covariant) inconsistency map is a map 
 	$$ii : PC_I(G) \rightarrow V_k$$
 	such that $ii(A)=0$ if $A$ is consistent. Moreover, we say that $ii$
 	is faithful if $ii(A) = 0$ implies that $A$ is consistent.  
 \end{Definition}
 In \cite{Ma2017-2}, we give arguments to consider only \textbf{normalized} inconsistency maps called inconsistency indicators.

 \vskip 12pt
 The main feature in applying this setting will be twofold,and these two points are far to be systematically solved with the present work:
 
 - define a comparisons group $G$ for which we can get at least one comparison coefficient $a_{i,j}$ between two states $s_i$ and $s_j$ (which means that the $G-$action needs to be transitive),
 
 - evaluate (and compute!) inconsistency, if possible generalizing the $\R_+^*$-setting, in a proper way to get safe decision making. This second point is linked with multiscale analysis.

 Examples from \cite{Ma2016-5} highlight these features, such as when $G$ is a matrix group $GL_n$ and when $G$ is an affine group. In this second example for $G,$ we find applications to modelization of perspective and error in tunnel building.

 \subsection{Algebraic properties, gauge group and generalization to graphs}
 The whole section is based on the work \cite{Ma2016-5}
 \begin{Proposition}
 	Any morphism of group $a:G \rightarrow G'$ extends to a map $\bar{a}:PC_I(G) \rightarrow PC_I(G')$ by action on the coefficients, and:
 	\begin{itemize}
 		\item If $A \in PC_I(G)$ is consistent, then $\bar{a}(A)\in PC_I(G')$is consistent.
 		\item If $Ker (a) = \{e_G\},$ then $A \in PC_I(G)$ is consistent, if and only if $\bar{a}(A)\in PC_I(G')$
 		
 	\end{itemize}	
 \end{Proposition}
 We call $G^I$ the \textbf{gauge group} of $G$. Then we get the following actions: 
 
 \begin{itemize}
 	\item a left action $L : G^I \times PC_I(G) \rightarrow PC_I(G)$ defined, for $(g_i)_I \in G^I$ and $(a_{i,j})_{I^2}\in PC_I(G)$ by $$L_{(g_i)_I}\left((a_{i,j})_{I^2}\right) = (b_{i,j})_{I^2}$$
 	with $$b_{i,j} = \left\{\begin{array}{ccl} 1 & \hbox{ if } & i=j\\
 	g_i a_{i,j} & \hbox{ if } & i<j\\ 
 	a_{i,j} g_{j}^{-1} & \hbox{ if } & i>j\\\end{array}\right.$$
 	\item a right action $R :  PC_I(G) \times G^I\rightarrow PC_I(G)$ defined, for $(g_i)_I \in G^I$ and $(a_{i,j})_{I^2}\in PC_I(G)$ by $$R_{(g_i)_I}\left((a_{i,j})_{I^2}\right) = (b_{i,j})_{I^2}$$
 	with $$b_{i,j} = \left\{\begin{array}{ccl} 1 & \hbox{ if } & i=j\\
 	a_{i,j} g_j & \hbox{ if } & i<j\\ 
 	g_i^{-1} a_{i,j}  & \hbox{ if } & i>j\\\end{array}\right.$$
 	\item an adjoint action $$Ad_{(g_i)_I}=L_{(g_i)_I}\circ R_{(g_i)_I^{-1}} =  R_{(g_i)_I^{-1}}\circ L_{(g_i)_I}$$
 	\item a coadjoint action $$\left((a_{i,j})_{I^2},(g_i)_I\right)\mapsto Ad_{(g_i)_I^{-1}}.$$
 \end{itemize}
 
 \begin{Theorem} \label{th1'}
 	Consistent PC-matrices are the orbits of the PC-matrix $(1)_{I^2}$ with respect to the adjoint action.
 \end{Theorem}
 After remarking that
 	$L$ and $R$ are effective actions, we have the following:
 
 \begin{Theorem}
 	If $I=\N_3,$ any orbit for the left action intersects $CPC_3(G).$ If $card(I)>3,$ there exists orbits for the left action which do not intersect $CPC_I(G).$ 
 \end{Theorem}
 
 Let us now turn to other properties inconsistency maps.
 
 \begin{Definition}
 	Let $ii$ be an inconsistency map. It is called:
 	\begin{itemize}
 		\item \textbf{normalized} if $\forall A \in PC_{I}(G), ||ii(A)||\leq 1.$
 		\item \textbf{Ad-invariant} if   $\forall A \in PC_{I}(G), \forall g \in G^I, ii\left(Ad_g(A)\right) =ii(A)$
 		\item \textbf{norm invariant} if $||ii(.)||$is Ad-invariant.
 	\end{itemize}
 \end{Definition}

 According to \cite{KS2015}, we give now the following definition:
 \begin{Definition}
 	An inconsistency indicator $ii$ on $PC_I(G)$ is a faithful, normalized inconsistency map with values in $\R_+$ such that there exists an inconsistency map $ii_3$ on $PC_3(G)$ that defines $ii$ by the following formula $$ ii(A) = \sup \left\{ ii_3(B) \ B \subset A ; \, \, \, B \in PC_3(G) \right\}.$$
 \end{Definition}	
 We remark here that since $ii$ is faithful, it is in particular (trivially) $Ad-$invariant on $CPC_I(G),$ but we do not require it to be $Ad-$invariant. Moreover, with such a definition, to show that $ii$ is $Ad-$invariant, it is sufficient to show that $ii_3$ is $Ad-$invariant.  However, we give the example 
 driven by Koczkodaj's approach. This is already proved that $Kii_3$ generates an inconsistency indicator \cite{KS2015} and we complete this result by the following property:
 \begin{Proposition} \label{Kad}
 	Let $n \geq 3.$ Koczkodaj's inconsistency maps $Kii_3$ and $Kii_n$ generate is $Ad-$invariant inconsistency maps on $PC_n(\R_+^*).$ 
 \end{Proposition}

 \subsection{Geometric aspects}
 
 We already mentionned the correspondence with holonomy. This correspondence is exaclty the one described in section \ref{ss:discreteholonomy}.
 \begin{Theorem}\cite{Ma2016-5}
 	If $G$ is a compact exponential finite dimensional Lie group, then for each $(\tilde{g}_i)_{i \in I} \in G^I,$ there exists $g \in  C^\infty(\Delta,G)$ such that $g(s_i)=g_i.$ Moreover, for any contravariant PC matrix $A$ which is the holonomy matrix of a connection $\theta,$ then $Ad_{(\tilde{g}_i)_{i \in I}}(A)$ is the holonomy matrix  of the connection $g^{-1}dg + Ad_{g^{-1}}\theta.$
 \end{Theorem}
 The gauge group also appears as a discretization of the classical gauge group of a trivial principal bundle. 
 We consider in this section a family of states $(s_i)_I$ such that any $s_i$ cannot be a priori compared directly with any other $s_j.$ This leads us to consider a graph $\Gamma_I$ linking the elements which can be compared. For example, in the previous sections, $\Gamma_I$ was the $1-$skeleton of the simplex. For simplicity, we assume that $\Gamma_I$ is a connected graph, and that at most one vertex connects any two states $s_i$ and $s_j.$ We note this (oriented) vertex by $<s_i,s_j>,$ and the comparison coefficient by $a_{i,j}.$ By the way, we get a pairwise comparisons matrix $A$ indexed by $I$ with ``holes'' (with virtual $0-$coefficient) when a vertex does not exist, and for which $a_{j,i}^{-1} = a_{i,j}.$
 
 Let us summarize the main correspondences that we have highlighted:
 \vskip 12pt
 \begin{tabular}{|c|c|}
 	\hline
 	& \\
 	discrete Yang-Mills formalism & Pairwise Comparisons (PC) \\
 	& \\
 	\hline
 	connection & PC matrix \\
 	& \\
 	flat connection & consistent PC matrix \\
 	& \\
 	curvature = loop holonomy & inconsistency \\
 	& \\
 	classical Yang-Mills functional & quadratic average of inconsistency \\
 	& on triads \\
 	& \\
 	``sup'' Yang-Mills functional & Koczkodaj's inconsistency indicator\\
 	& \\
 	\hline
 \end{tabular}
\vskip 12pt
 \subsubsection{Hierarchyless comparisons, ``hearsay'' evaluation and holonomy on a graph \cite{Ma2016-5}}
 In this model, the comparison between two states $s_i$ and $s_j$ can be performed by any path between $s_i$ and $s_j$ of any length. This model modelize the propagation of rumours, where validation of information is based on hearsay results. With this approach, the capacity of propagation of an evaluation is not controlled.
 We note by $$<s_{i_1} ,...,s_{i_k}>= <s_{i_1} ,s_{i_2}>\vee...\vee <s_{i_{k-1}},s_{i_k}>$$ the composition of paths along vertices.
 By analogy with the holonomy of a connection, we define:
 
 \begin{Definition}
 	Let $s=s_i$ and $s'=s_j$ be two states and let $$\mathcal{H}_{s,s'} = \left\{ a_{i,i_2}...a_{i_{k-1},i} | <s ,s_{i_2},...,s_{i_{k-1}},s'>  \hbox{ is a path from } s \hbox{ to } s'\right\}.$$
 	We note by $\mathcal{H}_s$ the set $\mathcal{H}_{s,s}.$
 	
 \end{Definition}
 By the way, we get the following properties, usual for classical holonomy and with easy proof:
 
 \begin{Proposition}
 	\begin{enumerate}
 		\item Let $s$ be a state, then $\mathcal{H}_s$ is a subgroup of $G.$ We call it holonomy group at $s.$
 		\item Let $s$ and $s'$ be two states. Then $\mathcal{H}_s$ and $\mathcal{H}_{s'}$ are conjugate subgroups of $G.$
 		\item  $$ \mathcal{H}_{s_i,s_j} = a_{i,j}.  \mathcal{H}_{s_j} = \mathcal{H}_{s_i} . a_{i,j}.$$
 	\end{enumerate}
 \end{Proposition}     
 
 \vskip 12pt
 \noindent

 \begin{Definition}
 	The PC matrix $A$ on the graph $\Gamma_I$ is \textbf{consistent} if and only if there exists a state $s$ such that $\mathcal{H}_s = \{1\}.$
 \end{Definition}   
 
 \subsubsection{Ranking the trustworthiness of indirect comparisons}
 The main problem with hierarchiless comparisons of two states $s $ and $s'$ is that paths of any length 
 give comparison coefficients which cannot be distinguished. An indirect comparison, given by a path with 3 vertices, has the same status as a comparison involving a path with 100 vertices. This is why we need to introduce a grading on $H_{s,s'}$ called order. This terminology will be justified by the propositions thereafter. 
 
 \begin{Definition}
 	Let $s$ and $s'$ be two states.
 	\begin{itemize}
 		\item Let $\gamma$ be a path on $\Gamma_I$ from $s$ to $s'$. The \textbf{length} of $\gamma,$ noted by $l(\gamma),$ is the number of vertices of $\gamma,$ and by $H(\gamma)$ its holonomy.
 		\item Let $h \in \mathcal{H}_{s,s'}.$ The \textbf{order} of $h$ is defined as $$ord(h) = \min \left\{ l(\gamma) \left| H(\gamma) = h \right. \right\}.$$ 
 	\end{itemize}
 \end{Definition}
 
 As a trivial consequence of the triangular equality, and as a justification of the terminology, we have:
 \begin{Proposition}
 	Let $s,s'$ a,d $s''$ three states. Let $(h,h') \in \mathcal{H}_{s,s'} \times \mathcal{H}_{s',s''} .$ Then $$ord(hh') \leq ord(h) + ord(h').$$
 \end{Proposition}
 
 Left action, right action and adjoint action of $G^I$ extend straightway to PC-matices on $\Gamma_I$ setting $$\forall g \in G, \quad g.0 = 0. g = 0.$$
 
 Adapting the proof of Theorem \ref{th1'} we get:
 
 \begin{Theorem}
 	Let $A = (a_{i,j})_{(i,j)\in I^2}$ be a PC matrix on $\Gamma$ Then $A$ is consistent if and only if there exists $(\lambda_i) \in G^I$ such that
 	$$ a_{i,j} = \lambda^{-1}_i \lambda_j$$
 	when $a_{i,j} \neq 0.$
 \end{Theorem}
 \subsubsection{Inconsistency maps ranked by trustworthiness \cite{Ma2016-5}}
 Let $A$ be a PC matrix on $\Gamma_I.$ Inconsistency will be given here by the holonomy of a loop. Let us recall that a trivial holonomy of a loop $<s_{i_1},s_{i_2},..., s_{i_{k}},s_{i_1}>$ implies that $$a_{i_1,i_k}\left(a_{i_1,i_2}... a_{i_{k-1},i_k}\right)^{-1} = 1.$$  
 The principle of ranking inconsistency with loop lengthgives the following:
 
 \begin{Definition}
 	Let $\mathcal{F} : G \rightarrow \R_+$ be a map such that $I(1)=0.$ Let $s$ be a basepoint on $\Gamma_I.$
 	The \textbf{ranked Koczkodaj's inconsistency map} associated to $\mathcal{F}$ the map $$ Kii_{\N} = \sum_{n \in \N} a_n X^n$$
 	where $$a_n = \sup \left\{ \mathcal{F}\left(H(\gamma)\right) \left| \gamma \hbox{ is a loop at } s \hbox{ and } \right. l(\gamma) = n \right\}.$$
 \end{Definition}
 One can easily see that $a_n$ generalize $Kii_n,$ and $Kii_{\N}$ is a $\R[[X]]-$valued inconsistency map. 
 \subsubsection{Holonomy versus distance}
 This section is based on \cite{Ma2016-5}.
 In this section, $G = \mathbb{R}_+^*.$
 Setting $$k_{i,j} = | log(a_{i,j}) |$$
 we get another matrix, that we define as the \textit{distance matrix} $$K = (k_{i,j})_{(i,j)\in I^2}.$$
 Notice that, if the coefficients of this matrix satisfy the triangle inequality
 $\forall (i,j,l) \in I^3, k_{i,l} \leq k_{i,j} + k_{j,l},$
 we get a curvature matrix for metric spaces \cite{Gro}. 
 Due to the absolute value, we have the following:
 
 \begin{Proposition}
 	Let $K$ be a non zero distance matrix on $\Delta_n.$ Let $N$ be the number of non zero coefficients in $K. $ Then $N$ is even and there exist $2^{N/2}$ corresponding PC matrices.   
 \end{Proposition}

 \noindent Therefore, we have the following results: 
 
 \begin{Proposition}
 	Let $K$ be the distance matrix on $\Delta_n$ associated to a consistent PC matrix $A,$ which is assumed to be non zero. Let 
 	$N'$ be the number of coefficients $k_{i,i+1}$ which are non zero. Then there exists $2^{N'}$ consistent PC matrices built with the coefficients $k_{i,i+1},$  but only $2$ consistent ones, $A$ and its transposition.   
 \end{Proposition}
 
 \vskip 6pt
 \noindent
 \underline{\bf Open problem:} The basic evaluation of inconsistency in pairwise comparisons \cite{KoSza2010} seems to be less efficient than holonomy evaluation. A common technical feature between the two approaches stems in potential analysis and Lagrangian theory. Such an approach still has to be developped, in order for example to include this part of decision theory as a part of quantum physics. 

 Back to inconsistency, the work led by physicists on discretized QFT may then apply to information theory. But we highlight the following features:
 \begin{itemize}
 	\item What would be the meaning in information theory of the partition function? of the continuum limit? 
 	\item What other geometric quantities than holonomy could be adapted for information theory?
 	\item Can there appear other "good" inconsistency indicators from physics? Can Feynman integration give an "ideal" approach to many-criteria decision?
 	\item Can physics apply directly to image processing through inconsistency? to faces/symbols recognition? to shape analysis? 
 \end{itemize}
\chapter{Pseudo-differential, Fourier integral operators with applications to geometry}

We now deal with non-formal pseudo-differential operators and Fourier integral operators. These operators have also applications and we focuse on some applications to the geometry of infinite dimensional manifolds. 

\section{On $\zeta-$renormalized traces} \label{s3}
$E$ is equipped this an Hermitian products $<.,.>$,
which induces the following $L^2$-inner product on sections of $E$: 
$$ \forall u,v \in C^\infty(S^1,E), \quad (u,v)_{L^2} = \int_{S^1} <u(x),v(x)> dx, $$
where $dx$ is the Riemannian volume. 
\begin{Definition} \cite{PayBook,Scott}
	$Q$ is a \textbf{weight} of order $s>0$ on $E$ if and only if $Q$ is a classical, elliptic,
	admissible pseudo-differential operator acting on 
	smooth sections of $E$, with an admissible spectrum.
\end{Definition}
Recall that, under these assumptions, the weight $Q$ has a real discrete spectrum, and that 
all its eigenspaces are finite dimensional. 
For such a weight $Q$ of order $q$, one can define the complex 
powers of $Q$ \cite{See}, 
see e.g. \cite{CDMP,Le,Asa2000,Pay,Scott}. 
The powers $Q^{-s}$ of the weight $Q$ 
are defined for $Re(s) > 0$ using with a contour integral,
$$ Q^{-s} = \int_\Gamma \lambda^s(Q- \lambda Id)^{-1} d\lambda,$$
where $\Gamma$ is an ``angular'' contour around the spectrum of $Q.$
Let $A$ be a log-polyhomogeneous pseudo-differential operator.  The map
$\zeta(A,Q,s) = s\in \mathbb{C} \mapsto \hbox{tr} \left( AQ^{-s} \right)\in \mathbb{C}$ , 
defined for $Re(s)$ large, extends
on $\mbc$ to a meromorphic function with a pole 
at
$0$ (\cite{Le}). When $A$ is classical, 
$\zeta(A,Q,.)$ has a simple pole at $0$  
with residue ${1 \over q} \res A$, where $\res$ is the Wodzicki
residue (\cite{W}, see also \cite{Ka}). Notice that the Wodzicki residue mimicks the Adler trace \cite{Adl} on formal symbols.
\begin{Definition} \label{d6} $tr^Q A = lim_{z \rightarrow 0} (\hbox{tr} (AQ^{-z})
	- {1 \over qz} res A)$.
\end{Definition}

%

If $A$ is trace class, $\hbox{tr}^Q{(A)}=\hbox{tr}{(A)}$.
The functional $\hbox{tr}^Q$ is of course not a trace on $Cl(M,E)$.
Notice also that, if $A$ and $Q$ are pseudo-differential operators
acting on sections on a real vector bundle $E$, they also act on
$E \otimes \mbc$.
Before giving our new developments, we need the following statements which are not so well applied in various contexts}
\vskip 12pt
\noindent 
$\bullet$ \underline{\it Development presented in the PhD thesis and published in \cite{CDMP,Ma2006},} {and hence not presented for as new results for the habilitation:}
\vskip 12pt

The Wodzicki residue res and the renormalized traces
$\hbox{tr}^Q$ have to be understood as functional defined on
pseudo-differential operators acting on $E \otimes \C$.
In order to compute $\tr^Q[A,B]$ and to differentiate $\tr^Q A$,
in the topology of classical pseudo-differential operators, we
need the following (\cite{CDMP} and references therein):

\begin{Proposition} \label{p6} \cite{CDMP}

(i)  Given two (classical) pseudo-differential operators A and B,
given a weight Q,
\begin{equation}\label{crochet}  \tr^Q[A,B] = -{1 \over q} \res (A[B,\log Q]). \end{equation}
(ii) Given  a differentiable family $A_t$ of pseudo-differential
operators, given a differentiable family $Q_t$ of weights of
constant order q,

\begin{equation}\label{deriv} {d \over dt} \left(tr^{Q_t}A_t\right) = tr^{Q_t} \left({d \over dt}A_t\right) -
	{1 \over q} \res \left( A_t ({d \over dt}\log Q_t) \right).
\end{equation}

\end{Proposition}

The following "covariance" property of $\hbox{tr}^Q$ (\cite{CDMP}, \cite{Pay})
will be useful to define renormalized traces on bundles of operators,

\begin{Proposition} \label{p7}

Under the previous notations, if C is a classical elliptic
injective operator of order 0, $tr^{C^{-1}QC}\left( C^{-1}AC
\right)$ is well-defined and equals $\tr^QA$.

\end{Proposition}

We moreover have specific properties for weighted traces of a more
restricted class of pseudo-differential operators (see
\cite{KV1},\cite{KV2},\cite{CDMP}), called odd class
pseudo-differential operators following \cite{KV1,KV2} :

\begin{Definition} \label{d7}

A classical pseudo-differential operator $A$ is called odd class
if and only if

$$ \forall n \in \Z, \forall (x,\xi) \in T^*M, \sigma_n(A) 
(x,-\xi) = (-1)^n  \sigma_n(A) (x,\xi).$$

We note this class $Cl_{odd}.$
\end{Definition}

Such a definition is consistent for pseudo-differential operators
on smooth sections of vector bundles, and applying the local
formula for Wodzicki residue, one can prove:

\begin{Proposition} \label{p8} \cite{CDMP}

If $M$ is an odd dimensional manifold, $A$ and $Q$ lie in the odd
class, then $f(s)=tr(AQ^{-s})$ has no pole at $s=0$. Moreover, if
A and B are odd class pseudo-differential operators, $\tr^Q \left(
[A,B] \right) =0$ and $\tr^QA$ does not depend on $Q.$

\end{Proposition}

This trace was first defined in the papers \cite{KV1} and \cite{KV2} by Kontesevich and Vishik. We remark that it is in particular a trace on $DO(M,E)$
when $M$ is odd-dimensional.

Let us now describe a class of operators which is, in some sense, complementary to odd class:
\begin{Definition} \cite{Ma2006}

A classical pseudo-differential operator $A$ is called even class
if and only if

$$ \forall n \in \Z, \forall (x,\xi) \in T^*M, \sigma_n(A) 
(x,-\xi) = (-1)^{n+1}  \sigma_n(A) (x,\xi).$$
We note this class $Cl_{even}.$
\end{Definition}
We precised that $Cl_{even}$ and $Cl_{odd}$ are "in some sense" complementary because these are not supplementary vector spaces: $Cl_{odd}\cap Cl_{even} = Cl^{-\infty}.$

\begin{Proposition} \cite{Ma2006}

$ Cl_{even} \circ Cl_{odd} = Cl_{odd} \circ Cl_{even} = Cl_{even}$
and 

$Cl_{even} \circ Cl_{even} = Cl_{odd} \circ Cl_{odd} = Cl_{odd} .$
\end{Proposition}

Now, following \cite{Ma2006}, we explore properties of $\tr^Q$ on Lie brackets.

\begin{Definition} \label{d8}

Let E be a vector bundle over M, Q a weight and $a \in \Z$. We define :
$$ \mathcal{A}^Q_a=\{B \in Cl(M,E); [B,\log Q] \in Cl^a(M,E)\}.$$

\end{Definition}

\begin{Theorem} \label{t1} \cite{Ma2006}

\begin{item}
	
	(i) $\mathcal{A}^Q_a \cap Cl^0(M,E) $ is an subalgebra of $Cl(M,E)$
	with unit.
\end{item}

\begin{item}
	
	(ii) Let $B \in Ell^*(M,E)$, $B^{-1}\mathcal{A}^Q_aB = A^{B^{-1}QB}_a.$
	
\end{item}

\begin{item}
	
	(iii) Let $A\in Cl^b(M,E)$, and $B \in \mathcal{A}^Q_{-dimM-b-1}$,
	then $\tr^Q[A,B]=0.$
\end{item}

\begin{item}
	
	(iv)  For $a< -{dimM \over 2}$, $\mathcal{A}^Q_a \cap Cl^{-dimM \over 2}(M,E) $
	is an algebra on which the renormalized trace is a trace (i.e. vanishes on the brackets).
	
\end{item}

\end{Theorem}

We now produce non trivial examples of operators that
are in $\mathcal{A}^Q_{a}$ when Q is scalar, and secondly we give a
formula for some non vanishing renormalized traces of a bracket.

\begin{Proposition} \label{p9} \cite{Ma2006}

Let Q be a scalar weight on $C^\infty_0(M, V)$. Then
$$Cl^{a+1}(M,V) \subset \mathcal{A}^Q_{a}.$$
Consequently,

\begin{enumerate}
	\item let B be a classical pseudo-differential operator of order $b$. Then $[B, \log Q]$ is a classical
	pseudo-differential operator of order $b-1$.
	\item if $ord(A)+ord(B)=-dimM,$ $\tr^Q [A,B]=0$.

	\item  when $M=S^1$, if A and B are classical pseudo-differential
	operators, if A is compact and B is of order 0, $\tr^Q[A,B]=0$.
	
	\item
	Let Q be a scalar weight on  $ C^\infty_0(M,V) $, and A, B two
	pseudo-differential operators of orders a and b on $
	C^\infty_0(M,V) $, such that $a+b=-m+1$ (m = dim M). Then
	$$ \tr^Q[A,B]=-{1 \over q} \res\left( A[B,\log Q] \right)=
	-{1 \over q(2\pi )^n} \int _M \int _{|\xi |=1} tr (\sigma _a( A
	)\sigma_{b-1}([B, \log Q])). $$
\end{enumerate}
\end{Proposition}

 Let $Q$ be a weight, the non-traciality of $\tr^Q$ defined a non-vanishing cocycle (which is a coboundary in Hochschild cohomology) express as:
$$(A,B) \in (Cl(M,E))^2\mapsto c^Q(A,B) = tr^Q[A,B],$$ whiich is proportionnal to $res(A[B,logQ]).$
This cocycle is a generalization to non formal pseudo-differential operators of the Kravchenko-Khesin-Radul cocycle
$$(a,b) \in (\mathcal{F}Cl(S^1,\mbbc))^2 \mapsto c_{KKR}(a,b)=\int_{S^1} \left(\sigma_{-1}\left(a[b,log \xi]\right)\right)_{\xi = 1} dx.$$
We notice that $c_{KKR}$ is not a coboundary \cite{KK}.
The two other cocycles that I would like to mention are due to a splitting: we need for this a sign operator $\epsilon,$ i.e. an operator such that $\epsilon^2 = Id,$ see e.g. \cite{Mickbook}. The classical case is when $E$ is a Clifford bundle over $M$, and $\epsilon$ is the sign of the Dirac operator \cite{Mick1994}, generalizing the classical $M=S^1$ see e.g. \cite{PS}. These works were motivated by \cite{RSF1985,SW1985,MR1988,Mick1988}. The splitting of $L^2(M,E)$ is given by the two eigenvalues $-1$ and $1$ of $\epsilon$ (the kernel of the Dirac operator is arbitrarily associated to the $-1$ or $1$ eigenspace). Let us note by $H_-$ and $H_+$, respectively, these two eigenspaces. The operators $A$ considered split blockwise $$A = \left(\begin{array}{cc} A_{++} & A_{-+} \\ A_{+-} & A_{--} \end{array} \right)$$ under the decomposition $L^2(M,E) = H_+ \oplus H_-=H.$ 
Defining the Banach Lie group \begin{eqnarray*}
	GL_{res}& = & \left\{ A \in GL(H) | [\epsilon, A] \hbox{ is Hilbert-Schmidt} \right\}\\&=&
	\left\{ A \in GL(H) | A_{+-} \hbox{ and } A_{-+} \hbox{ are Hilbert-Schmidt}\right\},
\end{eqnarray*}    
with Lie algebra 
\begin{eqnarray*}
	L_{res}& = & \left\{ A \in L(H) | [\epsilon, A] \hbox{ is Hilbert-Schmidt} \right\}\\
	&=&\left\{ A \in L(H) | A_{+-} \hbox{ and } A_{-+} \hbox{ are Hilbert-Schmidt}\right\}
\end{eqnarray*}
one can define \cite{SW1985,PS,Mickbook}, for $(A,B)\in (L_{res})^2$:

$$\lambda(A,B) = \tr\left([A_{++},B_{++}]-[A,B]_{++}\right)$$
and the Schwinger cocycle \cite{Sch}:
$$c_S(A,B)=\tr \left(\epsilon[\epsilon, A][\epsilon, B]\right).$$
These two cocycles are proportional and non trivial, and they naturally restrict to $L_{res}\cap Cl(M,E)$ \cite{CDMP}.
\subsection{A property, based on \cite{Ma2016-1}}

Let us now explore the action of $Diff(M)$ and of $Aut(E)$ on $tr^Q(A).$For this, we get: 

\begin{Lemma} \cite{Ma2016-1}
	Let $a \in \Z.$ Let $A \in Cl^a(M,E)$ and let $Q$ be a weight on $E.$ Let $B$ be an operator on $C^{\infty}(M,E)$ such that 
	\begin{enumerate}
		\item \label{a} $Ad_B(Cl^a(M,E)) \subset Cl^a(M,E)$
		\item \label{b} $Ad_BQ$ is a weight of the same order as $Q$
	\end{enumerate} 
	Then 
	\begin{itemize}
		\item $res(Ad_BA) = res(A)$
		\item $ tr^{Ad_BQ}(Ad_BA) = tr^Q(A).$
	\end{itemize}
	The properties \ref{a},\ref{b} are true in particular for operators $B\in Aut(E).$
\end{Lemma}
\subsection{Some cocycles on $PDO(M,E)$, based on \cite{Ma2006-2,Ma2008}}

  If one wants to extend this construction of the index or of the Schwinger cocycle to unbounded operators, one has to find a (non unitary) subalgebra $\mathcal{I}$ in the algebra of operators $\mathcal{A}$ considered such that

- $I^2$ is a set of trace-class operators

- $ [A,I] \subset I.$

Considering $\mathcal{A}=PDO(M,E),$ a natural choice for $\mathcal{I}$ is the ideal $Cl^{-\infty}(M,E)$ \cite{Ma2008}. But in that case, except when intrinsic geometric structures enable to find a sign operator (e.g. from a Dirac operator), it seems to be difficult to remain in an algebra of pseudo-differential operators. More precisely:

\begin{Proposition} \cite{Ma2008}
	When $E=M \times \mbbc,$ If $\epsilon \in Cl^{0}(M,\mbbc), $ $\epsilon = Id$ or $\epsilon = -Id$ up to a finite rank operator.
\end{Proposition}

Moreover, 

\begin{Proposition} \cite{Ma2008}
	The Schwinger cocycle $c_{S}$ on $\left(Cl(S^1,\mbbc)^{\otimes n}\right)_{res}$ related to the sign operator $\epsilon = (\epsilon(D))^{\otimes n},$ where $\epsilon(D)$ is defined in the section $M=S^1,$ is a non trivial cocycle.
\end{Proposition}

These results hae been obtained after those treating of the special case $M=S^1,$ principally given in \cite{Ma2006-2} and completed in \cite{Ma2008}. 

\vskip 6pt
\noindent\underline{Open question:} We notice that $Cl((S^1)^n, \mbbc) \neq \left(Cl(S^1,\mbbc)^{\otimes n}\right),$ and the question of an adequate choice of "maximal" natural algebra of operators for the Schwinger cocycle is still open when $M \neq S^1.$ The same question arise in \cite{R2018} where higher dimensional Kravchenko-Khesin cocycles are investigated. A natural, but trivial, extension on $Diff(M)-$pseudo-differential operators is given in \cite{Ma2016-1}, interesting results being for $M=S^1.$
\section{Specializing $M=S^1:$ a digression in algebraic structures, based on \cite{Ma2006-2,Ma2016-1,MRu2021-1}}
The operator $D = {-i} D_x$ splits $C^\infty(S^1, \mbbc^k)$ into three spaces :

- its kernel $E_0$, made of constant maps

- $E_+$, the vector space spanned by eigenvectors related to positive eigenvalues

- $E_-$, the vector space spanned by eigenvectors related to negative eigenvalues.

\noindent
The following elementary result will be useful for the sequel, 
see \cite{Ma2003} for the proof, and e.g. \cite{Ma2006,Ma2008}:  
\begin{Lemma} \label{l1'}
	
	(i) $\sigma(D) = {\xi }$
	
	(ii) $\sigma(|D|) = {|\xi| }$ where $|D| =
	\sigma \left(\int_\Gamma \lambda^{1/2} (\Delta - \lambda Id)^{-1}d\lambda\right)$, 
	with $\Delta = -D_x^2$. 
	
	(iii)  $\sigma(D|D|^{-1}) = {\xi \over |\xi|}$, where $ D|D|^{-1} = |D|^{-1}D$ is the sign of D, since $|D|_{|E_0}=Id_{E_0}.$
	
	(iv)  Let $p_{E_+}$ (resp. $p_{E_-}$) be the projection on $E_+$ (resp. $E_-$), then 
	$\sigma(p_{E_+}) ={1 \over 2}(Id + {\xi \over |\xi|})$ and $\sigma(p_{E_-}) = {1 \over 2}(Id - {\xi \over |\xi|})$.

\end{Lemma}

Let us now define two ideals of the algebra $\mathcal{F}PDO$, 
that we call $\mathcal{F}PDO_+$ and $\mathcal{F}PDO_-$, 
such that $\mathcal{F}PDO = \mathcal{F}PDO_+ \oplus \mathcal{F}PDO_-$. 
This decomposition is implicit in \cite{Ka}, section 4.4., p. 216,
for classical pseudo-differential operators 
and we furnish the explicit description 
given in \cite{Ma2003}, extended to the whole algebra of 
(maybe non formal, non classical) pseudo-differential symbols here.

\begin{Definition}
	
	Let $\sigma$ be a symbol (maybe non formal). Then, we define, for $\xi \in T^*S^1 - S^1$, 
	$$ \sigma_+(\xi) = \left\{ 
	\begin{array}{ll}
		\sigma(\xi) & \hbox{ if $ \xi > 0$} \\
		0 & \hbox{ if $ \xi < 0$} \\
	\end{array}
	\right. \hbox{ and }
	\sigma_-(\xi) = \left\{ 
	\begin{array}{ll}
		0 & \hbox{ if $ \xi > 0$} \\
		\sigma(\xi) & \hbox{ if $ \xi < 0$} . \\
	\end{array}
	\right.$$
	At the level of formal symbols, we also define the projections:  $p_+(\sigma) = \sigma_+$ and $p_-(\sigma) = \sigma_-$ .
\end{Definition}
The maps 
$ p_+ : \mathcal{F}PDO(S^1,\mbbc^k) \rightarrow \mathcal{F}PDO(S^1,\mbbc^k) $ 
{ and } $p_- : \mathcal{F}PDO(S^1,\mbbc^k) \rightarrow \mathcal{F}PDO(S^1,\mbbc^k)$ 
are clearly  algebra morphisms 
that leave the order invariant and are also projections 
(since multiplication on formal symbols is expressed 
in terms of pointwise multiplication of tensors). 

\begin{Definition} We define
	$  \mathcal{F}PDO_+(S^1,\mbbc^k) = Im(p_+) = Ker(p_-)$
	and $  \mathcal{F}PDO_-(S^1,\mbbc^k) = Im(p_-) = Ker(p_+).$ \end{Definition}
Since $p_+$ is a projection,  we have the splitting
$$ \mathcal{F}PDO(S^1,\mbbc^k) = \mathcal{F}PDO_+(S^1,\mbbc^k) \oplus \mathcal{F}PDO_-(S^1,\mbbc^k) .$$
Let us give another characterization of $p_+$ and $p_-$. 
Looking  more precisely at the formal symbols of $p_{E_+}$ and $p_{E_-}$ 
computed in Lemma \ref{l1'}, we observe that 
$$ \sigma( p_{E_+}) = \left\{ \begin{array}{ll}
	1 & \hbox{if }\xi > 0 \\
	0 & \hbox{if }\xi < 0 \\
\end{array} \right. \hbox{ and }
\sigma( p_{E_-}) = \left\{ \begin{array}{ll}
	0 & \hbox{if }\xi > 0 \\
	1 & \hbox{if }\xi < 0 \\
\end{array} \right. . $$
In particular, we have that $D^\alpha_x\sigma( p_{E_+}),$ 
$D^\alpha_\xi\sigma( p_{E_+}),$ $D^\alpha_x\sigma( p_{E_-}),$ $D^\alpha_\xi\sigma( p_{E_-})$ 
vanish for $\alpha > 0$. 
From this, we have the following result:

\begin{Proposition} \label{pag} \cite{Ma2003}
	Let $a \in \mathcal{F}PDO(S^1,\mbbc^k).$ 
	$ p_+(a) =  \sigma( p_{E_+}) \circ a = a \circ \sigma( p_{E_+})$ and 
	$  p_-(a) =  \sigma( p_{E_-}) \circ a = a \circ \sigma( p_{E_-}).$
\end{Proposition}
We then remind that if $V=\mbc^n$ and use the notations
$$ { DO}(S^1,V) = \bigcup_{o \in \N}\left\{ \sum_{0\leq k \leq o} a_k \partial^k \right\},\quad{ IO}(S^1,V) = \left\{ \sum_{k \leq -1} a_k \partial^k \right\}$$ 
we get also  the {vector space} decomposition 
\begin{eqnarray}
	\label{psi-DS} \Psi DO(S^1,V) = { DO}(S^1,V)\oplus { IO}(S^1,V).
\end{eqnarray}
{such that any (matrix) order $k$ pseudo-differential operator $A = \sum_{i=-\infty}^{k}a_i \partial^i$ is splitted  in two components $A = A_{D} + A_{S}$ with 
	$A_{D} =  \sum_{i=0}^{k}a_i \partial^i$ and $A_{S} =  \sum_{i=-\infty}^{-1}a_i \partial^i.$
	The Adler trace \cite{Adl} defined by $$ Tr: A = \sum_{k \leq o} a_k \partial^k \mapsto \int_{S^1} tr(a_{-1})$$ is the only non trivial trace on $ \Psi DO(S^1,V).$ Morover, see e.g. \cite{EKRRR1995} and \cite{KZ},
	\begin{Theorem}
		$(\Psi DO(S^1,V), { IO}(S^1,V), { DO}(S^1,V), Tr )$ is a Manin triple.
	\end{Theorem}
	The Wodzicki residue (\cite{W}, see e.g. \cite{Ka}) is usually known as an ``extension'' of the Adler trace to $\F Cl(S^1,V)$ and hence to $Cl(S^1,V).$ For the sake of deeper insight, we need to precise that the space of traces on $\F Cl(S^1,V)$ is 2-dimensional, generated by two functionals: $$ res_+: A \mapsto \int_{S^1} \sigma_{-1}(A)(x,1) |dx|$$
	and  $$ res_-: A \mapsto \int_{S^1} tr( \sigma_{-1}(A))(x,-1) |dx|.$$
	The functionals $res_\pm$ are the only non-vanishing traces on $\F Cl_\pm(S^1,V)$ (up to a scalar factor) and are vanishing on $\F Cl_\mp(S^1,V).$ The (classical) Wodzicki residue reads as $res = res_+ + res_-.$
\subsection{Radul, Schwinger and index cocycle on $PDO(S^1,E)$} 

Let $\pi : E \rightarrow S^1$ be a non trivial real vector bundle over $S^1$ of rank $k.$ 

\begin{Proposition} \cite{Ma2016-1}
	Let $\nabla$ be a Riemannian covariant derivative on the bundle $E \rightarrow S^1$ and let $\nabla \over dt$ be the associated first order differential operator, given by the covariant derivative evaluated at the unit vector field over $S^1.$  We modify the operator $\nabla \over dt$ into an injective operator $D = {\nabla \over dt} + p_{ker {\nabla \over dt}}$, where $p_{ker {\nabla \over dt}}$ is the $L^2$ orthogonal projection on $ker {\nabla \over dt} \subset C^\infty(S^1,E) \subset L^2(S^1,E),$ and we set $$\epsilon(\nabla) =  D \circ \left| D \right|^{-1}.$$
	Then the formal symbol of  $\epsilon(\nabla)$ is $i\xi \over |\xi |.$
\end{Proposition}

\begin{Proposition} \cite{Ma2006-2,Ma2008,Ma2016-1}
	For each $A \in PDO(S^1,E),$  $[A,\epsilon(\nabla)] \in PDO^{-\infty}(S^1;E).$
\end{Proposition}

The fiber bundle $T^*S^1 - S^1$ has two connected components and the 
phase function is positively homogeneous, so that
we can make the same procedure as in the case of the symbols. 
%
%
The main results gradually discovered in \cite{Ma2006,Ma2008,Ma2016-1} are now gathered. Here, $\epsilon(\nabla)$ is not a sign operator, but an operator such that $\epsilon(\nabla)^2 = Id$ up to a smoothing operator. :  

\begin{Theorem} \label{th1} \cite{Ma2006-2,Ma2008,Ma2016-1}
	For any $A \in PDO(S^1,E)$, $[A,\epsilon(\nabla)] \in PDO^{-\infty}(S^1,E).$ Consequently, 
	$$ c_s^\nabla : A,B \in PDO(S^1,E) \mapsto {1 \over 2}\tr \left( \epsilon(\nabla)[\epsilon(\nabla),A][\epsilon(\nabla),B]  \right) $$
	is a well-defined $\mathbb{R}$-valued 2-cocycle on $PDO(S^1, E).$ Moreover, $c_s^\nabla$ is non trivial on any Lie algebra $\mathcal A$ such that $C^\infty(S^1,\R)\subset \mathcal{A} \subset PDO(S^1,E).$
\end{Theorem}

\subsection{From $\Psi DO (S^1,V)$ to $\F Cl(S^1,V)$ based on \cite{MRu2021-1}} \label{s:Manin}

There exists a decomposition $\F Cl_+(S^1,V) = \F Cl_{+,D}(S^1,V) \oplus\F Cl_{+,S}(S^1,V) $ and another 
$\F Cl_-(S^1,V) = \F Cl_{-,D}(S^1,V) \oplus\F Cl_{-,S}(S^1,V), $ and setting $$ \F Cl_D(S^1,V) = \F Cl_{+,D}(S^1,V) \oplus\F Cl_{-,D}(S^1,V),  $$
$$ \F Cl_S(S^1,V) = \F Cl_{+,S}(S^1,V) \oplus\F Cl_{-,S}(S^1,V),  $$
we get the vector space decomposition analogous to (\ref{psi-DS}):
$$ \F Cl(S^1,V) = \F Cl_{D}(S^1,V) \oplus\F Cl_{S}(S^1,V),  $$
Let us consider the decomposition $$\F Cl(S^1,V) = \F Cl_{odd} (S^1,V) + \F Cl_{even}(S^1,V),$$
that we equip with the classical Lie bracket $[.,.]$ or with $[.,.]_{\epsilon(D)} = \frac{1}{2}\left([\epsilon(D) \circ .,.] + [.,\epsilon(D) \circ .] \right)$ and with the bilinear form $(A,B)=res(AB).$
\begin{Theorem} \label{1.25} \cite{MRu2021-1}
	res(AB) is a bilinear, non degenerate, symmetric and invariant form for both brackets, and $\F Cl_{odd} (S^1,V)$ as well as $\F Cl_{eo}(S^1,V)$ are isotropic vector spaces. Moreover, 
	\begin{itemize}
		\item for $[.,.],$ $\F Cl_{odd} (S^1,V)$ is a Lie algebra
		\item for $[.,.]_{\epsilon(D)},$  $\F Cl_{even}(S^1,V)$ is a Lie algebra.
	\end{itemize}
\end{Theorem}
\subsection{Extension of the classical Manin triple to $\F Cl(S^1,V)$  based on \cite{MRu2021-1}} \label{ss:manin}

Because the partial symbol $\sigma_{-1}(A)$ of an operator $A \in \Psi DO(S^1,V)$ is skew-symmetric in the $\xi-$variable, $res$ is vanishing on $\Psi DO(S^1,V) = \F Cl_{odd}(S^1,V), $ so that it is superficial to state that the Wodzicki residue is ``simply'' the extension of the Adler trace.  However the two linear functionals already described, namely 
$$ A \in \F Cl(S^1,V) \mapsto \sum_{k \in \Z} \sigma_{k}(A)(x,1) \partial^k$$
and  $$ A \in \F Cl(S^1,V) \mapsto \sum_{k \in \Z} \sigma_{k}(A)(x,-1) \partial^k,$$ identity $res_+$ and $res_-$ respectively with $Tr.$ By the way, we can state:
\begin{Theorem} We have three Manin triples: 
	$$(\F Cl_+(S^1,V), \F Cl_{+,S}(S^1,V), \F Cl_{+,D}(S^1,V), res_+ ), $$ 
	$$(\F Cl_-(S^1,V), \F Cl_{-,S}(S^1,V), \F Cl_{-,D}(S^1,V), res_- )$$ and $$(\F Cl(S^1,V), \F Cl_{S}(S^1,V), \F Cl_{D}(S^1,V), res ).$$
\end{Theorem}
\subsection{Injecting $\Psi DO(S^1,\K)$ in $\mathcal{F} Cl(S^1,\K).$} \label{s:inj}
We already mentionned the identification of 
$\Psi DO (S^1,\K)$ with $\F Cl_{odd}(S^1,\mbc).$ We claim here that this identification can be generalized to $$\Phi_{odd,\lambda} : \sum_{k \in \Z} a_{k} \left(\frac{d}{dx}\right)^k \in \Psi DO(S^1,\K) \mapsto \sum_{k \in \Z} a_{k} \left(\lambda\frac{d}{dx}\right)^k \in \F Cl_{odd}(S^1,\mbc) . $$
Similar to this identification, we have other { injections} { for } $\lambda \in \R^*:$ 
$$\Phi_{\epsilon(D),\lambda} : \sum_{k \in \Z} a_{k} \left(\frac{d}{dx}\right)^k \in \Psi DO(S^1,\K) \mapsto \sum_{k \in \Z} a_{k} \left(\lambda\epsilon(D)\frac{d}{dx}\right)^k \in \F Cl(S^1,\K) , \hbox{ and}$$ 
$$\Phi_{\lambda,\mu} : \sum_{k \in \Z} a_{k} \left(\frac{d}{dx}\right)^k \in \Psi DO(S^1,\K) \mapsto \sum_{k \in \Z} a_{k} \left(\lambda^k\left(\frac{d}{dx}\right)_+^k + \mu^k\left(\frac{d}{dx}\right)_-^k\right)  \in \F Cl(S^1,\K) $$
for {$(\lambda,\mu) \in \mbc^2 \backslash \{(0;0)\},$} with unusual convention $0^k = 0$ $\forall k \in \Z.$ 
\begin{rem}
	$\Phi_{1,1} = \Phi_{ee}$ and $\Phi_{1,-1} = \Phi_{\epsilon(D),1}.$
\end{rem}

\begin{rem}
	$Im \Phi_{1,0} = \F Cl_+(S^1,\K) $ and $\Phi_{1,0}$ is a isomorphism of algebras from $\Psi DO(S^1,\K)$ to $\F Cl_+(S^1,\K) .$ The same way, $\Phi_{0,1}$ identifies the algebras $\Psi DO(S^1,\K)$ and $\F Cl_-(S^1,\K) .$
\end{rem}
{
	\begin{rem}
		Wa have also to say that the maps $\Phi_{\lambda,\mu}$ are not algebra morphisms unless $(\lambda,\mu) \in \{(1;0),(0;1),(1;1)\}.$ For example, let $\lambda \in \mathbb{C}-\{0;1\}.$ the map $\Phi_{\lambda,0}$ pushes forward the multiplication on $\Psi DO(S^1,\K)$ to a deformed composition $*_k$ on $\F Cl_+(S^1,\K)$ that reads as
		$\sigma(A) *_k \sigma(B) = \sum_{\alpha \in \N} \frac{(-i)^\alpha}{ \alpha!.k^\alpha} D^\alpha_x\sigma(A) D^\alpha_\xi\sigma(B).$
\end{rem} }

From our previous remarks, we get:
\begin{Theorem}
	The map $$\Phi_{1,0} \times \Phi_{0,1} : \Psi DO(S^1,\K)^2 \rightarrow \F Cl_+(S^1,\K) \times \F Cl_-(S^1,\K) = \F Cl(S^1,\K)$$ is an isomorphism of algebra.
\end{Theorem}
We also remark a new subalgebra of $\F Cl(S^1,\K):$
\begin{Definition}
	Let $\F Cl_\epsilon(S^1,\K)$ be the image of $\Phi_{\epsilon(D),1}$ in $\F Cl(S^1,\K).$
\end{Definition}
We have the obvious identification $\F Cl_\epsilon(S^1,\K) = C^\infty(S^1,\K)((i|D|^{-1}))$ as a vector space. 
	\subsection{An integrable almost complex struxture on $\F Cl(S^1,\mbc)$ based on \cite{MRu2021-1} }\label{s:id}	Let $$ i \epsilon(D) = \left( \frac{d}{dx}\right).|D|^{-1} = |D|^{-1}.\left( \frac{d}{dx}\right) .$$ We define the operator $J_1 = i \epsilon(D) \circ (.)$ on  $\mathcal{F}Cl(S^1,V).$
\begin{Theorem}\label{th:J1}
	The operator $J_1$ defines an integrable almost complex structure on $\mathcal{F}Cl(S^1,V).$$\F Cl(S^1,V) = \Psi DO(S^1,V)\otimes\mbc$ as a real algebra, identifying $\F Cl_{odd}(S^1,V)$ with $\Psi DO(S^1,V)$ (real part) and  $\F Cl_{even}(S^1,V)$ with $i\Psi DO(S^1,V)$ (imaginary part).
\end{Theorem}

\vskip 6pt
\noindent
\underline{\bf Open problem:} study the obstruction of extending the almost complex structure $J_1$ from $\F Cl(S^1,V)$ to $Cl(S^1,V).$ 

 \section{Renormalized extension of the Hilbert-Schmidt Hermitian metric, based on \cite{Ma2021-1}}\label{s:HS}
The vector space $Cl^{-1}(S^1,V)$ is a space of Hilbert-Schmidt operators. As a subspace,  $Cl^{-1}(S^1,V)$ inherits a Hermitian metric from the classical Hilbert-Schmidt inner product. The renormalized trace $\tr^\Delta$ extends the classical trace $\tr$ of trace class operators to a smooth linear functional on $Cl(S^1,V).$ We investigate here the possible (maybe naive) extension of the classical Hilbert-Schmidt inner product to $Cl(S^1,V)$ via $\tr^\Delta.$
\subsection{Extension of the Hilbert-Schmidt metric  to $FCl.$}
Let  $(z^k)_{k \in \mathbb{Z}}$ is the Fourier $L^2-$orthonormal basis. Let us recall that there exists an ambiguity on $\epsilon(D)$ concerning its action on $z^0,$ which can be, or not, in the kernel of $p_+,$ or in the eigenspace of the eigenvalue $1$ or $-1.$ Depending on each of these three possibilities respectively, we set $\epsilon(k)$ as the eigenvalue of $\epsilon(D)$ at the eigenvector $z^k.$
\begin{Lemma} \label{calcultr}
	Let $X=u\frac{d}{dx},Y=v\frac{d}{dx}$ be two vector fields over $S^1,$ and let $a,b \in C^\infty(S^1,\mathbb{C}).$ 
	Then
	\begin{enumerate}
		\item \label{3.} $$\tr^{\Delta}(a\bar{b}) = 0$$
		\item \label{4.} $$ \tr^{\Delta}(XY^*) = 0 $$ 
		\item $$ \tr^{\Delta}(Xa)=\tr^\Delta(aX)=0$$ 
	\end{enumerate} 
\end{Lemma}

\begin{Theorem} \cite{Ma2021-1}
	The Hilbert-Schmidt definite positive Hermitian product $$ \left( A,B\right)_{HS} = \tr\left(AB^*\right)$$
	which is positive, definite metric on $Cl^{-1}(S^1,V)$ extends:
	\begin{itemize}
		\item to a Hermitian, non degenerate form on $Cl(S^1,V)$ by $(A,B) \mapsto (A,B)_{\Delta}=\tr^\Delta(AB^*)$
		\item to a Hermitian, non degenerate form on  $Cl_{odd}(S^1,V)$ by $(A,B) \mapsto (A,B)_{\Delta}=\tr^\Delta(AB^*)$
		\item to a $(\R-)$ bilinear, symmetric non degenerate form on $Cl(S^1,V) \oplus Vect(S^1) $ by $(A,B) \mapsto \mathfrak{Re}(A,B)_{\Delta} =\mathfrak{Re}\left(\tr^\Delta(AB^*)\right)$ where $A = a+u,$ $B = b + v,$ with $(a,b) \in Cl(S^1,V)$ and $(u,v) \in Vect(S^1).$ 
	\end{itemize}
\end{Theorem}

\begin{rem}
	We remark that $(.,.)_\Delta$ is bilinear, non degenerate but not positive. Indeed, from relation (\ref{3.}) of Lemma \ref{calcultr},  $C^\infty(S^1, M_n(\mathbb{C}))$ is an isotropic Lie subalgebra for $(.,.)_\Delta$ which proves that this $\R-$bilinear symmetric form is not positive. 
\end{rem}

From the Lie algebra $Cl(S^1,V) \oplus Vect(S^1), $ we then span by right-invariant action  of $FCl^{*}(S^1,V)$ on $TFCl^{*}(S^1,V)$ a right-invariant pseudo-metric. For this goal, the Lie algebra elements are identified as infinitesimal paths, and actions and Lie brackets are those derived from the coadjoint action (and right-Lie bracket) of $FCl^{*}(S^1,V)$ on $Cl(S^1,V) \oplus Vect(S^1), $ while we consider the trivial mapping defined by the sum $Cl(S^1,V) \oplus Vect(S^1) \rightarrow  Cl(S^1,V)  = Cl(S^1,V) + Vect(S^1) $ in order to compute $\mathfrak{Re}(.;.)_\Delta.$ The same constructions hold for the pseudo-Hermitian metric $(.;.)_\Delta$ on $Cl^*(S^1,V).$

\begin{Definition}
	Let $A \in FCl^{0,*}(S^1,V)$ and let $a \in Cl^0(S^1,V) \oplus Vect(S^1). $ We note by $R_A(a)$ the (right-)action by composition $$R_A(a) = a \circ A.$$
	Then, identifying $T_A FCl^{0,*}(S^1,V)$ with $R_A\left(Cl^0(S^1,V) \oplus Vect(S^1)\right)$ we set a smooth pseudo-Riemannian metric on $T FCl^{0,*}(S^1,V)$ by defining for $$(a,b) \in \left(Cl^0(S^1,V) \oplus Vect(S^1)\right)^2,$$ and hence for   $(R_A(a),R_A(b))  \in \left(T_A FCl^{0,*}(S^1,V)\right)^2,$ 
	$$ (R_A(a),R_A(b))_{\Delta, A} = (a,b)_\Delta.$$
\end{Definition}
\subsection{On bounded odd class $Diff(S^1)-$pseudo-differential operators} 

Let us finish our remarks with the group of ($L^2-$)bounded even-even $Diff(S^1)-$pseudo-differential operators. Its Lie algebra 
$$ Cl^0_{odd}(S^1,V) \rtimes Vect(S^1)$$ also reads as
$$ Cl^{-1}_{odd}(S^1,V) \oplus DO^0(S^1,V) \oplus (Vect(S^1)\otimes Id_V).$$
and the pseudo-Riemannian product $\mathfrak{Re}(.,.)_\Delta$ decomposes blockwise as
$$\left(\begin{array}{ccc}
	\mathfrak{Re}(.,.)_{HS} & * & * \\
	* & 0 & 0 \\
	* & 0 & 0 
\end{array}\right),$$ where $\mathfrak{Re}(.,.)_{HS} = \mathfrak{Re}\left((.,.)_{HS}\right)$ is the scalar product derived from the Hilbert-Schmidt Hermitian product $(.,.)_{HS}.$
\section{In search of pseudo-Hermitian connections for $(.,.)_{\Delta}$ based on \cite{Ma2021-1}} \label{s:conn}
There exists some difficulties in describing the whole space of connection 1-forms $\Omega^1(FCl_{Diff(S^1)}(S^1,V),Cl(S^1,V)\rtimes Vect(S^1)).$ Indeed the space of smooth linear maps acting on $Cl(S^1,V)$ is actually not well-understood to our knowledge. Finding an adjoint for $ad$ and for $(.,.)_\Delta$ fails apparently due to the non-traciality of $\tr^\Delta$ (and surprisingly not due to the lack of the classical arguments envolving strong metrics). We consider here a class of connections where this smooth linear endomorphism is defined by composition by a smoothing operator. The resulting technical simplifications enables us to get pseudo-Hermitian connections for $(.,.)_{\Delta}.$ Most of them can be easily adapted to get pseudo-Riemannian connections for $\mathfrak{Re}(.,.)_{\Delta}.$
\subsection{A class of connections}
Let us define now, for $w \in Cl(S^1,V)$ such that $\forall (a,b)\in Cl(S^1,V),$ 
$$\Theta^w_ab = b[a,w].$$
	The curvature of $\Theta^w$ reads as $$\Omega_{\Theta^w}(a,b)c =[wbw,a] - [waw,b] + [wa,wb]  - [wb,aw] -[[b,a]w] $$ for $(a,b) \in \F Cl(S^1,V).$
Let us analyze 
the connection $\Theta^w$ with $w = i\epsilon(D).$   
\begin{Theorem}
	$\Theta^{i\epsilon(D)}$ is a $Cl^{-\infty}(S^1,V)-$valued connection. 
\end{Theorem}

\subsection{Pseudo-Hermitian connections associated with a skew-adjoint pseudodifferential operator}

Let $w \in Cl(S^1,V)$ such that $w^* = -w.$ For example, one can consider the example $w = i\epsilon(D).$
\begin{Lemma}
	$\forall a \in Cl(S^1,V),$
	$\forall w \in Cl(S^1,V)$ such that $w^*=-w, $ $\Theta^w_{a^*}$ is the adjoint of $\Theta^s_a$ for $(.,.)_\Delta$
\end{Lemma}

Let us now analyze

$$ (a,b)\in Cl(S^1,{\mathbb{C}})^2 \mapsto \theta^{w}_ab=b[a-a^*,w] = (\Theta^w_a-\Theta^w_{a^*})(b).$$
\begin{Theorem}
	$\theta^{w}$ is the connection 1-form of a pseudo-Hermitian connection of $(.,.)_\Delta.$ Moreover, 

	$\theta^{i\epsilon(D)}$ is $Cl^{-\infty}(S^1,V)-$valued as $\Theta^{i\epsilon(D)}$ is.
\end{Theorem}

\subsection{On another class $Cl^{-\infty}(S^1,V)-$connections}
Motivated by the previous example of $Cl^{-\infty}(S^1,V),$ let us now give families of $Cl^{-\infty}(S^1,V)-$connections which a priori do not include the connections $\theta^{i\epsilon(D)}$ and $\Theta^{i\epsilon(D)}$.  Let us define now, for $s \in Cl^{-\infty}$ and $\forall (a,b)\in Cl(S^1,V),$ 
$$\Theta^{s,l}_ab = sas^*b,$$
$$\Theta^{s,r}_ab = bsas^*,$$
and 
$$\Theta^{s,[]}_ab = \left[sas^*,b\right].$$
Let us describe here their associated class of pseudo-Riemannian connections for $(.,.)_\Delta$ along the lines of the previous section.
Let $s \in Cl^{-\infty}(S^1,V)$ be a smoothing operator. Let $a,b \in  Cl(S^1,V)^2$ and let  $$\theta_a^{s,[]} b= \Theta_a^{s,[]}b - \Theta_{a^*}^{s,[]}b =\left[s(a-a^*)s^*,b\right] ,$$ 
$$\theta_a^{s,l} = \Theta_a^{s,l}b - \Theta_{a^*}^{s,l}b=s(a-a^*)s^*b,$$
and 
$$\theta_a^{s,r} = \Theta_a^{s,r}b - \Theta_{a^*}^{s,r}b= bs(a-a^*)s^*.$$
\begin{Lemma} \label{Thetas}
	$\forall a \in Cl(S^1,V),$
	$\forall s \in Cl^{-\infty}(S^1,V),$ 
	\begin{itemize}
		\item $ \Theta^{s,l}_{a^*}$ is the adjoint of $\Theta^{s,l}_a$ for $(.,.)_\Delta$
		\item $ \Theta^{s,r}_{a^*}$ is the adjoint of $\Theta^{s,l}_a$ for $(.,.)_\Delta$
		\item $ \Theta^{s,[]}_{a^*}$ is the adjoint of $\Theta^{s,l}_a$ for $(.,.)_\Delta$
	\end{itemize} 
\end{Lemma}

\begin{Theorem}
	
	Then  $\theta^{s,[]},$ $\theta^{s,l}$ and $\theta^{s,r},$ define three right-invariant pseudo-Hermitian connections on $FCl(S^1,V).$ 
\end{Theorem}

\vskip 6pt
\noindent
\underline{\bf Open problem:}
Determine the geodesic equations of these connections, determine the corresponding dynamical systems that one can deduce, and determine families of one parameter families of connections that link these systems. These ``homotopies'' can exibit or not symmetry breakings.

\subsection{The Schwinger cocycle and the connection $\Theta^{i\epsilon}$}
Let us make the two following remarks
\begin{Proposition}
	Let $(a,b) \in Cl(S^1,V)\rtimes Vect(S^1).$ Then 
	$$c_s(a,b) = -i\tr^\Delta(\Theta^{i\epsilon}_ab) = \tr^\Delta(\Theta^{i\epsilon}_a\Theta^{i\epsilon}_b \epsilon(D)).$$ 
\end{Proposition}

\begin{rem}
	When defining a smoothing connection $\theta$ on $Cl(S^1,V)\rtimes Vect(S^1),$ we define a map with values on the first component of the product $Cl(S^1,V)\times Vect(S^1).$
\end{rem}  

\begin{Theorem}
	The Schwinger cocycle $c_s$ has the same cohomology class as $$c_1^{i\epsilon} : (a,b) \in Cl(S^1,V)^2 \mapsto  \frac{1}{2}\tr^\Delta\left(\Omega^{i\epsilon}(a,b)\epsilon(D)\right)$$
	where $\Omega^{i\epsilon}$ is the curvature of $\Theta^{i\epsilon}.$
\end{Theorem}

\vskip 6pt
\noindent
\underline{\bf Open problem:} Determine the cohomology classes that can be obtained this way, by connections with  $Cl^{-\infty}-$valued curvatures, and implement an adequate holonomy bundle with suitable (maybe generalized) geometric structures. A partial answer is given in the non-presented (because non-published and only pre-published) work \cite{Ma2021-2}, that has to be generalized both for other classes of connections, and for changing $S^1$ for a higher dimensional manifold $M.$  
\subsection{On odd class $Diff(S^1)-$pseudo-differential operators}
Considering now $$ Cl^*_{odd}(S^1,V) \rtimes  Diff(S^1),$$
we remark that the renormalized trace $\tr^\Delta$ is tracial on its Lie algebra $Cl_{odd}(S^1,V) \rtimes  Vect(S^1),$ i.e. $$\forall (a,b)\in Cl_{odd}(S^1,V), \quad \tr^\Delta([a,b])=0$$
(representing $Cl_{odd}(S^1,V) \rtimes  Vect(S^1)$ in $Cl_{odd}(S^1,V)$ as in the rest of the text).This enables to state the following property:
\begin{Proposition}
	$\forall a \in Cl_{odd}(S^1,V),$ the adjoint map 
	$$ ad_a : b \mapsto ad_a b = [b,a]$$
	has an adjoint map for $(.,.)_\Delta$ given by 
	$$ ad_a^* = ad_{a^*}.$$
\end{Proposition}  

As a consequence, applying the arguments of \cite{F2} and especially those leading to \cite[Proposition 1.7]{F2} to \textbf{right-}invariant vector fields on $FCl_{odd,Diff(S^1)}(S^1,V)$, we get:
\begin{Theorem}
The pseudo-Riemannian metric $\mathfrak{Re}(.,.)_\Delta$ admits a unique pseudo-Riemannian, torsion-free (i.e. Levi-Civita) connection $\nabla^{\Delta}$ that reads as 
	$$\nabla^\Delta_ab = \frac{1}{2}\left( ad_a b  - ad_{a^*} b - ad_{b^*}a\right)$$
\end{Theorem}

\vskip 6pt
\noindent
\underline{\bf Open problem:} For a weak (pseudo-)metric in e.g. such a general framework, determine, depending on the chosen structure group, if there exist a torsion free (pseudo-)Riemannian or Hermitian connection. 

\section{Manifolds of mappings and embeddings}
Let $M$ be a compact boundaryless manifold and let $N$ be a finite dimensional Riemannian manifold. The basic structure of manifolds of maps 
{$C^\infty(M,N)$} is known since \cite{Ee}. For $f \in C^\infty(M,N),$ the tangent space $T_fC^\infty(M,N) = \gamma(M,f^*TN)$ and the $L^2-$ exponential map can be described pointwise, by the way, one can define a frame bundle over $C^\infty(M,N)$ defined by $0-$order differential operators. The space of embeddings $Emb(M,N)$  is an open submanifold of $C^\infty(M,N).$ 
On these spaces, we have to mention works from us (presented in our PhD thesis and hence not presented for as works for the habilitation evaluation) and others, which started with Freed's and Paycha's works \cite{F1,F2,Pay} on the development of Chern-Weil forms on mapping spaces. 
These construction were motivated by \cite{RSF1985,BR1987,BR1987-2}, where anomalies are identified as kind of extension of a Chern form "$\tr \Omega$", where $\tr$ needs to be understood only as a chosen summation over a. well-chosen orthonormal basis. This "well-chosen" summation has been interpreted, following the ideas of \cite{Pay,Mick1994}, see e.g. \cite{Ism}, as a renormalized trace $\tr^Q$ for $Q=\Delta.$ More precisely, re-interpreting and extending some results \cite{F2}, 

\begin{Theorem} \cite{CDMP}
	Let $G$ be a semi-simple compact Lie group, let $C^\infty_b(S^1,G)$ be the based loop group equipped with the almost complex structure $J = D/|D|,$ and let $\omega$ the K\"ahler form of the based loop group. Then $$\tr^\Delta(\Omega^{1,0}) = -i\omega,$$ where $\Omega^{1,0}$ is the holomorphic part of the $H^{1/2}-$Levi-Civita connection, associated to the $H^{1/2}$-metric $(\Delta^{1/2}.,.)_{L_2}.$ 
\end{Theorem} 

This theorem has been extended:

\begin{Theorem} \cite{Ma2006}
	With the same notations, if $\theta$ is a connection on $C^\infty_b(S^1,G)$ such that, read on left-invariant frames, $\forall X \in C^\infty_b(S^1,\mathfrak{g},)$ $\theta_X \in ad_X + Cl^{-1}(S^1,\mathfrak{g}),$ then the associated Chern form $$\tr(\Omega^{1,0})$$ is closed and it has the same cohomology class as $-i\omega.$
\end{Theorem}

Let us make two comments before going to other results:
\begin{itemize}
	\item Freed's results in \cite{F1} only extend the contruction of the first Chern form $\tr(\Omega)$ when $\Omega$ is trace-class, which is not the case here since $\Omega^{1,0}$ lies in the Dixmier ideal $L^{1,\infty}$ in Freed's framework \cite{F2} or in $Cl^{-1}(S^1,\mathfrak{g})$ in the framework of \cite{CDMP}. 
	\item With the infinite dimensional Ambrose-Singer theorem \ref{Ambrose-Singer}, and especially the part stated in \cite{Ma2004}, we have a principal bundle over $$C^\infty_b(S^1,G)_\mbbc,$$ modelled on $$GL^2 \cap Cl^0(S^1,\mathfrak{g}),$$ which is shown to be non trivial since its first Chern form has a non trivial cohomology class. This is a deep contrast with Kuiper's triviality results \cite{Kui} for $U(H)-$principal bundles, when $H$ is a Hilbert space.
\end{itemize}

Around this central example, 
natural questions raised:
\begin{itemize}

	\item \textit{What happens passing to $C^\infty(M,N)?$ } This is mostly the aim of \cite{Ma2006}. There are many cases when direct investigations on the properties of fields of weights cannot conclude if the obtained Chern-Weil forms are closed or not, and if this they are closed, when they belong to the same cohomology class. For example, the first Chern form of the $H^1-$Levi-Civita connection on $C^\infty_b(S^1\times S^1, \mathfrak{u}(n))$ is shown to vanish, bu with no comparison result with other $Cl^{-1,*}-$connections. The best result that one can state is the following, adapted from \cite{Ma2006}:
	
	\begin{Proposition}
		If $M$ is compact aand if $N$ is parallelizable, if $Q$ is a diagonal weight, then the Chern-Weil forms $\tr^Q(\Omega^k)$ lie in the same cohomology class for $Cl^{-m,*}-$connections.
	\end{Proposition}  
	We have similar results with odd class operators:
	\begin{Proposition}
		In $M$ is compact and odd-dimensional, if $Q$ is an odd-class weight, then then the Chern-Weil forms $\tr^Q(\Omega^k)$ lie in the same cohomology class for $Cl^{0,*}_{odd}-$connections.
	\end{Proposition}
	But these two results only enables one to get vanishing characteristic classes, or characteristic classes with undetermined cohomology class. 
	
	\item \textit{Can one choose other "traces"?} If one requires the traciality property $\tr[A,B]=0,$ the only trace on $Cl(M,\mbbc)$ is the Wodzicki residue \cite{W} when $dimM>1,$ with a quite similar situation for $dimM=1$ with a Wodzicki residue derived from two Adler traces \cite{W,Ka}, as suggested before with the splitting $\mathcal{F}Cl(S^1,\mbbc)= \mathcal{F}Cl_+(S^1,\mbbc)\oplus \mathcal{F}Cl_-(S^1,\mbbc).$  An investigation of the situation with the Wodzicki residue, initiated in \cite{Ma2006} where we showed that the Chern-Weil forms are trivial for $C^\infty(M,N),$ and extended in \cite{MRT} (without citing \cite{Ma2006}) to Chern-Simons forms, revealed potential applications. For bounded operators, the only traces on $Cl^0$ are spanned by the Wodzicki residue and the leading symbol trace \cite{PL2007}. This means that there is no tracial extension of the classical trace of trace-class operators on these algebras and all the traces vanish on trace-class operators. However, the discussion of the previous point is far from finished since some interesting traces can appear when restricting the considered structure groups. 
	
\end{itemize} 

Another manifold of interest is the space of embeddings, see e.g. \cite{BF}.  
The group of diffeomorphisms of $M$, $Diff(M)$,
acts smoothly and on the right on $Emb(M,N)$, by composition. Moreover,
$$B(M,N)=Emb(M,N)/ Diff(M)$$ is a smooth manifold \cite{BF,KM}, and $\pi: Emb(M,N)\rightarrow B(M,N)$
is a principal bundle with structure group $Diff(M)$ (see
\cite{KM,Mo}).
Let us now precise the vertical tangent space and a normal vector
space of the orbits of $Diff(M)$ on $Emb(M,N)$. $T_fPEmb(M,N)$,
the tangent space at $f$, is identified with the space of smooth
sections of $f^*TN$, which is the pull-back of $TN$ by $f$.
Let $\normal_f$ be the normal space to
$f(M)$ with respect to the metric $(.,.)$ on $N$. For any $x \in M$,
$T_{f(x)}N = T_{f(x)}f(M) \oplus \normal f(M)$. Hence, denoting $f^* \normal_f$ the
pull back of $\normal _f $ by $f$, we have that
$$ C^\infty(f^*TN) = C^\infty (TM) \oplus f^*\normal_f.$$
Moreover, for any volume form $dx$ on $M$,
if $$ <.,.> : X,Y \in C^\infty(f^*TN) \mapsto <X,Y> = \int_M (X(x),Y(x)) dx$$ is a
$L^2$-inner product on $C^\infty(f^*TN)$, this splitting is orthogonal for $<.,.>$.
We get here a fundamental difference between the inclusion $Emb(M,N) \subset C^\infty(M,N),$
where the model space of the type $C^\infty(f^*TN),$ and $Emb(M,N)$ as a $Diff(M)-$ principal bundle: 
sections of the vertical
tangent vector bundle read as order 1 differential operators, 
where as the operators acting on the normal vector bundle reads
as $0-$order differential operators, just like the structure group of $TC^\infty(M,N).$ 

Now, let $f \in Emb(M,N)$ and let us consider the map $$ \Phi^{U,f}: (f,v, X) \in TU \sim (1-p)TU \oplus pTU \mapsto \Xi^f(v). exp_{Diff(M)}(X) \in Emb(M,N). $$
This map gives a local (fiberwise) trivialization of the principal bundles $Emb(M,N) \rightarrow B(M,N)$ following \cite{HV,KM,Mo}, and we see that the changes of local trivializations have $Aut(\mathcal{N})$ as a structure group.

If $M$ is oriented, we note by $Diff_+(M)$ the group of orientation preserving diffeomorphisms and we have
$$\frac{Diff(M)}{Diff_+(M)} = \mathbb{Z}_2.$$

Then, defining $$B_+(M,N) = \frac{Emb(M,N)}{Diff^+(M)}$$
we get that
$B_+(M,N)$ is a 2-cover of $B(M,N).$

\subsection{On the structures of spaces of embeddings, based on \cite{Ma2016-1}}
Forgotten in \cite{GBV2014}, one can consider also based embeddings. By taking  basepoints $x_0 \in M$ and $y_0 \in N,$ we define the principal bundle of based embeddings.
\begin{Proposition} \cite{Ma2016-1}
	Let $Emb_b(M,N) = \{f \in Emb(M,N) | f(x_0) = y_0 \}.$
	Let $$Diff_b(M) = \{ g \in Diff(M) | g(x_0) = x_0 \} \hbox{ and } Diff_{b,+}(M) = Diff_b(M) \cap Diff_+(M).$$
	Let $$B_b(M,N) = Emb_b(M,N) / Diff_b(M,N) \hbox{ and }B_{b,+}(M,N) = Emb_{b}(M,N) / Diff_{b,+}(M,N).$$
	Then $Emb_b(M,N)$ is a principal bundle with base $B_b(M,N)$ (resp. $B_{b,+}(M,N)$) and with structure group $Diff_b(M)$ (resp. $Diff_{b,+}(M)$) 
\end{Proposition}
This completes \cite{GBV2014}, where structure groups such as volume preserving diffeomorphisms $Diff_\mu(M)$ were considered to build the quotient $B_\mu(M,N)= Emb(M,N)/Diff_\mu(M).$


\subsection{Chern forms in infinite dimensional geometry} 
\subsubsection{Chern forms in infinite dimensional setting, based on \cite{Ma2016-1}}
Surprisingly, the development of Chern-Weil forms using the full space of (adequate) polynomials, tstead of the ones derived from the standard polynomials $tr(A^k),$ was not present in the literature till \cite{Ma2016-1}. This remark is fundamental because, followin a remark given by C. Roger, there can exist, on infinite dimentional Lie algebras, other types of polynomials. Let us then sketch the general description of Chern-Weil forms.
  
Let $P$ be a
principal bundle, of basis $M$ and with structure group $G$. Let
$\mathfrak g$ be the Lie algebra of $G$. Recall that $G$ acts on
$P$, and also on $P \times \mathfrak g$ by the action $((p,v), g)
\in (P \times \mathfrak g)\times G \mapsto (p.g, Ad_{g^{-1}}(v))
\in (P \times \mathfrak g)$. Let $AdP = P \times_{Ad g} = (P \times
\mathfrak g )/ G $ be the adjoint bundle of $P$, of basis $M$ and
of typical fiber $\mathfrak g$, and let $Ad^kP = (Ad P)^{\times
	k}$ be the product bundle, of basis $M$ and of typical fiber
$\mathfrak g^{\times k}$.

\begin{Definition} Let $k$ in $\N^*$. We define
	$\pol^{k}(P)$, the set of smooth maps $Ad^kP \rightarrow \mbbc$ that are $k$-linear
	and symmetric on each fiber, equivalently as the set of smooth maps 
	$P \times \mathfrak{g}^k
	\rightarrow \mbbc$ that are $k$-linear symmetric in the second variable 
	and $G$-invariants with respect to the natural coadjoint action 
	of $G$ on $\mathfrak{g}^k .$
	
	Let $\pol(P) = \bigoplus_{k \in \N*}\pol(P)$.
\end{Definition}

Let $\C(P)$ be the set of connections on $P$. For any $\theta \in \C(P)$,
we denote, only for this section, by $F(\theta)$ its curvature and $\nabla^\theta$ 
(or $\nabla$ when it carries no ambiguity) its covariant derivation. 
Given an algebra $A$,
In this section, we study the maps, for $k \in \N^*$,
\begin{eqnarray*} \label{fomega}
Ch \quad : \quad \C(P) \times \pol^k(P) & \rightarrow & \Omega^{2k}(M,\mbbc)  \\
(\theta, f) & \mapsto& Alt (f(F(\theta),...,F(\theta))) \end{eqnarray*}
where $Alt$ denotes the skew-symmetric part of the form.
Notice that, in the case of the finite dimensional matrix groups
$Gl_n$ with Lie algebra $\mathfrak{gl}_n$, the set $\pol(P)$ 
is generated by the polynomials $A \in \mathfrak{gl}_n \mapsto \hbox{tr}(A^k),$ for $
k \in {0,...,n}$. This leads to classical definition of Chern forms, see e.g. \cite{KN}. 
However, in the case of infinite dimensional structure groups, most situations are 
still unknown and we do not know how to define a set of generators for $\pol(P).$ Moreover, since the usual trace $\tr$ of matrices is satisfies the "trace property" $\tr[A,B]=0 \quad \forall (A,B)\in M_n(\mbbc),$ the classical constructions of Chern forms, and the related proofs, are deeply simplified compared to what follows. We must also say that we have been surprised to find nowhere the following material, proved in \cite{Ma2016-1}.

\begin{Theorem} \label{chern-weil} \cite{Ma2016-1}
	Let $f\in \pol(P)$ for which there exists $\theta \in \mathcal{C}(P)$ such that $[\nabla^\theta,f]=0.$ 
	We shall note this set of polynomials by $\pol_{reg}(P).$
	Then, the map 
	$$Ch^f : \theta \in \C(P) \mapsto Ch^f(\theta)= Ch(\theta,f)\in \Omega^*(P,\mbbc)$$
	takes values into closed forms on $P$. Moreover, 
	
	(i) it is vanishing on vertical 
	vectors and defines a closed form on $M$.
	
	(ii) the cohomology class of this form does not depend on the choice of the
	chosen connexion $\theta$ on $P$. 
	
	Moreover,
	$\forall (\theta,f) \in\mathcal{C}(P)\times \pol_{reg}(P), [\nabla^\theta,f] = 0.$
\end{Theorem}

\begin{Proposition} \cite{Ma2016-1}
	Let $\phi : \mathfrak{g}^k \rightarrow \mbbc$ be a $k-$linear, symmetric, $Ad-$invariant form. Let $f:P\times \mathfrak{g}^k\rightarrow \mbbc$ be the map induced by $\phi$ by the formula:
	$f(x,g)= \phi(g).$ 
	Then $f \in \pol_{reg}.$ 
\end{Proposition}

 The problem of these infinite dimensional Chern forms was first raised by Freed \cite{F1,F2} where the connections considered where $Gl^p$ connections, i.e. with curvature valued in the Schatten ideal $L^p = \left\{A \in L(H) | \, |A|^p \hbox{ is trace class }\right\}.$ 

\subsubsection{Application to $Emb(M,N)$}
Mimicking the approach of \cite{Ma2006}, the cohomology classes of Chern-Weil forms should give rise to homotopy invariants.  
Applying Theorem \ref{chern-weil}, we get:

\begin{Theorem} \label{cantrace} \cite{Ma2016-1}
	The Chern-Weil forms $Ch^f$ is a $H^*(B(M,N))-$valued invariant of the homotopy class of an embedding, $\forall k \in \N^*.$ 
\end{Theorem}

When $M=S^1$, $Emb(S^1,N)$ is the space of (parametrized) smooth knots on $N$, and $B(S^1,N)$ is the
space of non parametrized knots. Its connected components are the homotopy classes
of the knots, through classical results of differential topology, see e.g. \cite{Hir}.
We now apply the material of the previous section to manifolds of embeddings. For this, we can define 
invariant polynomials of the type  
$$ A \mapsto \lambda(A^k) \in \pol^k_{reg},$$ where $\lambda = \tr^Q$ for a well-chosen weight $Q.$ 
Let us give the following example when $M$ is odd-dimensional: 
the Kontsevich and Vishik trace \cite{KV1,KV2} is a renormalized trace for which $tr^Q([A,B]) = 0$ for each differential operator $A,B$ and does not depend 
on the weight chosen in the odd class. For example, one can choose $Q = Id+  {\nabla } ^* {\nabla  }$, where $\nabla$ is a connection induced on $\mathcal{N}_f$ by the Riemannian metric, as described in \cite{Ma2006}. It is an order 2 injective elliptic differential operator (in the odd class), and the coadjoint action of $Aut(\mathcal{N}_f)$ will give rise to another order 2 injective elliptic differential operator \cite{Gil}. When $Q = Id+  {\nabla } ^* {\nabla  },$ this only changes $\nabla$ into another connection on $E.$ Thus,
$\phi(A,...,A) = tr^Q(A^k)\in \pol_{reg}.$
Let us now consider a connected component of $B(M,N),$ i.e. a homotopy class of an embedding among the space of embeddings.
Following \cite{Ma2016-1},

\begin{Theorem} Let $M$ be an odd dimensional boundaryless nanifold and let $Q = Id+  {\nabla } ^* {\nabla  }$ on $\normal.$
	The polynomial 
	$$ \phi :A \mapsto tr^Q(A^k)$$ is $Diff(M)-$invariant, and gives rise to an invariant of non oriented knots, i.e. a Chern form on the base manifold $$B(S^1,N) = Emb(M,N) / Diff(M)$$
\end{Theorem}

\vskip 6pt
\noindent
\underline{\bf Open problem:} Prove that these Chern-Weil forms give rise to non-vanishing characteristic classes. This problem seems related with the work \cite{MickPay2007} which is more or less heuristically linked with $FCl^{0,*}-$ connections in \cite{Pay2013}. The full link between these two settings still needs to be clarified, even if intuitionistic arguments given both in \cite{Pay2013} and in \cite{Ma2016-1} show that there are big similitudes between the two independently devolopped settings. 
\chapter{Contributions to the theory of KP hierarchies}
	Let  $R$ be an algebra of functions equipped with a derivation $\partial.$ For us, $R = C^\infty(S^1,\K)$ with $\K = \R, \mbc$ and  $\mathbb{H},$ and $\partial = \frac{d}{dx}.$ 
Let $T=\{t_n\}_{n \in \N^*}$ be an infinite set of formal (time) variables and let us consider the 
algebra of formal series 
$\Psi DO(S^1,\K)[[T]]$ with infinite set of formal variables $t_1,t_2,\cdot$ with $T-$valuation $val$ defined by $val_T(t_n) = n$ \cite{M1}. One can extend naturally on $\Psi DO(S^1,\K)[[T]]$ the notion of smoothness from the same notion on $\Psi DO(S^1,\K),$ see \cite{ER2013} for a more complete description.    
The Kadomtsev-Petviashvili (KP) hierarchy reads
\begin{equation} \label{eq:KP}
	\frac{d L}{d t_{k}} = \left[ (L^{k})_{D} , L \right]\; , \quad
	\quad k \geq 1 \; ,
\end{equation}
with initial condition $L(0)=L_0  \in \partial + \Psi^{-1} (R)$, where the subscrit $(.)_D$ means the algebraic canonical projection $$\Psi DO(S^1,\K)[[T]] = (\Psi DO^{-1}(S^1,\K) \oplus DO (S^1,\K))[[T]] \rightarrow DO (S^1,\K)[[T]]$$ . The dependent
variable $L$ is chosen to be of the form
$L = \partial + \sum_{\alpha \leq -1 } u_\alpha \partial^\alpha \in {\Psi}^1(S^1,\K)[[T]] \; .$
A standard reference on (\ref{eq:KP}) is L.A. Dickey's treatise \cite{Di}, see also 
\cite{KZ,M1,M3,RS} as well as \cite{Mickbook,PS,SW1985,Ue,W1,W2} for a link with infinite dimensional geometric structures (determinant bundles, restricted linear group $GL_{res}$) and with non linear partial differential equations such as KdV, Boussinesq, KP equations. The KP hierarchy produces a formal solution to all these systems, under mild assumptions on the initial conditions and constraints. It can be derived from the search of isospectral deformations of 
diferential operators, in link with the so-called Gelfand-Dickey hierarchy, and it recently appeared in various contexts such as Hodge theory and combinatorics.

\vskip 12pt

In order to solve the KP hierarchy, we need the following groups (see e.g. \cite{MR2016} for a latest adaptation of Mulase's construction \cite{M1,M3}):
$$ \bar{G} = 1 + \Psi DO^{-1}(S^1,\K)[[T]],$$
$$ \widehat{\Psi} = \left\{ P = \sum_{\alpha \in {\mathbb{Z}}}
a_{\alpha}\,\partial^{\alpha} \in {\Psi}(S^1,\K)[[T]] \; : \exists N \in \N,
\, val_T(a_\alpha)>_{\alpha \rightarrow +\infty} C \alpha-N  \hbox{ and } P|_{t = 0} \in 1 + \Psi DO^{-1}(S^1,\K)  \right\}$$ 
and 
$$\widehat{\D} = \left\{ P= \sum_{\alpha \in \mathbb{Z}}
a_{\alpha}\,\partial^{\alpha} : P \in\widehat{\Psi} \mbox{ and }
a_\alpha=0 \mbox { for } \alpha <0 \right\} \; .$$
We have a matched pair $ \widehat{\Psi} = \bar{G} \bowtie \widehat{\D}.$
The same constructions apply in various settings, see e.g. \cite{D1,D2,Ku2000,McI2011}, and all are envolved with purely algebraic arguments in contrast with our work. 
\section{Smoothness of Mulase decomposition and well-posedness of the KP hierarchy, based on \cite{ERMR2017,MR2019,MR2016}}

When algebras of functions $R = C^\infty(S^1,\K)$ are Fr\'echet algebras, a natural notion of differentiability occurs, making addition, multiplication and differentiation smooth. By the way, considering addition and multiplication in $\Psi DO(S^1,\K),$ one can say that addition and multiplication in $\Psi DO(S^1,\K)$ by understanding, under this terminology, that, if $A = \sum_{n \in \Z} a_n \partial^n $ and $ B = \sum_{n \in \Z} b_n \partial^n ,$
setting $A+ B = C = \sum_{n \in \Z} c_n \partial^n$
and $AB = D = \sum_{n \in \Z} d_n \partial^n$
the map $$ \left((a_n)_{n \in \Z},(b_n)_{n \in \Z}\right) \mapsto \left((c_n)_{n \in \Z},(d_n)_{n \in \Z}\right)$$ is smooth in the relevant infinite product, encoded into diffeological concepts in \cite{ERMR2017,MR2019,MR2016} where a fully rigorous framework for smoothness on these objects is described and used. Again following \cite{ERMR2017}, the same smoothness properties can be described for any differential unital algebra $(A,\partial),$ but in this general setting, diffeologies seem to be an essential tool. Let us describe more precisely the setting. 
Let $A^*$ be the group of invertible elements (or, {\em units}) of $A$.
Let $\xi$ be a formal variable not in $A$. The {\em algebra of symbols} over $A$ is the vector space
\[
\Psi_{\xi}(A) = \left \{ P_{\xi} = \sum_{\nu \in {\bf Z}} a_{\nu} \, \xi^{\nu} \; | \; a_{\nu}
\in A \, , \;  a_{\nu} = 0 \mbox{ for } \nu \gg 0 \right \} \;
\]
equipped with the associative multiplication $\circ\,$ given by
\begin{equation}  \label{alfa}
	P_{\xi} \circ Q_{\xi} = \sum_{k \geq 0} \frac{1}{k !} \, \frac{\partial^{k}P_{\xi}}
	{\partial \xi^{k}} \, \dd^{k} Q_{\xi} \; ,
\end{equation}
with the prescription that multiplication on the right hand
side of (\ref{alfa}) is standard multiplication of Laurent series in $\xi$
with coefficients in $A$, see \cite{Di}. The algebra $A$ is
included in $\Psi_\xi(A)$. The {\em algebra of formal pseudodifferential operators} over $A$ is the vector space
\[
\Psi (A) = \left \{ P = \sum_{\nu \in \mathbb{Z}} a_{\nu} \,
\partial^{\nu} \; | \; a_{\nu} \in A \, , \; a_{\nu} = 0 \mbox{ for } \nu \gg 0 \right \}
\]
equipped with the unique multiplication which makes the map
$\sum_{\nu \in \mathbb{Z}} a_{\nu} \, \xi^{\nu} \; \mapsto \;
\sum_{\nu \in \mathbb{Z}} a_{\nu} \, \partial^{\nu}$
an algebra homomorphism. The
algebra $\Psi (A)$ is associative but not commutative. It becomes
a Lie algebra over $K$ if we define, as usual,
\begin{equation}
	[ P , Q ] = P \, Q - Q \, P \; .  \label{liebrack}
\end{equation}

The {\em order} of $P \neq 0 \in \Psi (A)$, $P = \sum_{\nu \in \mathbb{Z}} a_{\nu} \,
\partial^{\nu}$, is $N$ if $a_{N} \neq
0$ and $a_{\nu} = 0$ for all $\nu > N$. If $P$ is of order $N$, the coefficient
$a_{N}$ is called the {\em leading term} or {\em principal symbol} of $P$. We note by $\Psi^N(A)$ the vector space 
of pseudodifferential operators $P$ as above satisfying
$k > N \Rightarrow a_k = 0.$ We note  $$\Psi^N(A)=\left\{a \in \Psi(A) \;|\; \forall m > N, a_m=0\right\}\; .$$ The special
case of $\Psi^0(A)$ is of particular interest, since it is an algebra. Adapting the notations used in the previous 
subsection, we write $\Psi^{0,*}(A)$ for its group of units,
i.e. the group of invertible elements of $\Psi^0(A).$ 
We assume now that the algebra $A$ is a \textbf{Fr\"olicher algebra}, and that therefore addition, scalar
multiplication and multiplication are smooth, and that inversion is a smooth operation on $A^*$, in which $A^*$ 
is equipped with the subset Fr\"olicher structure. We also assume that the derivation $\partial$ is smooth.
Then, identifying a formal operator $ P \in \Psi(A)$ with its sequence of partial symbols, we conclude that $\Psi(A),$ as a linear subspace of $A^\Z,$ carries a natural Fr\"olicher structure. We obtain:

\begin{Proposition} \label{Psi-F} \cite{ERMR2017}
	$\Psi(A)$ is a Fr\"olicher algebra.
\end{Proposition}

Now let us assume that $A^*$ is a Fr\"olicher Lie group with Lie algebra $\mathfrak{g}_A$,
and that $A$ is an integral Fr\"olicher vector space. Our Theorems \ref{regulardeformation} and \ref{exactsequence}
imply the following two results:

\begin{Lemma} \label{Psi-1} \cite{ERMR2017}
	$1+ \Psi^{-1}(A)$ is a regular Fr\"olicher Lie group with regular Lie algebra $\Psi^{-1}(A).$
\end{Lemma}

\begin{Theorem} \cite{ERMR2017} \label{groupabstrpdo}
	There exists a short exact sequence of groups:
	$$ 1 \longrightarrow 1 + \Psi^{-1}(A) \longrightarrow \Psi^{0,*}(A) \longrightarrow A^* \longrightarrow 1$$
	such that:
	\begin{enumerate}
		\item The injection $1 + \Psi^{-1}(A) \rightarrow \Psi^{0,*}(A)$ is smooth.
		\item The principal symbol map $\Psi^{0,*}(A) \rightarrow A^*$ is smooth and it has a global section which is 
		the restriction to $A^*$ of the canonical inclusion $A \rightarrow \Psi^{0}(A).$
	\end{enumerate}
	As a consequence, $A^*$ is a fully regular Fr\"olicher Lie group if and only if  $\Psi^{0,*}(A)$ is a 
	fully regular Fr\"olicher Lie group with Lie algebra $\mathfrak{g}_A \oplus \Psi^{-1}(A).$
\end{Theorem}

\vskip 6pt
\noindent
\underline{\bf Open problem:}
It is natural to wonder if we can extend Theorem \ref{groupabstrpdo} to the full algebra $\Psi(A),$ and to the Fr\"olicher Lie group $\Psi(A)^*$ separately. 
\begin{itemize}
	\item On one hand, in \cite{MR2016}, for $A=\R,$ we get that $\Psi(A) \sim \R((X))$ is the Lie algebra of a non-regular Lie group, which shows that the problem of enlargibility of $\Psi(A)$ cannot be solved by finding an underlying regular Fr\"olicher Lie group.
	\item The group $\Psi(A)^*$ has formally two candidates as Lie algebras. The first one is a Lie subalgebra of $\Psi(A)$ which has to be determined depending certainly on $\mathfrak{g}_A$ and the other one yields a formal correspondence with the group by a formal exponential described in e.g. \cite{KW} for $A = C^\infty(\R).$ In each of these more or less formal settings, the exact sequence envolving the group $1 + \Psi^{-1}(A)$ does not seem so straightforward to generalize.
\end{itemize}

\subsection{Smoothness of Mulase decomposition and refinements}
We can generalize our foregoing discussion. We hope this generalization will be of use, for instance, in the
development of $p$-adic KP theory. Our basic idea is to replace algebras of power series by general algebras
equipped with non-archimidean valuations. We only present the main points of this generalization and we refer to \cite{ERMR2017} for details.   
Let $A$ be an associative (but not necessarily commutative!) $K$--algebra with unit $1$,
in which $K$ is an arbitrary field of characteristic zero. We assume that $A$ is a
\textit{diffeological algebra} and that $A$ is equipped with a smooth derivation $D$. We consider $\Psi(A)$ as in the
previous subsection, but instead of assuming that $A$ is an algebra of series, we equip it with a valuation,
adapting an idea from \cite{D1,D2}. 
A valuation allows us to equip $A$ with a topology. For $\alpha \in \mathbb{Z}$ and $x_0 \in A$ we set
$$
V_\alpha (x_0) = \{ x \in A : \sigma(x - x_0) > \alpha\}\; ;
$$
we easily see that for each $x_0 \in A$, the collection 
$\{ V_\alpha (x_0) \}_{\alpha \in \mathbb{Z}}$ is a basis 
of neighborhoods for a first countable Hausdorff topology on $A$. It is a classical 
fact that this topology is metrizable: we define the absolute value classicaly 
and if we set $d(x,y) = |x-y|$ we obtain a metric $d$ on $A$ and the metric
topology coincides with the topology introduced above. Very importantly, in this topology the sets
$V_\alpha(x_0)$ coincide with the balls $\{ x \in A : | x | < c^\alpha \}$, and are both open and closed.
$A$ becomes a topological algebra and $D$ a continuous derivation. 
We also remark that compatibility of derivation and valuation implies that $|D(a)| \leq |a|$ for all $a \in A$. 

Now we let $\hat{A}$ be the completion of the metric space $(A , d)$, and we extend $D$
to a continuous derivation on $\hat{A}$. We continue denoting the extension of the
absolute value on $A$ by $| \cdot |$, and the extension of the continuous derivation on 
$A$ by $D$. 

In the previous subsection we used Fr\"olicher algebras and obtained Fr\"olicher structures on spaces of 
(regularized) formal pseudodifferential operators. It turns out that, because of the presence of quotients, the use 
of diffeologies is more natural than the use of Fr\"olicher spaces, since the latter structure does not pass so
easily to quotients and carry some problems of non-reflexivity (for a definition of reflexivity in diffeological 
spaces, see \cite{Wa}). At the present level of generality, we proceed as follows, after \cite{ERMR2017}:

For all $p \in \mathbb{Z},$ the quotient vector space projection  
$\pi_p: A \rightarrow A/A_p\,$, in which $A_p = \{a \in A\, : \, \sigma(a)\geq p\},$ extends to $\hat{A}$ in the following
way: for $\hat{a} \in \hat{A}$, we set $\pi_p (\hat{a}) = a + A_p$ if and only if $\sigma(a) = \hat{\sigma}(\hat{a})$
{\em and} $\hat{\sigma}(a-\hat{a}) \geq p$. It is possible to find such an $a \in A$ because of standard properties of valuations, as explained in \cite{ERMR2017}.

\begin{Definition}
	We equip the quotients $A/A_{p}$ with their quotient diffeology.
	The completion $\hat{A}$ is equipped with the pull-back diffeology with respect to  the family of maps 
	$\{\pi_p; p \in \mathbb{Z}\}$  (see {\rm \cite[p. 32]{Igdiff}}).
	The valuation $\sigma$ of $A$ is called a \textbf{diffeological valuation} if and only if the diffeology of $A$ 
	is the pull-back of the diffeology of $\hat{A}$  
	and all plots are continuous in the valuation topologies of $A$ and $\hat{A}$.
\end{Definition}

%

We assume that $A$ and $\hat{A}$ are equipped with diffeological valuations $\sigma$ and $\hat{\sigma}$, 
and that $D$ is smooth with respect to the product diffeology on $\hat{A},$ which implies that its restriction to $A$ 
is also smooth. The spaces of formal pseudodifferential and differential operators of infinite order now become:

\begin{Definition} \label{df} \cite{ERMR2017}
	The space of formal pseudodifferential and differential operators of infinite order are,
	respectively, 
	$\widehat{\Psi}(\hat{A})$ and $\widehat{\mathcal D}_{\hat{A}}\,$, in which
	\begin{equation} \label{aldef2}
		\begin{split}
			\widehat{\Psi}(\hat{A}) = 
			\left\{ P = \sum_{\alpha \in {\mathbb{Z}}}
			a_{\alpha}\,D^{\alpha} \; | \; a_{\alpha} \in \hat{A} \mbox { and } \exists
			A_P, B_P \in\mathbb{R}^+ \mbox{ and } M_P, N_P, L_P \in \mathbb{Z}^+ \mbox{ so that } \right.\\
			\left. M_P \geq N_P\,, \; |a_\alpha| < \frac{A_P}{\alpha - N_P}  \ \forall\ \alpha > M_P 
			\,,\mbox{ and } |a_\alpha| < B_P \ \forall\ \alpha < - L_P \right\}
		\end{split}
	\end{equation}
	and
	\begin{equation}
		\widehat{\mathcal D}_{\hat{A}} = \left\{ P= \sum_{\alpha \in \mathbb{Z}}
		a_{\alpha}\,D^{\alpha} \; | \; P \in\widehat{\Psi}(\hat{A}) \mbox{ and }
		a_\alpha=0 \mbox { for } \alpha <0 \right\} \; .
	\end{equation}
\end{Definition}

The definition of the absolute value 
$| \cdot |$ implies that $\hat{A}$ is contained in $\widehat{\Psi}(\hat{A})$ and,  as a by-product, 
we note that our assumptions on $A$ and $\hat{A}$ imply that $\widehat{\Psi}(\hat{A})$ and 
$\widehat{\mathcal D}(\hat{A})$ are diffeological spaces.

%

\begin{Lemma} {\bf \cite{ERMR2017}}
	The space $\widehat{\Psi}(\hat{A})$ has an algebra structure and
	$\widehat{\mathcal D}_{\hat{A}}$ is a subalgebra of $\widehat{\Psi}(\hat{A})$.
	{ Moreover, if $\hat{A}$ is a diffeological K-algebra, 
		$\widehat{\Psi}(\hat{A})$ and $\widehat{\mathcal D}_{\hat{A}}$ are diffeological 
		K-algebras.}
\end{Lemma}

\smallskip

Now we construct groups. There exist two standard structures associated to 
the valuation $\hat{\sigma}$ on $\hat{A}$. The subring  
$\mathcal{O}_{\hat{A}} = \{a\in \hat{A} : \hat{\sigma}(a) \geq 0\}$, 
and the two-sided ideal $\mathcal{P}_{\hat{A}} =\{a\in \mathcal{O}_{\hat{A}} : \hat{\sigma}(a)>0\}$. 
If $A$ is a diffeological algebra, these algebraic constructions carry natural underlying diffeologies. 
Since we can check that the derivation $D$ on $\hat{A}$ is compatible with $\hat{\sigma}$, 
we have $D ( \mathcal{P}_{\hat{A}} ) \subset \mathcal{P}_{\hat{A}}\,$; it follows that the 
derivation $D$ is well-defined on the quotient ring $\mathcal{O}_{\hat{A}}/\mathcal{P}_{\hat{A}}\,$. We let 
$\pi: \mathcal{O}_{\hat{A}} \rightarrow \mathcal{O}_{\hat{A}}/\mathcal{P}_{\hat{A}}$ 
be the canonical projection. Since $A$ is a diffeological algebra, the map $\pi$ is smooth. 

\smallskip

The set $G_{\hat{A}} = 1 +\mathcal{I}_{\mathcal{O}_{\hat{A}}/\mathcal{P}_{\hat{A}}}$ 
is a multiplicative group (see \cite{D2}), and a diffeological group according to Theorem \ref{regulardeformation}.  
For $P = \sum_{\nu \in \mathbb{Z}} a_\nu D^\nu \in \widehat{\Psi}(\mathcal{O}_{\hat{A}})$ we set 
$\pi(P) = \sum_{\nu \in \mathbb{Z}} \pi(a_\nu) D^\nu$. 

\begin{Definition} 
	We define the spaces
	\begin{equation} \label{345-1}
		\widehat{\Psi}(\hat{A})^{\times} = \{ P \in \widehat{\Psi}(\mathcal{O}_{\hat{A}}) \; | \; \pi(P) \in G_{\hat{A}} \}
	\end{equation}
	and
	\begin{equation} \label{346-1}
		\widehat{\mathcal D}_{\hat{A}}^{\times} = \{ P \in \widehat{\mathcal D}_{\mathcal{O}_{\hat{A}}} \; | \; \pi(P)=1 \} \; .
	\end{equation}
\end{Definition}

\begin{Proposition} \label{gr}
	The space $\widehat{\Psi}(\hat{A})^{\times}$ is a group: each element $P$ in $\widehat{\Psi}(\hat{A})^{\times}$ 
	has an inverse of the form
	\[
	P^{-1} = \sum_{n \geq 0} (1 - P)^{n} \; .
	\]
	In addition, the space $\widehat{\mathcal D}_{\hat{A}}^{\times}$ is a subgroup of 
	$\widehat{\Psi}(\hat{A})^{\times}$.
\end{Proposition}

We can prove the  following result on smoothness, see \cite{ERMR2017}. 

\begin{Proposition} \label{gr1}
	The group $G(\Psi(\hat{A})):=\widehat{\Psi}(\hat{A})^{\times}$ is a diffeological Lie group with Lie algebra 
	$\Psi(\mathcal{O}_{\hat{A}})$;
	the group $G_+(\mathcal{D}_{\hat{A}}) := \widehat{\mathcal D}_{\hat{A}}^{\times}$ is a diffeological Lie group with 
	Lie algebra $\mathcal{D}_{\mathcal{O}_{\hat{A}}}$;
	the group $G_-(\mathcal{I}_{\hat{A}}) := 1 + {\mathcal I}_{\mathcal{O}_{\hat{A}}}$ is a diffeological Lie group with 
	Lie algebra ${\mathcal I}_{\mathcal{O}_{\hat{A}}}$. 
	Moreover, the exponential map 
	$$ \exp: P \in  {\mathcal I}_{\mathcal{O}_{\hat{A}}} \mapsto 
	\sum_{n \in \mathbb{N}} \frac{(sP)^n}{n!}\in G_-(\mathcal{I}_{\hat{A}}) $$
	is one-to-one and onto, with inverse the classical logarithmic series $\log$, and both $\exp$ and $\log$ are smooth.
	As a consequence, the inversion in smooth in $G_-(\mathcal{I}_{\hat{A}})$.
\end{Proposition}

 Let us now specialize to $A=C^\infty(S^1,\R)$ as in  \cite{MR2019,MR2016},

\begin{Theorem} \label{Psihat}
	The following algebras are Fr\"olicher algebras:
	
	\smallskip
	
	\centerline{(1) $A_t\,$; \quad \quad (2) $\Psi(A_t)\,$; \quad \quad (3) $\widehat{\Psi}(A_t)\,$;
		\quad \quad (4) $\widehat{\D}_{A_t}\,$.}
\end{Theorem}


\begin{Theorem} \label{Ghat}
	
	\begin{enumerate}
		\item   $\widehat{\D}_{A_t}^{\times}$ is a Fr\"olicher group.
		\item $G_R = 1 + \Psi^{-1}(R)$ and $G_{A_t}=1+\Psi^{-1}(A_t)$ are Fr\"olicher Lie groups.
		\item $ G(\widehat{\Psi}(A_t))$ is a Fr\"olicher group.
	\end{enumerate}
\end{Theorem}

We stress the fact that we have not stated the existence of an exponential map. In fact, it seems difficult to show
the existence of the exponential map on $\widehat{\Psi}(A_t)$, and very difficult to determine 
the tangent space $T_1G(\widehat{\Psi}(A_t));$ even more, it is difficult to differentiate the possible adjoint
action of $G(\widehat{\Psi}(A_t))$ on it. Most of the difficulties come from the very general definition of 
$\widehat{\Psi}(A_t)$. This is why we construct a Fr\"olicher subalgebra 
$\overline{\Psi}(A_t)\subset\widehat{\Psi}(A_t)$ as follows:

	%

\begin{Definition}\label{barpsi}
	The regularized space of formal pseudo-differential and differential operators of infinite order are,
	respectively, 
	$\overline{\Psi}(A_t)$ and $\overline{\D}_{A_t}$, in which
	\begin{equation} \label{aldef1}
		\overline{\Psi}(A_t) = \left\{ \sum_{\alpha \in {\mathbb{Z}}}
		a_{\alpha}\,\dd^{\alpha} \in \widehat{\Psi}(A_t) \;|\; val_t(a_\alpha)\geq \alpha \right\}
	\end{equation}
	and
	\begin{equation}
		\overline{\D}_{A_t} = \left\{ P= \sum_{\alpha \in \mathbb{Z}}
		a_{\alpha}\,\dd^{\alpha}  \;|\; P \in\overline{\Psi}(A_t) \mbox{ and }
		a_\alpha=0 \mbox { for } \alpha <0 \right\} \; .
	\end{equation}
\end{Definition}

In addition, we define
\begin{equation} \label{345-bar}
	G(\overline{\Psi}(A_t)) = \{ P \in \overline{\Psi}(A_t) \;|\;
	P\vert_{t=0}\in G_{A_t}  \}
\end{equation}
and
\begin{equation} \label{346-bar}
	\overline{\D}_{A_t}^{\times} = \{ P \in \overline{\D}_{A} \;|\; 
	P\vert_{t=0}=1 \} \; ,
\end{equation}
and we can prove, see \cite{MR2016}, that $G(\overline{\Psi}(A_t))$ and $\overline{\D}_{A_t}^{\times}$
are fully regular Fr\"olicher Lie groups.

\subsection{Well-posedness of the KP hierarchy}
The following result  gives a synthesied statement of the main result of \cite{MR2016} and of \cite{ERMR2017} both on the KP hierarchy (\ref{eq:KP}), by considering two approaches of ``smoothness''
\begin{enumerate}
	\item In the general setting of \cite{ERMR2017}, smoothness is only with respect to $A-$coefficients of pseudodifferential operators, and the field of scalars is assumed to be equipped with the discrete topology (and discrete diffeology). this is classically what is needed to take safe limits in the ultrametric completions,
	\item In the classical setting of \cite{MR2016}, when $A = C^\infty(S^1,\R)$ principally and extended to $ A = C^\infty(S^1,\R)^\N$ in the same reference, the smoothness is understood with respect to $A$ \textbf{and} with respect to the variable $(t_1,t_2,...) \in \R^\infty = \bigcup_{n \in \N} \R^n.$ 
\end{enumerate}

It states smooth dependence on the initial conditions and on the values $t_1,t_2,...$ which justifies the terminology of well-posedness.

\begin{Theorem} \label{KPcentral} 
	\cite{MR2016}
	Consider the KP hierarchy 
	\ref{eq:KP}
	with initial condition $L(0)=L_0$. Then,
	\begin{enumerate}
		\item There exists a pair $(S,Y) \in \widehat{G} \times \widehat{\D}$ (resp. $\in \bar{G} \times \overline{\D}$ )
		such that the unique solution to Equation $(\ref{eq:KP})$ with $L|_{t=0}=L_0$ is
		$
		L(t_1,t_2,\cdots)=Y\,L_0\,Y^{-1} = S L_0 S^{-1}$.
		\item The pair $(S,Y)$ is uniquely determined by the smooth decomposition problem 
		$$exp\left(\sum_{k \in \N}\tau_k L_0^k\right) = S^{-1}Y$$ 
		and the solution $L$ depends smoothly on the initial condition $L_0$.
		
		\item The solution operator $L$ is smoothly dependent  on the 
		initial value $L_0.$ 
	\end{enumerate}
\end{Theorem}
\begin{rem}
	One may notice in \cite{MR2016} two proofs of Theorem \ref{KPcentral}. In the first proof of the third item of this Theorem, and more generally the first proof of this theorem; is inspired by Reyman and
	Semenov-Tian-Shansky approach to integrability via R-matrices and factorization theorems, see for instance
	\cite[Section 1.12, Theorem 7]{P}. However, our result is not exactly an instance of the 
	Reyman--Semenov-Tian-Shansky theory since in this paper we are not considering the hamiltonian content of 
	Equation $(\ref{jph})$. What we are observing here is that techniques appropriated for the study of 
	integrability of Hamiltonian systems can be adapted to prove well-posedness of the interesting equation 
	$(\ref{eq:KP})$. As in Mulase's papers \cite{M1,M3}, the crucial point of the proof is the existence of a 
	factorization of an infinite-dimensional Lie group, and not the possible hamiltonian character of the equation
	being investigated.
\end{rem}

\vskip 6pt \noindent
\underline{\bf Open question:} {sequences of approximation, pseudo-differential operators, Hamiltonnians and unstable solutions.}

Initiated in \cite{MR2016}, the procedure that consists in replacing non smooth functions by approximating sequences of smooth functions enable to enlarge the theory of non-smooth symbols initiated in \cite{BR1984}, see e.g. \cite{Marsch1988}, to symbols which approximation sequence converge in function spaces which are not embedded in $C^0.$ This embedding is the technical limitation of the actual theory. The classical results on pseuo-differential operators, such as boundedness, norm estimates, kernel analysis, spectral theory, then need to be analyzed in this generalized context, with in mind the main application of this theory which is the analysis of PDEs. 
Among these PDEs, the hamiltonnian equations can generalize straight way, again along the lines of the ideas announced in \cite{MR2016}, replacing the usual $\mbbc-$valued non degenerate pairing on regular polynomials on $DO(M)$ by a $\mbbc[[z]]-$valued pairing for hamiltonnians of the equation extended to approximation sequences. Thus first integrals of the (generalized) motion are $\mbbc[[z]]$-valued functionals, which convergence may intuitively depend on the stability of the solutions.

\section{A scaling for the KP hierarchy and the $h-$KP hierarchy, based on \cite{Ma2013}}
We make the following definition, along the lines of the theory developed in \cite{Ma2013} for formal 
pseudo-differential operators:
\begin{Definition} \label{dfs}
	Let $h$ be a formal parameter. The set of odd formal class $h-$pseudo-differential operators is the set of formal series 
	$$\Psi DO_{h}(S^1,V)=\left\{ \sum_{n \in \mathbb{N}} a_n h^n \, | \, a_n \in \Psi DO^{n}(S^1,V)  \right\}\; .$$
\end{Definition}

We state the following result on the structure of $\Psi DO_{h}(S^1,V)$: 
\begin{Theorem} \label{fs}
	The set $\Psi DO_{h}(S^1,V)$ is a Fr\'echet algebra, and its group of units 
	given by 
	$$\Psi DO_{h}^*(S^1,V)=
	\left\{ \sum_{n \in \mathbb{N}} a_n h^n \, | \, a_n \in \Psi DO^{n}(S^1,V), a_0\in \Psi DO^{0,*}(S^1,V)  \right\} \; ,
	$$
	is a regular Fr\'echet Lie group.
\end{Theorem}
This result is mostly an application of Theorem \ref{regulardeformation}.
Decomposition $L = L_S + L_D$, $L_S \in \Psi DO^{-1}(S^1,V)$,  $L_D \in DO^1(S^1,V)$ valid on $\Psi DO(S^1,V)$ extends straightforwardly to the algebra
$\Psi DO_{h}(S^1,V)$.
We now introduce the $h-$KP hierarchy with non-formal pseudo-differential operators. Let us assume that 
$t_1, t_2, \cdots, t_n , \cdots,$ are an infinite number of different formal variables. Then, again adapting work carried 
out in \cite{Ma2013}, we make the following definition:

\begin{Definition}
	Let $S_0 \in Cl^{-1,*}_{odd}(S^1,V)$ and let $L_0 = S_0 (h\frac{d}{dx})S_0^{-1}.$
	We say that an operator $$ L(t_1,t_2,\cdots) \in  \Psi DO_{h}(S^1,V)[[ht_1,...,h^nt_n...]]$$ satisfies the
	$h-$deformed KP hierarchy if and only if 
	\begin{equation} \label{jph}
		\left\{\begin{array}{cl} 
			L(0) = & L_0 \\
			\frac{d}{dt_n}L =& \left[(L^n)_D, L\right] \; .
		\end{array}
		\right.
	\end{equation}
\end{Definition}

We recall from \cite{Ma2013} that the $h-$KP hierarchy is obtained from the classical KP hierarchy by means of the 
$h-$scaling
$$\left\{ \begin{array}{ccc}t_n& \mapsto & h^nt_n \\
	\frac{d}{dx} & \mapsto &h\frac{d}{dx} \end{array}\right. \; ,$$
and we also recall that formal series in $t_1, \cdots , t_n, \cdots$ can be also understood as smooth functions on the 
algebraic sum 
$$T= \bigoplus_{n \in \mathbb{N}^*}(\mathbb{R}t_n)$$ for the product topology and product Fr\"olicher structure, 
see \cite{Ma2013}. Now we solve the initial value problem for (\ref{jph}).

	\begin{Theorem} \label{hKP}
		Let $U_h = \exp\left(
		\sum_{n \in N^*} h^nt_n (L_0)^n\right) \in Cl_h(S^1,V).$
		Then:
		\begin{itemize}
			\item There exists a unique pair $(S,Y)$ such that
			\begin{enumerate}
				\item $U_h = S^{-1}Y,$
				\item $Y \in \Psi DO_{h}^*(S^1,V)_D$
				\item $S \in \Psi DO_{h}^*(S^1,V)$ and $S - 1 \in Cl_{h,odd}(S^1,V)_S.$ 
			\end{enumerate}
			Moreover, the map 
			$$(S_0,t_1,...,t_n,...)\in Cl^{0,*}_{odd}(S^1,V)\times T \mapsto (U_h,Y)\in (\Psi DO_{h}^*(S^1,V))^2$$ is smooth.
			\item The operator $L \in \Psi DO_{h}(S^1,V)[[ht_1,...,h^nt_n...]]$ given by $L = S L_0 S^{-1} = Y L_0 Y^{-1}$, 
			is the unique solution to the hierarchy of equations 
			{\begin{equation}		\label{formalKP} 
					\left\{\begin{array}{ccl}
						\frac{d}{dt_n}L &=& \left[(L^n)_D(t), L(t)\right] =  -\left[(L^n)_S(t), L(t)\right]\\
						L(0) & = & L_0 \\
					\end{array}\right. \; ,
			\end{equation}}
			in which the operators in this infinite system are understood as formal operators and $A \mapsto(A)_D$ means projection 
			into the space of differential operators, {obtained by cutting the part of negative order along the lines of e.g. \cite{M1}, and which corresponds to the projection $A \mapsto A_D$ already defined on non formal, odd class pseudo-differential operators.} 
			
		\end{itemize}
	\end{Theorem}	
\begin{Definition}
	In what follows, $val_h$ denotes standard valuation of $h$ series.
	Let $$\Psi_h(R)=
	\left\{\sum_{\alpha \in \Z}a_{\alpha}\dd^\alpha \in \Psi(R)[[h]]\;|\; val_h(a_\alpha)\geq \alpha\right\}\; ,$$
	$$G\Psi_h(R)=
	\left\{\sum_{\alpha \in \Z}a_{\alpha}\dd^\alpha \in \Psi_h(R)\;|\; a_0 = 1 + b_0,\quad val_h(b_0)\geq 1\right\},$$
	$$G_{R,h} = \left\{A \in  G\Psi_h(R)\;|\; A = 1 + B, \quad B\in \Psi^{-1}(R)[[h]] \right\}\; ,$$
	$$\D_h(R) = 
	\left\{\sum_{\alpha \in \Z}a_{\alpha}\dd^\alpha \in \Psi_h(R)\;|\; a_\alpha = 0 \hbox{ if } \alpha < 0 
	\hbox{ and }a_0 = 1 + b_0, \quad val_h(b_0)\geq 1 \right\}\; .$$
\end{Definition}

\noindent The next Lemma is proved in \cite{Ma2013}; a shorter proof appears in \cite{MR2016}.

\begin{Lemma}
	$\Psi_h(R)$ is a Fr\"olicher algebra, an integral Lie algebra, and the groups $G\Psi_h(R),$ $G_{R,h}$ and 
	$\D_h(R)$ are regular Fr\"olicher Lie groups with Lie algebras given respectively by:
	$$\mathfrak{g}\Psi_h(R)=
	\left\{\sum_{\alpha \in \Z}a_{\alpha}\dd^\alpha \in \Psi_h(R) \;|\;  val_h(a_0)\geq 1\right\}\; ,$$
	$$\mathfrak{g}_{R,h}= \Psi^{-1}(R)[[h]]\; ,$$
	and
	$$\mathfrak{d}_h(R) = \left\{\sum_{\alpha \in \Z}a_{\alpha}\dd^\alpha \in D_h(R) \;|\; a_0 = 0 
	\hbox{ if } \alpha < 0 \quad val_h(a_0)\geq 1\right\}\; .$$
\end{Lemma}

\begin{Theorem} {\rm \cite{MR2016}}
	Mulase decomposition  holds for
	$W \in G\Psi_h(R), $ $S \in G_{R,h} $ and $Y \in \D_h(R)$ and, in particular,
	with $T = \cup_{n \in \N} \R^n,$ $R=C^\infty(T, C^\infty(S^1,\R)) = C^\infty(T \times S^1,\R)$ and $\partial = 
	\frac{d}{dx}$ on $S^1,$ recovering {\rm \cite{Ma2013}}. The map $U \mapsto (S,Y)$ is smooth.
\end{Theorem} 

\section{The KP hierarchy on an extended class of formal Pseudo-differential operators based on \cite{MRu2021-1}}

The algebra of operators that we intend to use in this section is the algebra of formal classical pseudo-differential operators $\F Cl(S^1,\K^n)$ that are obtained from classical pseudo-differential operators 
acting on smooth sections of the trivial vector bundle $S^1 \times \K^n$ over $S^1,$ for $K =  \C$ or $\mathbb{H},$ see e.g. \cite{Gil,PayBook}.  
	The key properties of $\epsilon(D)= D |D|^{-1} =  |D|^{-1} D$ that we use in our constructions are:
	\begin{itemize}
		\item the formal operator $\epsilon(D)\in \mathcal{F} Cl(S^1,\K^n)$ commutes with any formal operator $A \in  \mathcal{F} Cl(S^1,\K^n),$
		\item $\epsilon(D)^2=Id$
		\item the composition on the left $A \mapsto \epsilon(D) \circ A$ is an endomorphism of the algebra $ \mathcal{F} Cl(S^1,\K^n),$ which restricts to a bijective map from $\Psi DO(S^1,\K^n)=\F Cl_{odd}(S^1,\K^n)$ to an algebraic complement in $ \mathcal{F} Cl(S^1,\K^n)$ noted as $ \mathcal{F} Cl_{even}(S^1,\K^n)$ following the terminology of \cite{Scott}
		\item the restriction of the Wodzicki residue to $\Psi DO(S^1,\K^n)=\F Cl_{odd}(S^1,\K^n)$, which is similar to but not equal to the Adler functional, is vanishing.  
	\end{itemize}  
}
We have already stated the following: 
\begin{itemize}
	\item The space $\F Cl(S^1,\K^n)$ splits in various ways: one is derived from the splitting of $T^*S^1 - S^1$ into two connected components , the splitting with respect to $\Psi DO(S^1,\K^n)$ as a subalgebra , and the extension of the splitting related to the classical Manin triple on $\Psi DO(S^1,\K^n)$ to $\F Cl(S^1,\K^n)$. 
	\item The operator $\epsilon(D)$ is in the center of  $\F Cl(S^1,\K^n).$  It generates  then a polarized Lie bracket using it as a $\mathbf r-$matrix and an integrable almost complex structure on  $\F Cl(S^1,\K^n).$ 
\end{itemize}
These technical features enables us to state the announced main results of this section: existence and uniqueness of solutions of the KP hierarchy with various initial conditions (section \ref{ss:multKP}) and KP hierarchy with complex powers (section \ref{ss:complex}). 

\subsection{Multiple classical KP hierarchies on $\mathcal{F}Cl(S^1,\K)$} \label{ss:multKP}

The (classical) KP hierarchy on $\Psi DO(S^1,\K)$ can then push-forward on $\mathcal{F}Cl$-classes of operators by various ways. 
\paragraph{\bf Push-Forward via $\Phi_{\lambda,\mu}$ maps}
Let $\K =  \mbc$ or $\mathbb{H}.$ For each choice of $(\lambda,\mu) \in \C^2\backslash \{0;0\}$ identifies $\frac{d}{dx} \in \Psi DO(S^1,\K)$ with an operator in $\mathcal{F}Cl(S^1,\K)$ with the same algebraic properties. 
\vskip 6pt
\noindent
\textbf{Notation:} $\partial_{\lambda,\mu} = \Phi_{\lambda,\mu}\left(\frac{d}{dx}\right)$ and $\F Cl_{\lambda,\mu}(S^1,\K) = Im \Phi_{\lambda,\mu}.$
\vskip 6pt
Then we can develop the KP hierarchy on $\F Cl_{\lambda,\mu}(S^1,\K).$ We first remark that, since each map $\Phi_{\lambda,\mu}$ is a degree $0$ morphism of filtered algebras, each push-forward of the unique solotion $L$ of the KP hierachy (\ref{eq:KP}) generates a solution of the corresponding equation in $\F Cl(S^1,\K)$ which reads the same way: 
$$\frac{d L}{d t_{k}} = \left[ (L^{k})_{D} , L \right]\; , \quad
\quad k \geq 1 \; ,$$
where solutions operators now belong to $\F Cl^1(S^1,\K)[[T]]$ and where each initial value $\Phi_{\lambda,\mu}(L_0) \in \partial_{\lambda,\mu} + \F Cl_{\lambda,\mu}^{-1}(S^1,\K)$ with obvious extension of notations.
Therefore, for any initial value $L_0 \in \Psi DO(S^1,\K),$ we get a family of operators $$L_{\lambda,\mu} \in\F Cl_{\lambda,\mu}^1(S^1,\K)[[T]] \subset \F Cl^1(S^1,\K)[[T]]$$ parametrized by the complex parameters $\lambda$ and $\mu$ chosen as before, which satisfies the KP hierarchy in $\F Cl(S^1,\K)$ and with initial values $\Phi_{\lambda,\mu}(L_0).$ 
\paragraph{\bf Existence, uniqueness and well-posedness of the KP system in $\F Cl(S^1,\K).$}
We adapt here the $r-$matrix approach for the construction of the solutions, along the lines of \cite{ERMR2017} with the following specific choices:
\begin{itemize}
	\item The algebra of smooth coefficients for formal pseudo-differential operators is $  \mathcal{R} = C^\infty(S^1,M_n(\K)) \oplus \epsilon(D)C^\infty(S^1,M_n(\K))$ with multiplication rules inherited from $Cl(S^1,\K^n).$
	\item The differential operator is $\partial = \frac{d}{dx}.$
\end{itemize}
\begin{Proposition}\label{prop:R-Kn}
	$\Psi DO(\mathcal{R}) = \F Cl(S^1,\K^n)$ and there is an identification of the Manin triples $(\Psi DO(\mathcal{R}), DO(\mathcal{R}), IO(\mathcal{R}))$ with $(\F Cl(S^1,\K^n),\F Cl_D(S^1,\K^n),\F Cl_S(S^1,\K^n)).$
\end{Proposition}

Hence, applying the main result of \cite{RS} completed, for well-posedness, by \cite[Theorem 4.1]{ERMR2017} or by \cite[Theorem 4.1]{MR2016} when $\mathcal R = C^\infty(S^1,\K) = M_1(C^\infty(S^1,\K))$ is a commutative algebra, we can state the following: 

\begin{Proposition} \label{prop:KP-FCl-I}
	The Kadomtsev-Petviashvili (KP) hierarchy (\ref{eq:KP}) on $\Psi DO (\mathcal R)$ (resp. $\F Cl(S^1,\K^n)$) 
	with initial condition $L(0)=L_0  \in \partial + \Psi DO^{-1} (\mathcal R)$ (resp. $\in \partial + \F Cl^{-1}(S^1,\K^n)$) satisfies Theorem \ref{KPcentral}.
\end{Proposition}
\begin{rem}
	We have used here, intrinsically, the integrable almost complex structure $J_1.$ Indeed, $\mathcal R = C^\infty(S^1,M_n(\K)) + J_1 C^\infty(S^1,M_n(\K))$ is an algebra.
\end{rem}
\begin{rem}
	There exists another way to justify Proposition \ref{prop:KP-FCl-I}. One can use alternatively the splitting $$\F Cl(S^1,\K^n) = \F Cl_+(S^1,\K^n) \oplus \F Cl_-(S^1,\K^n).$$ Then Equation (\ref{eq:KP}) on $\F Cl(S^1,\K^n)$ splits into two independent equations, similar to Equation (\ref{eq:KP}) on $\F Cl_\pm(S^1,\K^n).$ Through the identification maps $\Phi_{1,0}$ and $\Phi_{0,1}$ of $\F Cl_\pm(S^1,\K^n)$ with $\Psi DO(S^1,\K^n),$ we get existence, uniqueness and well-posedness for  Equation (\ref{eq:KP}) on $\F Cl(S^1,\K^n)$ with initaial value $L_0 \in \partial + \F Cl^{-1}(S^1,\K^n).$ 
\end{rem}

From this last remark, we can generalize the  identification procedure, changing the maps $\Phi_{ee},$ $\Phi_{1,0}$ and $\Phi_{0,1}$ by the family of maps $\Phi_{\lambda,\mu}.$


\begin{Theorem} \label{th:KPlambdamu}
	Let $(\lambda,\mu) \in (\mbc^*)^2 .$ Then the KP equation (\ref{eq:KP}) in $\F Cl(S^1,\K^n)$ with initial value $L_0 \in \partial_{\lambda,\mu} + \F Cl^{-1}(S^1,\K^n)$ has an unique solution $L$ in $\partial_{\lambda,\mu} + \F Cl^{-1}(S^1,\K^n)[[T]]$ and the problem is well-posed: the solution $L$ depends smoothly on $L_0.$ 
\end{Theorem}
\paragraph{\bf Twisted KP hierarchy}
Let us now change the standard multiplication on $\F Cl(S^1,\K)$ by 
$ (A,B) \mapsto \epsilon A B$
where $ \epsilon = \epsilon(D)$ or $a\epsilon(D)$ for any $a \in \C^*.$ Since $\epsilon(D)$ commutes with any element of $\F Cl(S^1,\K)$ for the  standard multiplication, this new multiplication defines a new algebra structure on $\F Cl(S^1,\K).$ When necessary we note by $\circ$ the standard multiplication, and by $\circ_\epsilon$ the twisted one.  Associated to this multiplication, we get the deformed Lie bracket $[.,.]_\epsilon.$
Then we get again and equation similar to (\ref{eq:KP})
\begin{equation} \label{eq:KPepsilon}
	\frac{d L}{d t_{k}} = \left[ \epsilon^{k-1}(L^{k})_{D} , L \right]_\epsilon =\epsilon^{k}\left[ (L^{k})_{D} , L \right] \; , \quad
	\quad k \geq 1 \; ,
\end{equation}
where powers in this equation are taken with respect to $\circ.$

\begin{Theorem} \label{th:KPepsilon}
	The Let $L_0$ such that $L_0 \in \partial_{\lambda,\mu} + \F Cl^{-1}(S^1,\K^n),$ with $(\lambda,\mu)\in (\C^*)^2.$ Then the $\epsilon-$KP hierarchy (\ref{eq:KPepsilon}) with initial value $L_0$ has an unique solution. Moreover, the problem is well-posed.
\end{Theorem}


\subsection{KP hierarchies with complex powers} \label{ss:complex}
We finally extend all the constructions of the last section to complex powers, along the lines of \cite{EKRRR1995}. Let $\K = \C $ or $\mathbb{H}.$
We consider an operator $L_0$ of complex order $\alpha$ such that 
\begin{equation} \label{eq:complex1} L_0 \in  \left(\frac{d}{dx}\right)^\alpha + \F Cl^{\alpha-1}(S^1,\K) \end{equation}
or
\begin{equation} \label{eq:complex2} L_0 \in  \left|D\right|^\alpha + \F Cl^{\alpha-1}(S^1,\K) \end{equation}
For each setting (\ref{eq:complex1}) and (\ref{eq:complex2}), we define the complex KP hierarchy on $\F Cl^\alpha(S^1,\K)$ by \begin{equation} \label{eq:KPalpha}
	\frac{d L}{d t_{k}} = \left[ (L^{k/\alpha})_{D} , L \right]_\epsilon =-\left[ (L^{k})_{S} , L \right] \; , \quad
	\quad k \geq 1 \; ,
\end{equation}
where $L^{k/\alpha} = \exp\left(\frac{k}{\alpha}\log L\right)$
and the solution $L \in \F Cl^\alpha(S^1,\K)[[T]].$
\begin{Theorem} \label{th:KPalpha}The KP hierarchy (\ref{eq:KPalpha}) with initial value $L_0$ defined along the lines of (\ref{eq:complex1}) or (\ref{eq:complex2}) has an unique solution in $\F Cl^\alpha(S^1,\K)[[T]].$ Moreover, the prblem is well-posed.
\end{Theorem}

\vskip 6pt
\noindent
\underline{\bf Open question:} The extension of the KP hierarchy from $\F Cl(S^1, \mbc)$ to $Cl(S^1, \mbc)$ is a non trivial open problem, due to the lack of local slice $\F Cl(S^1, \mbc)\rightarrow Cl(S^1, \mbc)$ in the short exact sequence
$$ 0 \rightarrow Cl^{-\infty}(S^1, \mbc) \rightarrow Cl(S^1, \mbc) \rightarrow \F Cl(S^1, \mbc) \rightarrow 0.$$ 

\backmatter


\begin{thebibliography}{200}
	
	
	\bibitem[-]{pres}{\bf Presented works}
	
		\bibitem{Ma2004} Magnot,J-P.; Structure groups and holonomy in infinite dimensions, \textit{Bull. Sci. Math.}
	\textbf{128} no6, 513-529 (2004)
	
	\bibitem{Ma2006-2}  Magnot, J-P.; Renormalized traces and cocycles on the algebra 
	of $S^1$-pseudo-differential operators; \textit{Lett. Math. Phys.} \textbf{75} no2, 111-127 (2006)
	
	\bibitem{Ma2008} Magnot, J-P.; The Schwinger cocycle for algebras with unbounded operators; \textit{Bull. Sci. Math.} \textbf{132} no. 2, 112-127 (2008).
	
	\bibitem{Ma2013} Magnot, J-P.; Ambrose-Singer theorem on diffeological bundles and complete integrability of the KP equation; \textit{Int. J. Geom. Meth. Mod. Phys.} \textbf{ 10}, No. 9, Article ID 1350043, 31 p. (2013). (2013)
		
\bibitem{MaICM} Magnot, J-P.; Infinite dimensional integrals beyond Monte Carlo methods: yet another approach to normalized infinite dimensional integrals; IC-MSQUARE 2012: International Conference on Mathematical Modelling in Physical Sciences, 3-7 September 2012, Budapest, Hungary; \textit{J. Phys.: Conf. Ser.} {\bf 410}, 012003 (2013)

	\bibitem{Ma2016-1} Magnot, J-P.; On $Diff(M)-$pseudo-differential operators and the geomery of non-linear grassmannians; \textit{Mathematics} \textbf{4} no 1; articleID 1 (2016)
		\bibitem{Ma2016-2}  Magnot, J-P.; Differentiation on spaces of triangulations and optimized triangulations. \textit{ J. Phys.: Conf. Ser.}, \textbf{738}, No 1 articleID 012088 (2016)

	\bibitem{Ma2016-3} Magnot, J-P.; The mean value for infinite volume measures, infinite products and heuristic infinite dimensional Lebesgue measures; \textit{J. Math.} Vol. 2017 (2017), Article ID 9853672, 14 pages

	\bibitem{Ma2017-2} W.W. Koczkodaj, J.-P. Magnot, J. Mazurek, J.F. Peters, H. Rakhshani, M. Soltys, D. Strzałka, J. Szybowski, A. Tozzi; On normalization of inconsistency indicators in pairwise comparisons \textit{Int. J. Approx. Reasonning} \textbf{86}  73 - 79 (2017)
	
		

	\bibitem{ERMR2017} Eslami-Rad, A.; Magnot, J-P.; Reyes, E. G.; The Cauchy problem of the Kadomtsev-Petviashvili hierarchy with arbitrary coefficient algebra \textit{J. Nonlinear Math. Phys.} \textbf{24} Supp. issue no 1, 103-120 (2017)
	

	
	\bibitem{Ma2018-1} Magnot, J-P.; A Mathematical Bridge between Discretized Gauge Theories in Quantum Physics and Approximate Reasoning in Pairwise Comparisons \textit{Adv. Math. Phys.} \textbf{2018} articleID 7496762, 5 pages (2018)
	
	\bibitem{Ma2013-2} Magnot, J-P.; The group of diffeomorphisms of a non compact manifold is not regular.\textit{ Demonstr. Math.} \textbf{51} 8-16 (2018)
	
	\bibitem{Ma2018-3} Magnot, J-P.; Remarks on a New Possible Discretization Scheme for Gauge Theories \textit{Int. J. Theoret. Phys.} \textbf{57} 2093–2102 (2018)
		
	
		
		\bibitem{Ma2016-5} Magnot, J-P.;On mathematical structures on pairwise comparisons matrices with coefficients in an abstract group arising from quantum gravity \textit{Heliyon} \textbf{5} e01821 (2019)
		
		\bibitem{MR2019} Magnot, J-P. Reyes, E.G.; The Cauchy problem of the Kadomtsev-Petviashvili hierarchy and
		infinite-dimensional groups.  in \textit{ Nonlinear Systems and Their Remarkable Mathematical Structures, Volume 2;} Norbert Euler and  Maria Clara Nucci Editors, CRC press (2019) section B6
		

		
	\bibitem{MR2016} Magnot, J-P.; Reyes, E.G.; Well-posedness of the Kadomtsev-Petviashvili hierarchy, Mulase factorization, and Fr\"olicher Lie groups \textit{Annales Henri Poincar\'e}
	{\bf 21}, 1893-1945 (2020)
	
	\bibitem{Ma2020-3} Magnot, J-P.; On the domain of implicit functions without extra norm estimates \textit{Demonstr. Math.} {\bf 53} no 1 112-120 (2020)
	
	
	\bibitem{Ma2016-4} Magnot, J-P.; {On the differential geometry of numerical schemes and weak solutions of functional equations.} \textit{Nonlinearity} 33, no. 12, 6835-6867 (2020)
	. 
	\bibitem{Ma2021-1} Magnot, J-P.; On the geometry of $Diff(S^1)-$pseudodifferential operators based on renormalized traces \textit{Proceedings of the International Geometry Center} {\bf 14} No. 1, 19-48 (2021)
	
	\bibitem{MRu2021-1} Magnot, J-P.; Rubtsov, V.; On the Kadomtsev-Petviashvili hierarchy in an extended class of formal pseudo-differential operators  \textit{Theoret. Math. Phys.} {\bf 207}, No. 3, 799-826 (2021)

\bibitem[-]{nonpres} {\bf Non presented works: works from the PhD thesis}

\bibitem{CDMP} Cardona, A.; Ducourtioux, C.; Magnot, J-P.; Paycha, S.;
Weighted traces on pseudo-differential operators and geometry on loop groups;
\textit{Infin. Dimens. Anal. Quantum Probab. Relat. Top.} \textbf{5} no4 503-541 (2002)

\bibitem{Ma2003}  Magnot, J-P.; The K\"ahler form on the loop group and the Radul cocycle on Pseudo-differential Operators; \textit{GROUP'24: Physical and Mathematical aspects of symmetries}, Proceedings of the 24th International Colloquium on Group Theorical Methods in Physics, Paris, France, 15-20 July 2002; Institut of Physic conferences Publishing \textbf{173}, 671-675, IOP Bristol and Philadelphia (2003)

\bibitem{Ma2006} Magnot, J-P.; Chern forms on mapping spaces,
\textit{Acta Appl. Math.} \textbf{91}, no. 1, 67-95 (2006).

\bibitem[-]{npres2} {\bf Other non presented works}

	\bibitem{Ma2006-3} Magnot, J-P.; Diff\'eologie sur le fibr\'e d'holonomie d'une connexion en dimension infinie
\textit{C. R. Math. Acad. Sci., Soc. R. Can.} \textbf{28}, no. 4, 121-128 (2006)


	\bibitem{Ma2015} 
Magnot, J-P., q-deformed Lax equations and their differential geometric background, Lambert University Publishing, Saarbruken, Germany, 2015.

\bibitem{Ma2015-1} Magnot, J-P; From infinitesimal symmetries to deformed symmetries of Lax-type equations; \textit{J. Of Physics: Conference series} \textbf{633} (2015) 012013


	\bibitem{Ma2017-1}  Kakiashvili, T.; Koczkodaj, W.W; Magnot, J-P.; Approximate reasoning by pairwise comparisons: “Topodynamics of metastable brains” by Arturo Tozzi, et al.
\textit{Phys. of Life reviews} {\bf 21} July 2017, Pages 37-39 (2017)


\bibitem{Ma2017-0} Magnot, J-P. ; From configurations to branched configurations and beyond \textit{Res. Rep. Math.}  \textbf{1} no1 art. ID 1000105 (2017)
	\bibitem{MW2016} Magnot, J-P.; Watts, J.; The diffeology of Milnor's classifying space \textit{Top. Appl.} \textbf{232} 189-213 (2017) 

	\bibitem{Ma2018-0} Magnot, J-P.; On multidisciplinary applications of gauge theories \textit{Biostat. Biomet. Open access  J.} \textbf{7} no1, art. ID.555701 (2018)
	
		\bibitem{Ma2015-2} Magnot, J-P.; Remarks on the geometry and the topology of the loop spaces $H^{s}(S^1, N),$ for $s\leq 1/2.$ ; \textit{International Journal of Maps in Mathematics} \textbf{2} no 1 14-37 (2019) 
		\bibitem{SSCM}	Singh, N.; Son, L.H.; Chiclana, F.; Magnot, J-P.; A new fusion of salp swarm with sine cosine for optimization of non-linear functions. \textit{Engineering with Computers} \textbf{36} 185–212 (2020)
	
	\bibitem{Ma2015-3} Magnot, J-P.; On algebras and groups of formal series over a groupoïd and application to some spaces of cobordism;  	\texttt{ArXiv:1507.00932} 
	
\bibitem{MR2020} Magnot, J-P. ; Reyes, E. ; On the Cauchy problem for a Kadomtsev-Petviashvili hierarchy on non-formal operators and its relation with a group of diffeomorphisms \texttt{arXiv:1808.03719}

\bibitem{Ma2021-2} Magnot, J-P.; On a class of closed cocycles for algebras of non-formal, possibly unbounded, pseudodifferential operators \texttt{ArXiv:2012.06941}

\bibitem[-]{autres} {\bf Other references}

\bibitem[ACMM1989]{ACMM1989} Abbati, M. C.; Cirelli, R.; Mania, A.; Michor, P.; The Lie group of automorphisms of a principal bundle. \textit{J. Geom. Phys.} \textbf{6}, 215-235 (1989)
\bibitem[AM2007]{AM2007}  Abbati, M. C.; Mania, A. A geometrical setting for geometric phases on complex Grassmann manifolds. J. Geom. Phys. 57 (2007), no. 3, 777–797
\bibitem[ARS1986-1]{ARS} M. Adams, T.Ratiu, R. Schmid {\it A Lie group structure for pseudodifferential operators}
Math. Ann. \textbf{273} no4 (1986), 529-551

\bibitem[ARS1986-2]{ARS2} Adams, M.; Ratiu, T.; Schmidt, R.; A Lie group structure for Fourier integral operators;
\textit{Math. Annalen} \textbf{276}, no 1, 19-41 (1986)

\bibitem[Ad1979]{Adl} Adler, M.; On a trace functionnal for formal pseudo-differential
operators and the symplectic structure of Korteweg-de Vries type equations
\textit{Inventiones Math.} \textbf{50} 219-248 (1979)
	\bibitem[AFKHKL1986]{AFKHKL1986} Albeverio, S.; Hoegh-Krohn, R.; Fenstad, J. E.; Lindstrom, T.;
	{\it Nonstandard methods in stochastic analysis and mathematical physics.} Reprint of the 1986 original ed. Dover Publications 514 p. (2009). 
	
	
	
	
	
	
	\bibitem[AB1994]{AB} Albeverio, S; Brze\'zniak, Z.; Feynman path integrals as infinite dimensional oscillatory integrals: some new developpements; \textit{Acta Appl. Math.} \textbf{35}; 5-27 (1994)
	
	\bibitem[AHKM2005]{AHM} Albeverio, S.; Hoegh-Krohn, R.; Mazzuchi, S.; \textit{Mathematical theory of Feynman Path Integrals; an introduction} 2nd edition;  Lecture Notes in Mathematics \textbf{523} , Springer (2005)
	
	\bibitem[AKLU2000]{A}  Albeverio, S.; Kondratiev, Y., Lytvynov, E.; g. F. Us, G.F.; Analysis and geometry on marked configuration spaces;
	\textit{Infinite Dimensional Harmonic Analysis (Kyoto, September 20-24, 1999)}, H. Heyer et al., eds, Gr\"abner, Altendorf, 1-39 (2000) 
	
	
	\bibitem[AM2005]{AMaz1} Albeverio, S.; Mazzucchi, S.; Generalized Fresnel integrals \textit{Bull. Sci. Math.} \textbf{129}, no1, 1-23 (2005)
	
	
	
	\bibitem[AM2016]{AM2016}
	Albeverio, S.; Mazzucchi, S.;
	A unified approach to infinite-dimensional integration. 
	{\it Rev. Math. Phys.} {\bf 28}, No. 2, Article ID 1650005, 43 p. (2016)
	
	\bibitem[AM2018]{AM2018}
	Albeverio, S.; Mazzucchi, S.;
	A unified approach to infinite dimensional integrals of probabilistic and oscillatory type with applications to Feynman path integrals  in:
	Gesztesy, Fritz (ed.) et al., {\it Non-linear partial differential equations, mathematical physics, and stochastic analysis. The Helge Holden anniversary volume on the occasion of his 60th birthday.} Based on the presentations at the conference ‘Non-linear PDEs, mathematical physics and stochastic analysis’, NTNU, Trondheim, Norway, July 4–7, 2016. (2018)
	 
	\bibitem[AZ1990]{AZ1990} Albeverio, S.; Zegarlinski, B.; Construction of convergent simplicial approximations of quantum fields on Riemannian manifolds \textit{Comm. Mat. Phys.} \textbf{132} 39-71 (1990)
	
	
	
		\bibitem[ADL2001]{Alb} Albeverio, S.; Daletskii, A.; Lytvynov, E.;
		De Rham cohomology of configuration spaces with Poisson measure
	\textit{J. Funct. Anal.} 185, No.1, 240-273 (2001). 
	
	\bibitem[AS1953]{AS} Ambrose, W.; Singer, I.M.; { A theorem on holonomy} {\it Trans. Amer. Math. Soc.}; \textbf{75} (1953)                                                                                                                                                                                                                                                                                                                                                                                                                                                                                                                                                                                                                                                                                                                                                                                                                                                                             428-443 

\bibitem[Arn1966]{Arn1966} Arnold, V.I.; Sur la g\'eom\'etrie diff\'erentielle des groupes de Lie de dimension infinie et ses applications \`a l’hydrodynamique des fluides parfaits, {\it Ann. Inst. Fourier} {\bf 16}, 316-361 (1966). 

	\bibitem[Asa2000]{Asa2000} Asada, A; regularization of differential operators on a Hilbert space and geometric meaning of Zeta-regularization. \textit{Steps in differential geometry, proceding of the colloquium on differential geometry 25-30 July 2000, debrecen, Hungary}
	
%
	
	\bibitem[BH2011]{BH}
	 Baez, J. C.; Hoffnung, A.E.; Convenient categories of smooth spaces. \emph{Trans. Amer. Math. Soc.}, \textbf{363}, 5789--5825 (2011).
	\bibitem[Bak1991]{Ba1} Baker, R. L.; Lebesgue measure on $\mathbb{R}^\infty$. \textit{Proc. Amer. Math. Soc.} \textbf{113}  no4, 1023-1029 (1991)
	
	\bibitem[Bak2004]{Ba2} Baker, R. L.; Lebesgue measure on $\mathbb{R}^\infty$. II. \textit{Proc. Amer. Math. Soc.} \textbf{132}  no9, 2577-2591 (2004)
	

	
	
	
	
	
	
	\bibitem[BBHM2012]{BBHM2012} Bauer,M.; Bruveris, M.; Harms, P.; Michor, P.; Vanishing geodesic distance for the
	Riemannian metric with geodesic equation the KdV-equation. \textit{Ann. Global Anal. Geom.}  \textbf{41} no 4 (2012) 461--472
	
	
	\bibitem[BR1984]{BR1984} Beals, M.; Reed, M.; Microlocal regularity theorems for nonsmooth pseudodifferential operators and applications to nonlinear problems; \textit{Trans. Amer. Math. Soc.} {\bf 285} (1984), 159-184
\bibitem[BFR2018]{R2018} Beltran, J.; Farinati, M.; Reyes, E. G.;
Central extensions of the algebra of formal pseudo-differential symbols via Hochschild (co)homology and quadratic symplectic Lie algebras. 
{\it J. Pure Appl. Algebra} {\bf 222} no. 8, 2006-2021 (2018). 
	\bibitem[BGV1992]{bgv} Berline, N.; Getzler, E.; Vergne, M.; \textit{Heat kernels and Dirac operators}; Grundlehren der Mathematischen Wissenschaften [Fundamental Principles of Mathematical Sciences], 298.
	Springer-Verlag, Berlin, (1992)
	\bibitem[BDIP1996]{BDIP1996} Bertin, J.; Demailly, J-P.; Illusie, L.; Peters, C.; \textit{Introduction \`a la th\'eorie de Hodge} SMF Panoramas et Synth\`eses \textbf{3} (1996)
	\bibitem[BF1981]{BF} E. Binz, H.R. Fischer: The manifold of embeddings of a closed manifold. Proceedings of Differential geometric methods in theoretical physics, Clausthal 1978, Lecture Notes in Physics 139, Springer-Verlag 1981, pp. 310--329
%
		\bibitem[Bog2010]{Bog} Bogachev, V.I.; {\it Differentiable measures and the Malliavin calculus} AMS Math. Surveys and monographs \textbf{164} (2010) 
	\bibitem[BS2017]{BS2017} Bogachev, V. I.; Smolyanov, O. G.;
	\textit{Topological vector spaces and their applications.} Expanded and revised translation, originally published 2012. 
	Springer Monographs in Mathematics, 456 p. (2017). 
%
	
	\bibitem[BK1969]{BK}  Bokobza-Haggiag, J.; { Op\'erateurs pseudo-diff\'erentiels sur une vari\'et\'e diff\'erentiable}; {\it Ann. Inst. Fourier, Grenoble} \textbf{19,1}  125-177 (1969)
	
	\bibitem[Bou]{Bou}  Bourbaki, N.; {\it El\'ements de math\'ematiques}; Masson, Paris (1981)
	
	\bibitem[BR1987-1]{BR1987} Bowick, M.; Rajeev, S.; String theory as the K\"ahler geometry of the loop space \textit{Phys. Rev. Lett.} \textbf{58}, 535 (1987)
	
	\bibitem[BR1987-2]{BR1987-2} Bowick, M.; Rajeev, S.; Anomalies as curvature in complex geometry \textit{Nucl. Phys. B} \textbf{296} 1007 (1987)
\bibitem[BS1994]{BS1994} Brenner, S. C.; Scott, L. R.
\textit{The mathematical theory of finite element methods. }
Texts in Applied Mathematics. {\bf 15.}  Springer-Verlag.  294 p. (1994). 
	\bibitem[Bry2008]{Br}
	Brylinski, J-L.;
	\textit{Loop spaces, characteristic classes and geometric quantization.} Reprint of the 1993 edition. 
	Modern Birkhäuser Classics. Basel: Birkhäuser  (2008).
	
		\bibitem[CS1983]{CS1983} Cameron, R.H.; Storvick, D.A.; \textit{A simple definition of the Feynman integral, with applications} Mem. of the AMS \textbf{46} no 288 (1983)
		\bibitem[Can2015]{Can2015} Canarutto, D.; Fr\"olicher-smooth geometries, quantum jet bundles and BRST symmetry. 
		{\it J. Geom. Phys.} {\bf 88} 113-128 (2015).
		\bibitem[Can2020]{Can2020}Canarutto, D.;
		\textit{Gauge field theory in natural geometric language. A revisitation of mathematical notions of quantum physics.} Oxford University Press  368 p. (2020)
	
	
	\bibitem[CdAS2010]{Cd'AS2010} Cavallo, B.; D’Apuzzo, L.; Squilllante,M.; Pairwise Comparison Matrices over abelian Linearly Ordered Groups: A Consistency Measure and Weights for the Alternatives; in : "Multicriteria and Multiagent Decision Making with Applications to Economics and Social Sciences"
	\textit{Studies in Fuzziness and Soft Computing} {\bf 305} (2010) 49-64 
	
%
%
%
	\bibitem[CW2014-2]{CW2014-2}Christensen, J. D.; Wu, E.;
	The homotopy theory of diffeological spaces.
	\textit{New York J. Math.} \textbf{20}, 1269-1303 (2014)
%
\bibitem[CvS2021]{CvS2021} Connes, A.; van Suilekom, W.D.; Tolerance relations and operator systems \texttt{arXiv:2111.02903}
	\bibitem[CrW1985]{CW1985} Crawford, G.; Williams, C.; The analysis of subjective judgment matrices; \textit{A project AIR FORCE report prepared for united states air force} report number \textbf{R-2572-1-AF} (may 1985) 
	
	  
	\bibitem[Dem1995-1]{D1} Demidov, E.E.; On the Kadomtsev-Petviashvili hierarchy with a noncommutative
	timespace. {\em Funct. Anal. Appl.} 29, no. 2, (1995), 131--133.
	\bibitem[Dem1995-2]{D2} Demidov, E.E.; Noncommutative deformation of the Kadomtsev-Petviashvili hierarchy.
	In ``Algebra. 5, Vseross. Inst. Nauchn. i Tekhn. Inform. (VINITI)'', Moscow, 1995.
	(Russian). {\em J. Math. Sci. (New York)} 88, no. 4, (1998), 520--536 (English).
	\bibitem[Dic2003]{Di} L.A. Dickey, \textit{Soliton equations and Hamiltonian systems, second edition} (2003).
	Advanced Series in Mathematical Physics $12$, World Scientific Publ. Co., Singapore.
	\bibitem[Di1968]{Dieu2} Dieudonn\'e, J.; \'El\'ements d'analyse I,II,III; Gauthier-Villars (1968)
	\bibitem[Dui1974]{Dui} Duistermaat, J.J.; Oscillatory integrals, Lagrange immersions and unfoldings of singularities; \textit{Comm. Pure Appl. Math.} \textbf{27} 207-281 (1974)
	
	\bibitem[DGV2015]{DGV} Dodson, C.; Galanis, G.; Vassiliou, E.; \textit{Geometry in the Fr\'echet context: a projective limit approach} London Mathematical Society Lecture Notes Series {\bf 428}, Cambridge University Press (2015)
	
	
	
	\bibitem[Duc2001]{D} Ducourtioux, C.; {\it Weighted traces on pseudo-differential operators and associated determinants}; Ph.D thesis, Clermont-Ferrand, France (2001) 
	
	
	
	\bibitem[Dup1978]{Dup1978} Dupont, J. L.;
	{\it Curvature and characteristic classes.} 
	Lecture Notes in Mathematics. \textbf{640} (1978). 

	\bibitem[Ee1966]{Ee}Eells, J.;  A setting for global analysis
	\textit{Bull. Amer. Math. Soc.} {\bf 72} 751-807 (1966)
	
	\bibitem[ET1984]{ET} Elworthy, D.; Truman, , A.; Feynman maps, Cameron-Martin formulae and anharmonic oscillators; \textit{Ann. Inst. H. Poincar\'e Phys. Theo.} \textbf{41} (2) 115-142 (1984)
	
		\bibitem[EKRRR1995]{EKRRR1995} Enriquez, B.; Khoroshkin, S.; Radul, A.; Rosly, A.; Rubtsov, V.; Poisson-Lie aspects of classical W-algebras; in: \textit{The interplay between Differential geometry and differential equations} \textit{Amer.  Math. Soc. Transl. (2)} \textbf{167}, 37-59 (1995)
	\bibitem[ER2013]{ER2013} Eslami Rad, A.; Reyes, E. G.; The Kadomtsev-Petviashvili hierarchy and the Mulase factorization of
	formal Lie groups {\it J. Geom. Mech.} {\bf 5}, no 3 (2013) 345--363.
	\bibitem[vEK1964]{vEK1964} van Est, W. T.; Korthagen, Th. J.;
	Non-enlargible Lie algebras. 
	\textit{Nederl. Akad. Wet., Proc., Ser. A}
	{\bf 67}, 15-31 (1964). 
	\bibitem[FH2001]{Str} Fadell, E.R.; Husseini, S.Y.; \textit{Geometry and topology of configuration spaces} 
	Springer, Berlin (2001)
	
	\bibitem[Fed1969]{Fed}  Federer, H.; \textit{Geometric measure theory}, Springer (1969)
	
	\bibitem[Fr1988-1]{F1} Freed, D.; 
	An index theorem for families of Fredholm operators parametrized by a group\textit{Topology} \textbf{27}, No.3, 279-300 (1988)
	\bibitem[Fr1988-2]{F2} Freed, D.;
	{The geometry of loop groups}, \textit{J. Diff. Geom.} {\bf 28}
	(1988) 223-276
	
	
	\bibitem[FK1988]{FK} Fr\"olicher, A; Kriegl, A; \textit{Linear spaces and differentiation
		theory} Wiley series in Pure and Applied Mathematics, Wiley Interscience
	(1988)

	\bibitem[Fu2017]{F2017} Fujiwara, D.;
	{\it Rigorous time slicing approach to Feynman path integrals. }
	Mathematical Physics Studies.  Springer, 333 p. (2017). 
		
	\bibitem[GV1997]{GV} Galanis, G. and Vassiliou, E.; A generalized frame bundle for certain
	Fr\'echet vector bundles and linear connections. \textit{Tokyo J. Math.} {\bf 20}  no.1 (1997), 129--137
		
\bibitem[GBV2014]{GBV2014} Gay-Balmaz, F; Vizman, C.; Principal bundles of embeddings and non linear grassmannians; {\it Annals of Global Analysis and Geometry},{\bf 46},  293–312 (2015)
 
	\bibitem[Gil1984]{Gil} Gilkey, P; \textit{Invariance theory, the heat equation
		and the Atiyah-Singer index theorem} Publish or Perish (1984)
	
	
	\bibitem[Gl2002]{Glo2002} Gl\"ockner, H; Algebras whose groups of the units are
	Lie groups \textit{Studia Math. } \textbf{153}, no2 (2002), 147-177
	
	\bibitem[Gl2007]{Glo2007} Gl\"ockner, H; Direct limits of infinite-dimensional Lie groups
	compared to direct limits in related categories.
	 {\it J. Funct. Anal.} 
	 \textbf{245} 19-61 (2007)
	
	\bibitem[GN2012]{GN2012} Gl\"ockner,, H.; Neeb, K-H.; When unit groups of continuous inverse algebras are regular Lie groups \textit{Studia Math.} \textbf{211} no2, 95-109 (2012)
	
	\bibitem[GN2017]{GN2017} Gl\"ockner,, H.; Neeb, K-H.;
	Diffeomorphism groups of compact convex sets. 
	{\it Indag. Math. (N.S.)} {\bf 28} no. 4, 760–783 (2017).
	
	
	 \bibitem[GK1971]{GK1971}  Gelfand, I.; Kazhdan, D.; Certain questions of differential geometry and the computation of the cohomologies of the Lie algebras of vector fields. \textit{Soviet Math. Doklady} {\bf 12}, 1367–1370 (1971)
		\bibitem[Gr1997]{Gro} 
		Gromov, M., Metric structures in Riemannian and non-Riemannian spaces, 2nd ed. Birkauser, 1997.
	
	 \bibitem[GR2007]{GR} Guieu, L. and Roger, C.; \textit{L' Alg\`ebre et le
		groupe de Virasoro: Aspects g\'eom\'etriques et
		alg\'ebraiques, generalisations}. Centre de Recherches Mathematiques, Universit\'e de
	Montreal, (2007).
	\bibitem[Hah2004]{Hah2004} Hahn, A.;
	The Wilson loop observables of Chern-Simons theory on
	$\mathbb{R}^3$
	in axial gauge
	,
	\textit{Comm. Math. Phys.}
	\textbf{248}
	, no. 3, 467?499 (2004)
	\bibitem[HV2004]{HV} Haller, S.; Vizman, C.; Non-linear grassmannian as coadjoint orbits \textit{Math. Ann.} \textbf{139}, no4, 771-785 (2004)
	
	
	\bibitem[Ham1984]{Ham}  Hamilton, R.S.; { The inverse function theorem of Nash and Moser}; 
	{\it Bull. Amer. Math Soc. (NS)} \textbf{7} (1984) 65-222 
	\bibitem[HL2020]{HL2020} Herzog, R.; Loayza-Romero, E.; A manifold of planar triangular meshes with complete Riemannian metric \texttt{arXiv:2012.05624}

	\bibitem[Hir1976]{Hir} Hirsch, M.; \textit{Differential topology}; GTM \textbf{33}, Springer (1976)
	
	
	\bibitem[HN1971]{HN1971} Hoghe-Nlend, H.; \textit{Th\'eorie des bornologies et applications} Lect. Notes in Math. \textbf{273} (1971)
	
	\bibitem[Ho1971]{Horm} H\"ormander,L.; Fourier integral operators. I; \textit{Acta Mathematica} \textbf{127} 79-189 (1971) 
	\bibitem[HCR2015]{HCR2015} Huber, M.; Campagnari, D.; Reinhardt, H.; Vertex functions of Coulomb gauge Yang--Mills theory \textit{Phys. Rev. D} \textbf{91}, 025014 (2015) 
	\bibitem[Lim2012]{Lim2012} Non-abelian gauge theory for Chern-Simons path integral on
	$\mathbb{R}^3$
	, \textit{Journal of Knot
		Theory and its Ramifications}
	\textbf{21}
	no. 4., articleID1250039 (24p) (2012)
	
	
	\bibitem[HM1983]{Marsden} Hughes, T.; Marsden, J.E.; \textit{Mathematical foundations of elasticity}
	Prentice-Hall Civil Engineering and Engineering Mechanics Series. Englewood Cliffs, New Jersey: Prentice-Hall, Inc. XVIII, (1983).

	\bibitem[IZ1985]{IgPhD} Iglesias-Zemmour, P.; \textit{Fibrations diff\'eologiques et homotopie},  PhD thesis, universit\'e de Provence (1985)
	
	\bibitem[IZ1987]{Ig} Iglesias-Zemmour, P.; Connexions et diff\'eologie 
	\textit{Aspects dynamiques et topologiques des groupes infinis de transformation de la m\'ecanique}
	Travaux en cours \textbf{25}, Hermann (1987), 61-78
	
	\bibitem[IZ2013]{Igdiff} Iglesias-Zemmour, P.; \textit{Diffeology} Mathematical Surveys and Monographs
	\textbf{185} (2013)
	\bibitem[McI2011]{McI2011} Mc Intosh, I.; The quaternionic KP hierarchy and conformally immersed 2-tori on the 4-sphere. \textit{Tohoku Math. J.} \textbf{63} 183-215 (2011)  
	
	\bibitem[Ism1996]{Ism} Ismaginov, O.V; \textit{Representations of infinite dimensional groups}
	Translations of mathematical monographs; American mathematical society (1996)
	
	
	\bibitem[Ka1989]{Ka} Kassel, Ch.;
	Le r\'esidu non commutatif (d'apr\`es M. Wodzicki)  S\'eminaire
	Bourbaki, Vol. 1988/89. \textit{Ast\'erisque} {\bf 177-178},
	Exp. No. 708, 199-229 (1989)
	
	\bibitem[KW2009]{KW} Khesin, B.A. and Wendt, R.; ``The geometry of infinite-dimensional groups" (2009). Springer-Verlag, Berlin.
	\bibitem[KZ1995]{KZ} Khesin, B.A. and Zakharevich, I.; Poisson-Lie groups of pseudodifferential
	symbols, {\em Comm. Math. Phys.} 171, no. 3 (1995), 475--530.
	 
	\bibitem[KN63-69]{KN} Kobayashi, S.; Nomizu,  K.; \textit{Fundations of differential geometry}; {\bf 1,2} , Wiley interscience (1963-1969)
	\bibitem[Ko1993]{K1993} 
	Koczkodaj, W.W., A new definition of consistency of pairwise comparisons, {\it Mathematical and Computer Modelling} {\bf 7} 79-84 {(1993)} 
	
	
	\bibitem[KKL2014]{KKL2014}
	Koczkodaj, W.W.; Kulakowski, K.; Ligeza, A.,
	On the quality evaluation of scientific entities in Poland Supported by consistency-driven pairwise comparisons method, \textit{Scientometrics,} {\bf 99} no3 911-926 (2014)	
	
	\bibitem[KoSza2010]{KoSza2010} Koczkodaj, W.W.; Szarek, S.; On distance-based inconsistency reduction algorithms for pairwise comparisons \textit{Logic J. of the IGPL} \textbf{18} no 6 859-869 (2010)
	\bibitem[KoSzw2014]{KS2014} 
	Koczkodaj, W.W.; Szwarc, R.;
	Axiomatization of Inconsistency Indicators for Pairwise Comparisons, Fundamenta Informaticae, 132(4): 485-500, 2014.
	\bibitem[KoSzy2015]{KS2015} Koczkodaj, W.W.; Szybowski, J.; Axiomatization of Inconsistency Indicators for Pairwise Comparisons Matrices Revisited , \texttt{arXiv:1509.03781}
	\bibitem[KMS1993]{KMS} Kolar, I.; Michor, P.W.; Slovak, J.; \textit{Natural operations in differential geometry}; Springer (1993)
	
	\bibitem[KV1994]{KV1} Kontsevich, M.; Vishik, S.;
	{ Determinants of elliptic pseudo-differential operators} Max
	Plank Institut fur Mathematik, Bonn, Germany, preprint n. 94-30
	(1994)
	
	\bibitem[KV1995]{KV2}  Kontsevich, M.; Vishik, S.; Geometry of determinants of elliptic operators.
	Functional analysis on the eve of the 21st century, Vol. 1 (New Brunswick, NJ, 1993), 
	\textit{Progr. Math.} \textbf{131},173-197
	(1995)
	
	
	
	\bibitem[KK1991]{KK} Kravchenko, O.S.; Khesin, B.A.; A central extension of the algebra of pseudo-differential 
	symbols \textit{Funct. Anal. Appl.} \textbf{25} 152-154 (1991) 
	
	
	\bibitem[KM2000]{KM} Kriegl, A.; Michor, P.W.; \textit{The convenient setting
		for global analysis} Math. surveys and monographs \textbf{53}, American
	Mathematical society, Providence, USA. (2000)
	
	
	\bibitem[KMR2015]{KMR}  Kriegl, A.;  Michor, P. W.; Rainer, A.; An exotic zoo of diffeomorphism groups on $\R^n.$ \textbf{Ann. Glob. Anal. Geom.} \textbf{ 47}, 2 (2015), 179-222
	
	\bibitem[Ku1998]{Ku}  Kubo, F.; Non-commutative Poisson algebra structures on affine Kac-Moody algebras. {\em J. Pure and Applied Algebra} 126 (1998),
	267--286.
	\bibitem[Kub2020]{Kub2020} Kubrusly, C. S.;
	{\it Spectral theory of bounded linear operators.} Birkhauser 249 p. (2020).
	\bibitem[K1965]{Kui} Kuiper, N.; The homotopy type of the unitary group of  Hilbert spaces \textit{Topology} \textbf{3} 19-30 (1965)  
	\bibitem[Ku2000]{Ku2000} Kuperschmidt, B.A.; \textit{KP or mKP: Mathematics of Lagrangian, Hamiltonian and Integrable Systems.} Math. Surv. Monographs \textbf{78} (2000)
	\bibitem[LLA2012]{LLA2012} Lahby, M.; Leghris, C.; Adib, A.; Network Selection Decision based on handover history
	in Heterogeneous Wireless Networks; \textit{International 
		Journal of 
		Computer 
		Science and 
		Telecommunications} {\bf 3} no 2,  (2012) 21-25
	\bibitem[Lan1985]{Lang} Lang, S.; {\it Differentiable manifolds} 2nd edition, Springer-Verlag, New-York (1985)
	
	\bibitem[Lax1968]{Lax} Lax, P.; Integrals of nonlinear equations of evolution and solitary waves.
	 \textit{Comm. Pure Appl. Math.} 
	 \textbf{21} no5 467-490 (1968)
	
	\bibitem[dLS2009]{dLS1}C. De Lellis, L. Sz\'ekelyhidi Jr.;
	The Euler equations as a differential inclusion.
	\textit{Ann. of Math. (2)} {\bf 170} no. 3, 1417–1436 (2009)
	
	\bibitem[dLS2010]{dLS2}C. De Lellis, L. Sz\'ekelyhidi Jr.;
	On admissibility criteria for weak solutions of the Euler equations.
	{\it Arch. Ration. Mech. Anal.} {\bf 195}  no. 1, 225–260 (2010)
	
	
	\bibitem[dLS2012]{dLS3} C. De Lellis, L. Sz\'ekelyhidi Jr.;
	The h-principle and the equations of fluid dynamics.
	{\it Bull. Amer. Math. Soc.} {\bf 49}, 347-375 (2012)
	
	\bibitem[dLS2021]{dLS2021}  C. De Lellis, L. Sz\'ekelyhidi Jr.; Weak stability and closure in turbulence \texttt{arXiv:2108.01597}
	
	\bibitem[Le1998]{Le} Lesch, M.; On the non commutative residue for pseudo-differential operators 
	with log-polyhomogeneous symbol \textit{Ann. Glob. Anal. Geom.} \textbf{17} 151-187 (1998)
	\bibitem[LP2007]{PL2007} Lescure, Jean-Marie; Paycha, Sylvie
	Uniqueness of multiplicative determinants on elliptic pseudodifferential operators. 
	{\it Proc. Lond. Math. Soc. (3)} {\bf 94}  No. 3, 772-812 (2007)
	\bibitem[Lesl2003]{Les} Leslie, J.; On a Diffeological Group Realization of
	certain Generalized symmetrizable Kac-Moody Lie Algebras \textit{J.
		Lie Theory} \textbf{13} (2003), 427-442	
	\bibitem[Lich1956]{Li} Lichnerowicz, A.  {\it Th\'eorie globale des connexions et des groupes d'holonomie}  ed. Cremonese, Roma (1956)
	\bibitem[Marsch1998]{Marsch1988} Marschall, J.; Pseudo-differential operators with coefficients in Sobolev spaces;
	{\em Trans. AMS } \textbf{307}, no1 (1988) 335-361.
	\bibitem[MRT2014]{MRT} Maeda, Y.; Rosenberg, S. Torres-Ardila, F. The geometry of loop spaces II: characteristic classes \texttt{arXiv:1407.2491} (2015 for v3)
	\bibitem[Maj1995]{Maj} Majid, S.; {\it Foundations of quantum groups} Cambridge University Press (1995)
	\bibitem[Mal1972]{Mal} Malliavin, P.; {\it G\'eom\'etrie diff\'erentielle intrins\`eque} Hermann (1972) 
	\bibitem[MaRob1995]{MR1995} Marion, J.; Robart, T.; Regular Fr\'echet Lie groups of invertible elements in some inverse limits of unital involutive Banach algebras \textit{Georgian Math. J.} \textbf{2} no 4 425-444 (1995)
	\bibitem[MR1999]{MR} Marsden, J. and Ratiu, T.; ``Introduction to mechanics and symmetry.
	A basic exposition of classical mechanical systems. Second edition'' (1999).
	Texts in Applied Mathematics, 17. Springer-Verlag, New York.
	
	
	\bibitem[Mich1980]{Mich1980} Michor, P.; Manifolds of smooth maps III: The principal bundle of embeddings of a non compact smooth manifold. \textit{Cahiers Topologie Geometrie Differentielle} \textbf{21}  325-337 (1980)
	
	
	

	
	
	\bibitem[Mick1994]{Mick1994} Mickelsson, J.; Wodzicki residue and anomalies on current algebras \textit{ Integrable models and strings} A. Alekseev and al. eds. \textit{Lecture notes in Physics} \textbf{436}, Springer (1994)
	
	\bibitem[Mick1989]{Mickbook} Mickelsson, J.; {\it Current algebras and groups}
	Plenum monographs in non linear physics (1989)
	
	\bibitem[Mick1988]{Mick1988} Mickelsson, J.; Current algebra representation for the 3+1 dimensional Dirac-Yang-Mills theory \textit{Comm. Math. Phys.} \textbf{117} 261 (1988)
	
	\bibitem[MP2007]{MickPay2007} Mickelsson, J.; Paycha, S.; Renormalized Chern-Weil forms associated with families of Dirac operators \textit{J. Geom. Phys.} \textbf{57}, no9, 1789-1814 (2007)
	
	\bibitem[MR1988]{MR1988} Mickelsson, J.; Rajeev, S.; Current algebras and Determinant bundles for infinite dimensional grassmannians \textit{Comm. Math. Phys.} \textbf{116} 365 (1988)
	\bibitem[Mil1984]{Mil} Milnor, J.; { Remarks on infinite dimensional Lie groups}; {\it Proc. Summer school on Quantum Gravity (Les Houches, 1983)}, ed. DeWitt, B., North-Holland, Amsterdam, (1984) 1008-1056
	
	\bibitem[Mil1963]{Mil1963} Milnor, J.; \textit{Morse theory} 
	Based on lecture notes by M. Spivak and R. Wells.; 
	Annals of Mathematics Studies. \textbf{51}; Princeton University Press (1963)
	
	\bibitem[Mil1956-1]{milnor1}
	Milnor, J.; Construction of univeral bundles, I \emph{Ann.\ of Math.\ (2)}, \textbf{63} (1956), 272--284.
	
	\bibitem[Mil1956-2]{milnor2}
	John Milnor, ``Construction of universal bundles, II'', \emph{Ann.\ of Math.\ (2)}, \textbf{63} (1956), 430--436.
	
	
\bibitem[Mis1997]{Mis1997}	Misiołek, G.;
	The exponential map on the free loop space is Fredholm. 
	\textit{Geom. Funct. Anal.} \textbf{7}, No. 5, 954-969 (1997). 
\bibitem[Mis1999]{Mis1999}	Misiolek, G.;
	Exponential maps of Sobolev metrics on loop groups. 
	\textit{Proc. Am. Math. Soc.} {\bf 127}, No. 8, 2475-2482 (1999)
	
	. 
	\bibitem[MP2009]{MP2009} Misiołek, G.; Preston, S. C.;
	Fredholm properties of Riemannian exponential maps on diffeomorphism groups. 
	\textit{Invent. Math.} {\bf 179}, No. 1, 191-227 (2010). 
	\bibitem[Mo2008]{Mo} Molitor, M.;  	
	La Grassmannienne Non-lin\'eaire comme Vari\'et\'e Fr\'ech\'etique Homog\`ene;
	\textit{J. Lie Theory} \textbf{18} no. 3, 523--539 (2008)
	\bibitem[MS2016]{MoSm2016} Montaldi, J.; Smolyanov, O.G; Feynman path integrals and Lebesgue-Feynman measures Dokl. Math. 96, No. 1, 368-372 (2017); translation from Dokl. Akad. Nauk, Ross. Akad. Nauk 475, No. 5, 490-495 (2016). 
	\bibitem[Mu1984]{M1} Mulase, M.; {Complete integrability of the Kadomtsev-Petviashvili equation} \textit{Adv. Math.} {\bf 54} (1984) 57-66 
	
	\bibitem[Mu1988]{M2} Mulase, M.; { Solvability of the super KP equation and a generalization of the Birkhoff decomposition} \textit{Invent. Math.} {\bf 92} (1988), 1-46 
	
	\bibitem[Mu1983]{M3} Mulase, M.; Geometry of soliton equations; MSRI preprint (1983)
	
	
	
	\bibitem[Nee2007]{Neeb2007} Neeb, K-H.; Towards a Lie theory of locally convex groups \textit{Japanes J. Math.} \textbf{1} (2006), 291-468
	
	
	
	
	\bibitem[Olv1993]{Olv} Olver, P.J.; \textit{Applications of Lie groups to differential equations (2nd edition)} GTM \textbf{107}, Springer (1993)
	
	\bibitem[Om1997]{Om} Omori, H.; \textit{Infinite dimensional Lie groups} AMS translations of mathematical monographs \textbf{158} (1997)
	
	\bibitem[Om1970]{Om2} Omori, H.; On the group of diffeomorphisms on a compact manifold; 1970 Global Analysis (Proc. Sympos. Pure Math., Vol. XV, Berkeley, Calif., 1968) 167-183 Amer. Math. Soc., Providence, R.I.
	\bibitem[Om1981]{Om3} Omori, H.; A remark on non-enlargable Lie algebras \textit{J. Math. Soc. Japan} \textbf{33} no4 707-710 (1981)
	\bibitem[Om2007]{Om2007} Omori, H.; Toward geometric Quantum theory \textit{Progress in Math} \textbf{252} 213-252 (2007)
	\bibitem[OmMMY2007]{OmMMY2007} Omori, H.; Maeda, Y.; Myazaki, N. Yoshioka, A.; Geometric objects in an approach to quantum geometry \textit{Progress in Math} \textbf{252} 303-324 (2007)
	\bibitem[OMY1]{OMY1} Omori, H.; Maeda, Y.; Yoshioka, A.; On regular Fr\'echet Lie groups, I; Some differential geometric expressions of Fourier integral operators on a Riemannian manifold \textit{Tokyo J. Math.} \textbf{3}, no 2, 353-390 (1980)
	
	\bibitem[OMY1994]{OMY1994} Omori, H; Maeda, Y; Yoshioka, A.; A Poincar\'e-birkhoff-Witt theorem for infinite dimensional Lie algebras
	\textit{J. Math. Soc. Japan} \textbf{46} no1 25-50 (1994)
	\bibitem[OMY2]{OMY2} Omori, H; Maeda, Y; Yoshioka, A.; On regular Fr\'echet Lie groups II; Composition rules of Fourier integral operatorson a Riemannian manifold \textit{Tokyo J. Math.} \textbf{4}, no 2, 221-253 (1981)
	
	
	\bibitem[OMYK3]{OMYK3} Omori, H; Maeda, Y; Yoshioka, A.; Kobayashi, O.; On regular Fr\'echet Lie groups III; A second cohomology class related to theLie algebra of pseudo-differential oprators of order 1 \textit{Tokyo J. Math.} \textbf{4}, no 2, 255-277 (1981)
	
	\bibitem[OMYK4]{OMYK4} Omori, H; Maeda, Y; Yoshioka, A.; Kobayashi, O.; On regular Fr\'echet Lie groups IV; Definition and fundamentaltheorems \textit{Tokyo J. Math.} \textbf{5}, no 2, 365-398 (1981)
	
	\bibitem[OMYK5]{OMYK5} Omori, H; Maeda, Y; Yoshioka, A.; Kobayashi, O.; On regular Fr\'echet Lie groups V; Several basic properties \textit{Tokyo J. Math.} \textbf{6}, no 1, 39-64 (1983)
	
	\bibitem[OMYK6]{OMYK6} Omori, H; Maeda, Y; Yoshioka, A.; Kobayashi, O.; Infinite dimensional Lie groups that appear in general relativity \textit{Tokyo J. Math.} \textbf{6}, no 2, 217-246 (1983)
	
	\bibitem[OMYK7]{OMYK7} Omori, H; Maeda, Y; Yoshioka, A.; Kobayashi, O.; On regular Fr\'echet Lie groups VII; The group generated by pseudo-differential operators of negative order \textit{Tokyo J. Math.} \textbf{7}, no 2, 315-336 (1984)
	
	\bibitem[OMYK8]{OMYK8} Omori, H; Maeda, Y; Yoshioka, A.; Kobayashi, O.; Primordial operators and Fourier integral operators  \textit{Tokyo J. Math.} \textbf{8}, no 1, 1-47 (1985)
	
	
	
	
	\bibitem[Pat1988]{Pa} Paterson, A.T.; \textit{Amenability} Math. surveys and Monographs
	\textbf{29}, Amer. Math. Soc., Provodence, R.I. (1988) 
	
	
	\bibitem[Pay2001]{Pay} Paycha, S.;
	{Renormalized traces as a looking glass into infinite dimensional
		geometry} \textit{Infin. Dimens. Anal. Quantum Probab. Relat.
		Top.} {\bf 4},  no2,  221-266 (2001)
	
	\bibitem[PR2004]{PR2004} Paycha, S.; Rosenberg, S.; Traces and Characteristic classes on loop spaces in \textit{IRMA lectures in mathematics and theorical physics} ed. T. Wurzbacher (2004)
	
	. 
	\bibitem[Pay2012]{PayBook} Paycha, S; 
	\textit{Regularised integrals, sums and traces. An analytic point of view.}
	University Lecture Series \textbf{59}, AMS (2012).
	\bibitem[Pay2013]{Pay2013} Paycha, S.; Paths towards an extension of Chern-Weil calculus to a class of infinite 
	dimensional vector bundles. {\it Geometric and topological methods for quantum field theory}, 81–143, 
	Cambridge Univ. Press, Cambridge, (2013)
	
	\bibitem[Pes2006]{Pes} Pestov, V.; \textit{Dynamics of Infinite-Dimensional Groups : the
		Ramsey-Dvoretzky-Milman phenomenon}  University Lecture
	Series \textbf{40},Amer. Math. Soc. (2006)
	
	
	\bibitem[Pen1970]{Pen}  Penot, J.-P.; {Sur le th\'eor\`eme de Frobenius }; {\it Bull. Soc. math. France } \textbf{98} (1970),
	47-80 
	\bibitem[Per1990]{P} Perelomov, A.M.; ``Integrable systems of classical mechanics and Lie algebras'' (1990)
	Birkh\"{a}user Verlag, Berlin.
	
		
		
		
	\bibitem[PS2021]{PS2021}	Pitcho, J.; Sorella, M.; Almost everywhere non-uniqueness of integral curves for divergence-free Sobolev vector fields. \texttt{arXiv:2108.03194}
	\bibitem[PS1988]{PS} Pressley, A.; Segal, G.; {\it Loop Groups} Oxford Univ. Press (1988)
	\bibitem[Prok1956]{Prok} Prokhorov, Yu.V.; Convergence of random	processes and limit	theorems in	probability theory. \textit{Theor. Prob. Appl.} \textbf{1},\'ez\'e\'e
	157-214
	(1956)
	\bibitem[Rad1991]{Rad} O.A.Radul; {Lie albegras of differential operators, their central extensions, and W-algebras} \textit{Funct. Anal. Appl.} \textbf{25}, 25-39 (1991)
	
	
	\bibitem[RatSc1981]{RaS} Ratiu, T.; Schmid, R.; The differentiable structure of three remarkable diffeomorphism groups; \textit{Math Z.} \textbf{177}, 81-100 (1981) 
	\bibitem[RS1981]{RS} Reyman, A.G.; Semenov-Tian-Shansky, M.A.; Reduction of Hamiltonian Systems,
	Affine Lie Algebras and Lax Equations II. {\em Invent. math.} 63 (1981), 423--432.
	\bibitem[Re1997]{Re1997} Reinhardt, H.; Yang-Mills in axial gauge Phys.Rev. D \textbf{55}  2331-2346 (1997)
	\bibitem[RSF1985]{RSF1985} Reyman, A.; Semenov-Tyan-Shanskii, M.A.; Faddeev, L.; Quantum anomalies and cocycles on gauge groups \textit{Funct. Anal. Appl. } \textbf{18} 319 (1985)
	
	\bibitem[Rob1997]{Rob} Robart, T.; Sur l'int\'egrabilit\'e des sous-alg\`ebres de
	Lie en dimension infinie; \textit{Can. J. Math.} \textbf{49} (4) (1997),
	820-839
	\bibitem[RV2014]{RV2014} Rovelli, C. Vidotto, F. {\it Covariant Quantum Loop Gravity} Cambridge University Press (2014)
	\bibitem[Saa1977]{S1977} Saaty, T.; A scaling methods for priorities in hierarchical structures; \textit{J. Math. Psychol.} \textbf{15} (1977) 234-281
	\bibitem[SGO2021]{SGO2021} di Salvo, R.; Gorgone, M.; Oliveri, F.; A consistent approach to approximate symmetries \textit{Nonlinear Dyn.} {\bf 91}, 371--386 (2018)
	\bibitem[SavSt2013]{SS} Savin, A.Yu.; Sternin, B.Yu.
	Uniformization of nonlocal elliptic operators and KK-theory;
	\textit{Russ. J. Math. Phys.} \textbf{20}, no. 3, 345-359 (2013).
	\bibitem[Sche1993]{Sch1993}  Scheffer, V.;  An inviscid flow with compact support in space-time. {\it J.	Geom. Anal.} {\bf 3} no 4 , 343–401 (1993)
	
	\bibitem[Shn1997]{Sh1} Shnilerman, A.; On the nonuniqueness of weak solution of the Euler
	equation. \textit{Comm. Pure Appl. Math.} {\bf 50} no 12, 1261–1286 (1997)
	
	\bibitem[Shn2000]{Sh2} Shnilerman, A.; Weak solutions with decreasing energy of incompressible
	Euler equations. \textit{Comm. Math. Phys.} \textbf{210} no 3, 541–603 (2000)
	\bibitem[Sch1959]{Sch}  Schwinger, J.; Field theory of commutators; 
	\textit{Phys. Rev. Lett.} \textbf{3}, 296-297 (1959)
	
	\bibitem[Sco2010]{Scott} Scott, S.;
	\textit{Traces and determinants of pseudodifferential operators}; 
	OUP (2010)
	
	\bibitem[SchW1999]{Sc} Schaefer, H. H.; Wolff, M. P.; {\it Topological vector spaces} Second edition. Graduate Texts in Mathematics {\bf 3} Springer-Verlag, New York  (1999)
	
	
	\bibitem[See1967]{See} Seeley,  R.T.; {Complex powers of an elliptic operator}; {\it Proc. Symp. Pure Math.} {\bf 10} (1967) 288-307 
	
	\bibitem[SW1985]{SW1985} Segal, G.; Wilson, G.; Loop groups and equations of KdV type \textit{Publ. Math. IHES} \textbf{61} 5 (1985)
	\bibitem[SSSA2000]{SSSA} Sen, S.; Sen, S.; Sexton, J.C.; Adams, D.H.; A geometric discretisation scheme applied to the Abelian Chern-Simons theory \textit{Phys.Rev. E} \textbf{61}  3174-3185 (2000)
	
	
	
	\bibitem[Sou1985]{Sou} Souriau, J.M.; Un algorithme g\'en\'erateur de structures quantiques; 
	\textit{Ast\'erisque}, Hors S\'erie, (1985) 341-399 
	
	
	\bibitem[T2006]{T2006} Terng, Chuu-Lian
	Applications of loop group factorization to geometric soliton equations. 
	\textit{Sanz-Sol\'e, Marta (ed.) et al., Proceedings of the international congress of mathematicians (ICM), Madrid, Spain, August 22–30, 2006. Volume II: Invited lectures.} Zürich: European Mathematical Society (EMS)  927-950 (2006). 
	
	\bibitem[Ue1985]{Ue} Ueno, K.; Analytic and algebraic aspects of the Kadomtsev-Petviashvili hierarchy from the viewpoint of the universal Grassmann manifold; \textit{ Infinite Dimensiona groups with applications, MSRI publications} \textbf{4}, Springer (1985), 335-353
	
	\bibitem[Vas1978]{Vas1978} Vassiliou, E, On the infinite dimensional holonomy theorem, {\it Bull. Soc. Roy. Sc. Li\`ege}, {\bf 9-10}, 223-228, (1978)
	
	\bibitem[Vil2006]{Vil2006} Villani, C.; \textit{Optimal transport, old and new} Springer (2006) 
	
	\bibitem[Vil2008]{Vil} Villani, C.; Paradoxe de Scheffer-Shnirelman revu sous l’angle de l’int\'egration convexe (d’après C. De Lellis, L. Sz\'ekelyhidi). {\it S\'eminaire Bourbaki} Exp. 1001 (November 2008)
	
	\bibitem[Vin2013]{Vin2013} Vinogradov, A.; What are symmetries of nonlinear PDEs and what are they themselves? \texttt{arXiv:1308.5861}
	
	\bibitem[Wan1983]{W1} Watanabe, Y.; Hamiltonian structure of Sato's hierarchy of KP equations and a
	coadjoint orbit of a certain formal Lie group. {\em Lett. Math. Phys.} 7 (1983),
	99--106.
	\bibitem[Wan1984]{W2} Watanabe, Y.; Hamiltonian structure of M. Sato's hierarchy of
	Kadomtsev--Petviashvili equation. {\em Ann. Mat. Pura Appl.} (4) 136 (1984), 77--93.
	\bibitem[Wa2012]{Wa} Watts, J.; \textit{Diffeologies, differentiable spaces
		and symplectic geometry} PhD thesis, university of Toronto (2012) arXiv:1208.3634
	
	\bibitem[We2021]{We2017} Welker, K.; Suitable Spaces for Shape Optimization. \textit{Appl. Math. Optim.} {\bf 84}, Suppl. 1, S869-S902 (2021)
		\bibitem[Wh1957]{Wh} 
		Whitney, H., Geometric Integration Theory, Princeton University Press, Princeton, NJ, 1957.

	\bibitem[Wid1980]{Wid} Widom, H.; { A complete symbolic calculus for pseudo-differential operators}; {\it Bull. Sc. Math. 2e serie} \textbf{104} (1980) 19-63 
	
	\bibitem[Wod1984]{W} Wodzicki, M.; {Local invariants in spectral asymmetry}
	\textit{Inv. Math.} {\bf 75}, 143-178  (1984)
	
	
	\bibitem[Wur1995]{Wu} Wurzbacher, T.;
	{Symplectic geometry of the loop space of a Riemannian manifold}
	\textit{Journal of Geometry and Physics} {\bf 16}, 345-384  (1995)
	
	\bibitem[Y1999]{Y1999} Young, K.; Foreign exchange markets as a lattice gauge theory, {\it Am. J. Phys.} {\bf 67} no 10
	862–868 (1999)
	
\end{thebibliography}
\end{document}